\documentclass[reqno]{amsart}
\usepackage{graphicx,amsmath,amsaddr}
\usepackage{float}
\usepackage[utf8]{inputenc}
\usepackage{color}
\usepackage{cite}
\usepackage{fancyhdr}
\usepackage[margin=2cm]{geometry}
\usepackage{appendix}
\usepackage{multirow}
\usepackage{hyperref}
\usepackage{verbatimbox} 
\usepackage{accents}
\usepackage{multicol}
\usepackage{etoolbox}

\usepackage{tikz}
\def\checkmark{\tikz\fill[scale=0.4](0,.35) -- (.25,0) -- (1,.7) -- (.25,.15) -- cycle;} 

\usepackage{pifont}
\newcommand{\xmark}{\ding{55}}%

\makeatother

\let\oldtocsection=\tocsection

\let\oldtocsubsection=\tocsubsection

\let\oldtocsubsubsection=\tocsubsubsection

\renewcommand{\tocsection}[2]{\hspace{0em}\oldtocsection{#1}{#2}}
\renewcommand{\tocsubsection}[2]{\hspace{1em}\oldtocsubsection{#1}{#2}}
\renewcommand{\tocsubsubsection}[2]{\hspace{2em}\oldtocsubsubsection{#1}{#2}}

\makeatletter
\def\@tocline#1#2#3#4#5#6#7{\relax
  \ifnum #1>\c@tocdepth 
  \else
    \par \addpenalty\@secpenalty\addvspace{#2}%
    \begingroup \hyphenpenalty\@M
    \@ifempty{#4}{%
      \@tempdima\csname r@tocindent\number#1\endcsname\relax
    }{%
      \@tempdima#4\relax
    }%
    \parindent\z@ \leftskip#3\relax \advance\leftskip\@tempdima\relax
    \rightskip\@pnumwidth plus4em \parfillskip-\@pnumwidth
    #5\leavevmode\hskip-\@tempdima #6\nobreak\relax
    \ifnum#1<0\hfill\else\dotfill\fi\hbox to\@pnumwidth{\@tocpagenum{#7}}\par
    \nobreak
    \endgroup
  \fi}
\makeatother

\newcommand{\dotr}[1]{%
#1\accentset{\mbox{\bfseries .}}{\vphantom{#1}}}

\newenvironment{changemargin}[2]{%
	\begin{list}{}{%
			\setlength{\topsep}{0pt}%
			\setlength{\leftmargin}{#1}%
			\setlength{\rightmargin}{#2}%
			\setlength{\listparindent}{\parindent}%
			\setlength{\itemindent}{\parindent}%
			\setlength{\parsep}{\parskip}%
		}%
		\item[]}{\end{list}}

\title{Heat equations beyond Fourier: from heat waves to thermal metamaterials}

\author{R. Kovács}
\address{
$^1$Department of Energy Engineering, Faculty of Mechanical Engineering, Budapest University of Technology and Economics, Műegyetem rkp. 3., H-1111 Budapest, Hungary \\
$^2$Department of Theoretical Physics, Wigner Research Centre for Physics, Institute for Particle and Nuclear Physics, Budapest, Hungary \\
$^3$Montavid Thermodynamic Research Group, Budapest, Hungary
}
\date{\today}

\begin{document}

\begin{changemargin}{-0.5cm}{-0.5cm}
\maketitle
\begin{abstract}
In the past decades, numerous heat conduction models beyond Fourier have been developed to account for the large gradients, fast phenomena, wave propagation, or heterogeneous material structure, such as being typical for biological systems, superlattices, or thermal metamaterials. It became a challenge to orient among the models, mainly due to their various thermodynamic backgrounds and possible compatibility issues. Additionally, in light of the recent findings on the field of non-Fourier heat conduction, it is not even straightforward how to interpret, solve and then utilize a particular non-Fourier heat equation, primarily when one aims to thermally design the material structure to construct the new generation of thermal metamaterials. Adding that numerous modeling strategies can be found in the literature accompanying different interpretations even for the same heat equation makes it even more difficult to orient ourselves and find a comprehensive picture of this field of research.

Therefore, this review aims to ease the orientation among advanced heat equations beyond Fourier by discussing properties concerning their possible practical applications in light of experiments. The observed phenomena and the need to model them act as a guiding principle throughout this paper. We start from the simplest model with basic principles and notions, then proceed toward the more complex, coupled phenomena such as ballistic heat conduction.
We do not involve the often complicated technical details of each thermodynamic framework and do not aim to compare each approach from a methodological point of view. Despite that, we still briefly present the models' background to highlight their origin, the limitations acting on the models, and the corresponding stability conditions, if any.
Additionally, the field of non-Fourier heat conduction has become quite segmented, and that paper also aims to provide a common ground, a comprehensive mutual understanding of the basics of each model, and what phenomenon they can be applied to.
\end{abstract}

\begin{multicols}{2}
	\small
\tableofcontents
\end{multicols}
\end{changemargin}

\section{Introduction}
Various thermodynamic approaches have appeared in the past decades, and each has begun to develop its methodology to generalize the classical transport laws. Here, we mainly focus on the Fourier law of heat conduction and its extensions, but we also include the fluid equations when necessary. Experimental findings induce this research and stand as motivations to find proper extensions for Fourier's law, including further time or space derivatives, size dependence, or a particular nonlinearity of state variables.
For instance, the low-temperature observations are the so-called second sound and ballistic heat propagation modes \cite{McNEta70a, JosPre89, DreStr93a}. In a room temperature environment, the behavior of rarefied gases, heterogeneous (e.g., porous materials, foams) \cite{Jiang03, VirtoEtal09, GeigerEmmanuel10, Botetal16}, one-, and two-dimensional materials \cite{FujiEtal05, ShiMar06, GuEtal18} in micro and nanoscale \cite{Chen21} are typical examples. Furthermore, the so-called piston effect, the sudden thermal expansion of the boundary layer in a fluid, also appears to be a challenging task from both theoretical and numerical points of view \cite{ZappEtal90, ZappEtal15b}. Therefore, the potential for practical applications is wide such as advanced electronic devices \cite{VerMey08, ChenZhan19}, manufacturing technologies \cite{OaneEtal21}, biological systems \cite{XuSeffen08, NazmEtal21}, and heat foam-based heat exchangers and thermal storage technologies \cite{ChenEtal21, NematEtal22, ZhangEtal22}.

The increasing number of models makes the field of non-Fourier heat conduction more diverse and more difficult to overview, especially for someone who wants to find a proper model for a particular task without trying to understand multiple thermodynamic concepts, their differences, but more importantly, wants to see their interpretation and limitations immediately. While the review of Joseph and Preziosi \cite{JosPre89} summarized the most important findings about heat waves and collected state-of-the-art knowledge about heat equations, the results of the previous 30 years showed considerable novel possibilities of how we can think about non-Fourier equations. Additionally, as the generalized models are not yet standard, they cannot be solved easily with commercial software since the conventional techniques are not necessarily working in the same way as used for Fourier's law. The initial and boundary conditions, the various nonlinearities, and their analytical and numerical treatment are not straightforward. With the present paper, we aim to provide a general insight into the essential attributes for the most frequently used non-Fourier heat equations, a systematic guide that can be helpful for anyone who encounters such advanced heat conduction problems. While we intend to provide a comprehensive overview of each topic as much as possible, inevitably, we cannot sink into the details everywhere. Therefore we will also refer to some additional reviews or books wherever appropriate.

Furthermore, non-Fourier modeling does not end at heat waves, it is much more than that. The recent experimental findings promote the possibility of how to adjust the heat conduction properties to achieve the optimal setting, i.e., developing a new class of thermal metamaterials. At the same time, the modeling background remains efficient, reliable, and environmentally friendly due to the much less computational demand. There is substantial untapped potential in non-Fourier equations, and the present review also aims to ease the understanding and open future discussions in this respect. In the following, before immersing into the world of heat equations, we briefly visit some essential aspects to lay the foundations of common notions and mutual understanding.

\subsection{The role of the second law.} In all thermodynamic approaches, the second law stands as a central theorem, which can be formulated accurately as a balance equation of the entropy density for continuum models \cite{Gyarmati70b}:
\begin{align}
	\rho \dot s + \nabla \cdot \mathbf J_s = \sigma_s \geq 0, \label{slaw}
\end{align}
in which $\rho$ is the mass density, $\mathbf J_s$ denotes the current density of entropy, and $\sigma_s$ is the positive semidefinite entropy production. Furthermore, the upper dot presents the material time derivative, and $\nabla \cdot$ is the divergence. In the kinetic theories, Eq.~\eqref{slaw} usually referred to as $h$-theorem \cite{Cercignani82, Kaniadakis01}, where $h=-s$, i.e., while $s$ is a concave potential function of the state variables, $h$ is a convex potential. In both situations, the entropy inequality \eqref{slaw} is exploited as a constraint in order to derive constitutive equations. Models with proper thermodynamic background, i.e., compatibility with the second law of thermodynamics, have asymptotically stable equilibrium solutions. This is what we call thermodynamic compatibility from here on. Not all heat equations satisfy this property and thus must be supplemented with additional conditions to remain in the physically admissible region. One representative example is related to the popular dual-phase-lag (DPL) equation. For further discussion on the second law of thermodynamics and its role in different fields, we refer to \cite{LiebYng99, Mato04b, CapShee05b, Van15a}.

The continuum thermodynamic approaches mainly differ in constructing the state space and the entropy current. This is most visible between Classical Irreversible Thermodynamics (CIT) \cite{GrooMaz63non} and Extended Irreversible Thermodynamics (EIT) \cite{Lebon89, JouEtal10b}. Therefore, after presenting the basic concept of the example of the Fourier equation, we will continue with the non-Fourier models by briefly discussing their origin in this regard. In the case of a kinetic theory-based Rational Extended Thermodynamics (RET) \cite{MulRug98, RugSug15}, the model construction strongly depends on the particular situation, including numerous prior assumptions on the heat conduction mechanisms; thus, the situation becomes more complicated. Wherever it is necessary, we will briefly discuss them. From that point of view, GENERIC (General Equation for Non-Equilibrium Reversible–Irreversible Coupling) \cite{GrmOtt97, OttGrm97, Ott05b, Grmela2018b, PavEtal18b} is unique due to its construction as dividing the models into reversible and irreversible parts. That decision adds further aspects, particularly in obtaining numerical solutions. Furthermore, we must discuss shortly how the non-equilibrium thermodynamic background with internal variable (NET-IV) \cite{Verhas96, Verhas97, VanEtal08, BerVan17b} approach fits the extended heat conduction models, primarily since NET-IV provides the most freedom for the models either about their interpretation or their utilization.
However, the present review does not aim to provide a detailed comparison between various thermodynamic approaches. On the contrary, we want to overview what assumptions are needed and how these assumptions influence the understanding and the applicability of each model. We must remember that all thermodynamic approaches are models, consequently, they do not valid for all situations, and thus it is only a matter of choice what to use for a particular problem. We feel it essential to present the corresponding advantages and disadvantages. Even the DPL \cite{Tzou95} or a thermomass \cite{Guo2010general, SellCimm15} model can be helpful in spite of their mathematical and physical issues if one clearly knows their limitations and correctly interprets the solutions in the given framework. 
Throughout this paper, we want to draw attention to the crucial aspects and clarify the properties of non-Fourier heat equations.

\subsection{Different levels of modeling.} This is not considered a classification of heat equations but the part of the decision procedure which model we need to solve a specific problem. All heat conduction models can either be explanatory or descriptive depending on what physical setting we start from. For instance, a kinetic theory-based approach needs a detailed description of the transport mechanisms. Therefore it could provide additional insight into the transport process, but in parallel, it also restricts the modeling possibilities. It remains a model, a simplified mathematical picture of what one observes in experiments. Nevertheless, the level of detail differs between the thermodynamic approaches, which could be a decisive property but not necessarily. A model with so many details could easily be computationally demanding. In this respect, the reader is referred to the excellent review of Bargmann et al.~\cite{BargEtal18} about various techniques to determine the representative volume element of a complex heterogeneous structure, which also aims to reduce the computational need by introducing a certain level of homogenization. Unfortunately, determining the effective thermal properties of solids, even with known microstructure, is only experimentally possible in most cases, especially when their temperature or pressure (stress) dependence comes into the relevant aspects. As an example that interests many researchers, for a biological system, the artery-vein geometry, the porosity of each type of tissue, the blood and tissue properties, and heat transfer modes contribute to the overall transport mechanism. However, every detail cannot be taken into account without the loss of generality, and also the modeling errors accumulate.

Although it would be very insightful if all these details were known and could be implemented into a heat transfer model, due to the numerous uncertainties in each factor, it seems to be a more difficult modeling decision, which restricts the validity as well; therefore it is natural that uncertainties remain. For heterogeneous, porous materials, the work of Vafai et al.~is remarkable, focusing on the diffusion, particle migration, and convection effects with averaged quantities in a stationary environment \cite{Vafai84, KhalVaf03, Vafai15b}. For transient situations, the uncertainties have a more substantial impact, and it becomes even more challenging to find proper representative averaging without losing the needed details.

We strongly agree with Bargmann et al. \cite{BargEtal18} that developing efficient homogenization techniques is of great importance. In many situations, especially in transient problems, a more viable option could be the use of an effective, `homogenized' model in the sense that substituting the Fourier law with a generalized one can be a possible choice for which the new parameters can effectively describe the overall thermal behavior. However, based on the study of Auriault \cite{Auriault91}, the substituted properties must occur on different spatial (or time) scales. We emphasize that in such a sense, one cannot interpret the observed temperature history as wave propagation of heat, as such a phenomenon would require a completely different physical setting. For instance, during the widely known experiment of Mitra et al.~\cite{MitEta95}, the temperature evolution of frozen meat is investigated. Since there was an apparent delay in the temperature history, they could fit a hyperbolic non-Fourier equation and interpret it as a first observation of heat waves on a macroscale at room temperature. No one could repeat that experiment, and it was refuted \cite{HerBec00, HerBec00b, Ant05meat}, the phase change caused that delay. This is an excellent example of effective modeling. It is a matter of choice what model we choose, but it must be clear that the excellent fit of a non-Fourier equation does not mean that one can interpret the data in the same way as for superfluid helium, for instance, and then having solid statements about factually observing wave propagation. Furthermore, an effective modeling strategy is also usual for the Fourier heat equation. Thermal conductivity is often used as an effective thermal parameter (and can only be measured in that way).
This attribute endows any heat equation with such universality that makes them applicable for various thermal problems and makes it more challenging to unify different frameworks and approaches as well.

While it seems relatively straightforward that the physical background is entirely distinct between heat waves and phase change, the non-Fourier heat conduction literature suffers from such misunderstandings, and thus it is essential to emphasize these aspects. Nonetheless, effective modeling can be a helpful tool, and this approach has already been tested for rocks, foams \cite{Vanetal17}, but this should not be the first option when Fourier's law seems inadequate at the first attempt, various heat sources can also ease the modeling (when appropriate), especially in biological systems \cite{SudEtal21}. However, in parallel, it opens new opportunities for how to characterize and design heterogeneous materials from a thermal point of view, and that could form an entirely new class for thermal metamaterials, heavily including 3D printing.

Depending on the particular approach, these factors can be realized in various ways. In the kinetic theory, one starts from the Boltzmann equation as the most in-depth level of modeling, and as it is usually too complicated, the problem is downsized, and a finite number of moments substitutes the original model and characterizes the level of modeling. This is an analogous procedure with the state space reduction in GENERIC \cite{PavEtal18b}.  
In a continuum framework, these aspects emerge when constructing the state space together with the corresponding Gibbs relation and balances. The state space extension with non-equilibrium variables introduces additional time evolution equations. The appearance of gradient terms leads to nonlocal, higher-order spatial derivatives. Altogether, these enable the modeling of complex phenomena on various temporal and spatial scales. 

\subsection{What non-Fourier phenomena do we know?} 
\begin{figure}[]
	\centering
	\includegraphics[width=15.5cm,height=8cm]{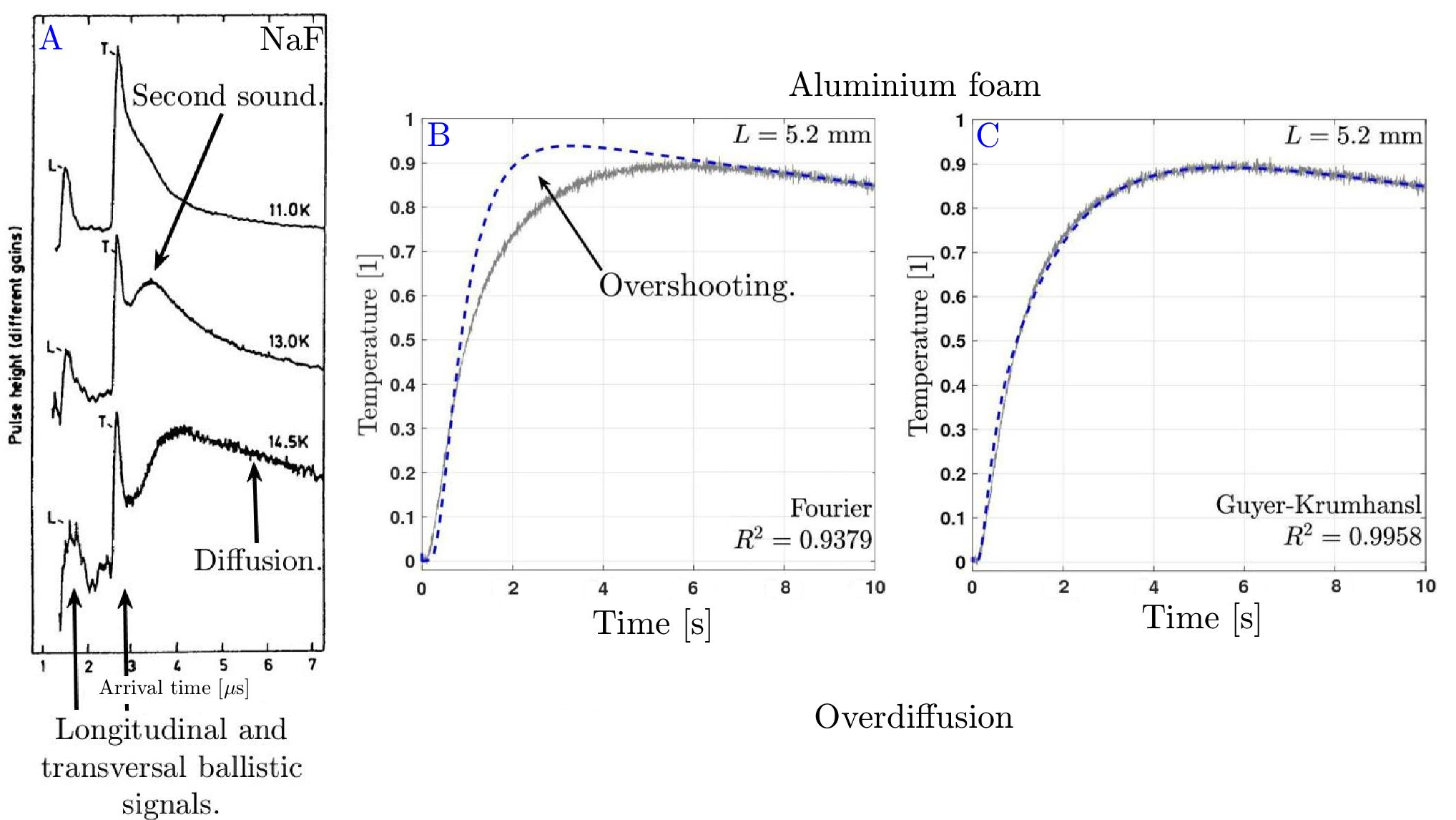}
	\caption{Measured rear side temperature history for a NaF sample at low-temperature (A) \cite{McN74t}, and for a metal foam sample at room temperature, showing the best Fourier (B) and Guyer-Krumhansl (C) fits \cite{FehKov21}.}
	\label{fig1}
\end{figure}

According to the best of our knowledge, there are four characteristic experimentally observed modes of heat conduction: diffusion, over-diffusion, second sound, and ballistic propagation, detailed below. Figure \ref{fig1} provides further insight into how these phenomena are observed experimentally. 

Diffusion is the classical, well-known heat conduction mode described by Fourier's law. It has limitations towards fast phenomena (in the scale of microseconds or even faster), small spatial scales (micrometers or even smaller), or heterogeneous material structures. 
In more detail, heterogeneous materials conduct heat through diffusion, mainly (convection and radiation can be present, too), but there inevitably are multiple heat transfer channels due to the complexity of the material structure. Fourier's law with an effective thermal conductivity can adequately characterize the material when the coupling of different heat transfer channels becomes negligible on large enough spatial and time scales, i.e., merely one time scale dominates the process. However, either for shorter periods or smaller samples, Fourier's law fails, the effect of parallel heat transfer channels is apparent, and the interaction of multiple heat transfer channels and their outcome is called over-diffusion. This phenomenon requires a model with at least two diffusion time scales, and there is no wave propagation in this case. Figure \ref{fig1} presents different situations. These temperature histories are obtained as an outcome of a standard heat pulse experiment \cite{ParEtal61, James80, Van16}, where one side of the sample is excited with a single heat pulse, and the temperature is recorded on the opposite side. On the left one (Fig.~\ref{fig1}/A), a typical low-temperature observation of heat waves is visible \cite{JacWalMcN70, McN74t, BeardoEtal21}. In the middle (Fig.~\ref{fig1}/B), we can observe the over-diffusion, the experimental appearance of the interaction of multiple heat transfer channels. The best Fourier fit is compared to the experimental data \cite{FehKov21}. Its remarkable characteristics are noteworthy: in the beginning, Fourier's prediction is slower, but at the top, the temperature overshoots the measured signal. That double-diffusive behavior becomes apparent only after performing a Fourier fit and comparing the fitted temperature history to the observed one. It indicates that at least two intrinsic heat transfer time scales are simultaneously present (besides cooling). Furthermore, Fig.~\ref{fig1}/C is an example of effective modeling, that complex heat transfer phenomenon can be modeled, e.g., with the Guyer-Krumhansl equation \cite{Vanetal17, FehKov21}. Later, we will provide a more detailed reason for how it is possible.

The second sound, however, clearly shows the wave nature of heat as well as the ballistic mode. Their propagation speeds differentiate between them. While the ballistic mode always propagates with the speed of sound, the second sound is slower. The second sound can be interpreted as a damped wave propagation of heat, first observed in liquid helium by Peshkov in 1944 \cite{Pesh44}, and also measured in solids with macroscopic size (about 5-8 mm) in a low-temperature environment, usually below 20 K.
The ballistic mode is discovered 20 years later by Jackson, Walker, and McNelly \cite{JacWalMcN70, JacWal71} in extremely pure crystals under low-temperature conditions. As it always propagates with the speed of sound, it refers to a mechanical coupling, so more appropriately, it is a thermo-mechanical phenomenon, at least from a continuum thermodynamic point of view. In phonon hydrodynamics, ballistic propagation means `propagation without interaction', only boundary scattering is present. This is analogous to rarefied gases, which produce observable ballistic effects in the low-pressure state. It is also reflected by the mathematical structure of the evolution equations; however, their experimental background is entirely different.

We must mention that the second sound and ballistic modes are also measurable on the nanoscale in a room-temperature environment due to the small spatial scales. Again, the kinetic picture can connect these situations through the Knudsen number, which is the ratio of the mean free path of phonons ($l$) and the characteristic length of the system ($L$) (Kn$=l/L$). There is also a critical length, proportional to the ratio of thermal diffusivity and speed of the heat wave, on which the wave propagation effects are notable \cite{Rubin92}. For a room temperature situation, this is usually below $1$ $\mu$m, but under low-temperature conditions, that could be much higher, even millimeters. 

\subsection{Model uniqueness.} When one encounters a heat conduction problem, it does not mean that there is only one approach, i.e., only one model would be capable of providing the necessary information and insight. In general, multiple models could be appropriate for the same task. One interesting example is related to ballistic heat conduction. The needed structure of evolution equations can be constructed in almost any thermodynamic approach, although with different assumptions and restrictions. For example, the DPL approach can provide the exact temperature representation (when only the temperature is used as the mere field variable) of the model as NET-IV or RET but cannot provide a more detailed structure for the state space as tensorial quantities are missing from the DPL model but present in the others. They differ in interpretation but still provide the same (or nearly identical) temperature history. This can sometimes be confusing, but this review will clarify these attributes.

\subsection{Parabolicity vs.~hyperbolicity.} According to the usual reasoning, the Fourier heat equation is parabolic, hence showing infinite propagation of temperature signal, i.e., if a sudden change of temperature is made at some point on the body, it will be felt instantly everywhere, though with exponentially small amplitudes at distant points \cite{JosPre89}. Although that property depends on the initial and boundary conditions and strictly speaking, it contradicts our physical feeling and does not forbid its use in practice \cite{Fich92}. Fourier's law still provides a reliable basis for most engineering problems. The appearance and fast spreading of modern manufacturing technologies (e.g., nanostructures) induce the research for a proper extension of Fourier's law, but parabolicity cannot be an exclusionary reason. As a glaring example, the outstanding result of Guyer and Krumhansl in the 1960s to find the so-called window condition that helped find the second sound in solids was based on a parabolic model. 
Furthermore, while hyperbolic equations describe finite wave speeds, neither relativistic nor non-relativistic theory can provide an upper bound for propagation speed. Therefore, strictly speaking, it still can be higher than the speed of light, violating causality. However, since the material parameters uniquely determine the propagation speed, the lack of an upper bound should not be a crucial shortcoming for realistic settings.
In the following, we do not investigate the space-time aspects of heat conduction in detail, hence restricting ourselves to non-relativistic situations. In this regard, we want to refer to the works of Ruggeri and his coworkers \cite{LiuEtal86, PenRugg16, RuggSug21b} and Van \cite{Van17gal} for further reading. Either way, hyperbolic equations carry advantageous mathematical properties, such as particular numerical techniques developed for that family of equations, but these can also be transferred to parabolic equations. Strict hyperbolic models are used in the framework of RET \cite{MulRug98} and conservation-dissipation formalism (CDF) developed by Wen-An Yong \cite{WenAn08, LiuEtal01b}. 
It is also worth mentioning Godunov's approach \cite{Godunov61} in which the form of evolution equations is constrained to follow
\begin{align}
	\mathbf A(p) \frac{\partial p}{\partial t} + \mathbf B_k(p)\frac{\partial p}{\partial x_k} = 0,
\end{align}
where $p$ can be any (conservative) state variable. Moreover, $\mathbf A^{\textrm{T}} = \mathbf A$, $\mathbf B_k^{\textrm{T}} = \mathbf B_k$ is also satisfied, i.e., this is a form of a symmetric hyperbolic set of equations for which local well-posedness is proved. This approach is called SHTC formalism (symmetric hyperbolic thermodynamically compatible), and due to the structure of evolution equations, there is a close connection between RET, GENERIC, and SHTC. To gain further insight, we refer to the paper of Peshkov et al. \cite{PeshkovEtal18}.

\vspace{0.3cm}

As apparent from the above-discussed aspects, it is difficult to orient ourselves among the various models, as even the same equation can bear different properties depending on our chosen approach. Moreover, the literature can sometimes be self-contradictory, especially about the validity of specific models. The present review aims to help in this process, providing a guide from the most straightforward situations to the more complicated, less-known models explained in connection with experiments. Hence we first start our review with the Fourier heat equation as some specific properties are the easiest to present on that classical equation. We do not follow their chronological order in history as it would mix up the kinetic and continuum-based theoretical results and hide the practical aspects. Thus we continue the overview with increasingly complicated models, for which `complicated' means that more and more terms appear in the constitutive equation. Additionally, wherever necessary, we comment on the analytical and numerical solution techniques, as the literature can also be misleading. That systematic presentation of heat conduction models is closed by mentioning further concepts such as the DPL, thermomass, and fractional derivatives. For a recent book about the non-Fourier equations, we also mention \cite{Zhmakin23b}, in which one can find considerable interesting references, although the presentation feels less cohesive and sometimes contradictory.

\section{Fourier's law}

\subsection{Continuum background.} The well-known and widely used standard heat conduction model, the Fourier law \cite{Fou822}, reads
\begin{align}
	\mathbf q =-  \lambda \nabla T, \label{f1}
\end{align}
in which the heat flux $\mathbf q$ is proportional with the temperature gradient $\nabla T$, and the thermal conductivity $\lambda$ is scalar for isotropic materials. In CIT, it is possible to use the same state variable in both equilibrium and out of equilibrium, i.e., the entropy density depends only on the internal energy density $e$, $s=s(e)$ and $\textrm{d} e = T \textrm{d} s$. 
The derivation requires the balance of the internal energy density,
\begin{align}
	\rho \dot e + \nabla \cdot \mathbf J_e = Q_v, \label{ebal}
\end{align}
where $\mathbf J_e$ is the current density of the internal energy, and $Q_v$ is a volumetric heat source. For a purely heat conduction phenomenon, i.e., when thermal effects are decoupled from mechanics assuming zero thermal expansion coefficient, $\mathbf J_e = \mathbf q$ and $e=c_v T$ with $c_v$ being the isochoric specific heat. Using $\mathbf J_s = \mathbf q/T$, $\sigma_s$ can be easily calculated using \eqref{slaw}:
\begin{align}
	\sigma_s = \mathbf q \cdot \nabla \frac{1}{T} \geq 0. \label{f2}
\end{align}
The solutions of the entropy inequality are constitutive functions, connecting the thermodynamic forces and fluxes \cite{Onsager31I, Onsager31II}. One can find infinitely many solutions for the inequality \eqref{f2} as a function of $\nabla (1/T)$, e.g., in a polynomial form \cite{SzucsEtal20}. This is inherited for non-Fourier models and appears to be an open question of what further possibilities exist or what other functions would be physically meaningful. Nevertheless, it is treated as a less important question since even the linear Onsagerian relations can result in a complex model. To obtain \eqref{f1}, the simplest linear solution is enough, with isotropy,
\begin{align}
	\mathbf q = - \frac{l}{T^2} \nabla T = - \lambda \nabla T, \quad \lambda = \frac{l}{T^2}, \quad l\in\mathbb R^+.
\end{align}
For an anisotropic situation, $l$ becomes a second-order tensor, and $\lambda$, too. The second law of thermodynamics \eqref{slaw} provides closure for \eqref{ebal}, also using \eqref{ebal} as a constraint. The resulting system of equations is often used in $T$-representation, that is, using $\rho c \dot T = \nabla \cdot(\lambda \nabla T)$, without heat sources, and the thermal diffusivity is formed $\alpha = \lambda/(\rho c_v)$ in the linear, temperature-independent thermal conductivity case. The $T$-representation does not only imply that $\lambda$ is $T$-independent but restricts what type of initial and boundary conditions are meaningful. For example, the $q$-representation $\dot {\mathbf q} = \alpha \nabla\nabla\cdot \mathbf q$ is also meaningful for which the temperature boundary conditions are entirely excluded. Both forms represent the same system; one must choose the most convenient form according to the situation. For example, a time-dependent heat flux boundary condition is much easier to solve analytically in a $q$-representation, and the temperature evolution is reconstructed using the balance law Eq.~\eqref{ebal}. In general, the system \eqref{f1}-\eqref{f2} is the best choice for numerical solutions, together with an appropriate caloric equation of state for $e$.

\subsection{Kinetic background}
It is first developed for monatomic dilute gaseous materials, statistically describing the molecules' state in a state space with a quantity $f$, called single particle probability distribution, for which the Boltzmann equation prescribes the time evolution. Although there are numerous assumptions (hence, restrictions) on the molecule structure and how they interact \cite{Klimontovich95b, TrueMunc80b, Liboff03b}, there are many situations for which that description is helpful, especially in decreasing the number of parameters to be fitted. Namely, one notable property is that the transport coefficients can be estimated prior to any measurements. For a dilute gas the thermal conductivity $\lambda$ and the shear viscosity $\eta$ are
\begin{align}
	\lambda = \frac{1}{3} \bar v c_v n m  l; \quad \eta=\frac{1}{3} \bar v n m l, \label{+eq1}
\end{align}
where the mean velocity $\bar v$ is proportional with $\sqrt{T}$, n is the particle number density, $m$ being the molecule mass, and $l$ is the mean free path. For clarity, Eq.~\eqref{+eq1} does not require the utilization of the Boltzmann equation. However, in the lack of such interacting molecules in solids, the kinetic theory utilizes phonons as quasi-particles (lattice vibrations), and their hydrodynamic model describes heat conduction \cite{DreStr93a}. That approach is restricted to relatively large Knudsen numbers (about Kn$>0.001$), which forbids its applications for usual engineering tasks. Nevertheless, it provides useful insight into low-temperature heat conduction phenomena. In the case of diffusive propagation mode, the so-called resistive processes dominate the phonon interactions, i.e., processes that do not conserve phonon momentum on a time scale of $\tau_R$; thus their frequency is characterized by $1/\tau_R$, and its contribution to the thermal conductivity is
\begin{align}
	\lambda = \frac{1}{3} c_v \tau_R c^2, \label{macrok}
\end{align}
in which $c^2$ is the Debye speed of phonons and the method how Eq.~\eqref{macrok} estimates the thermal conductivity is referred to as Debye's law in the kinetic theory literature. The second sound occurs when the so-called normal processes dominate, where the phonon momentum is conserved. Its modeling requires the extension of Fourier's law, which is discussed in the following Section about the Maxwell-Cattaneo-Vernotte (MCV) equation.

\subsection{Functionally graded materials.} 
\begin{figure}[]
	\centering
	\includegraphics[width=17.5cm,height=5cm]{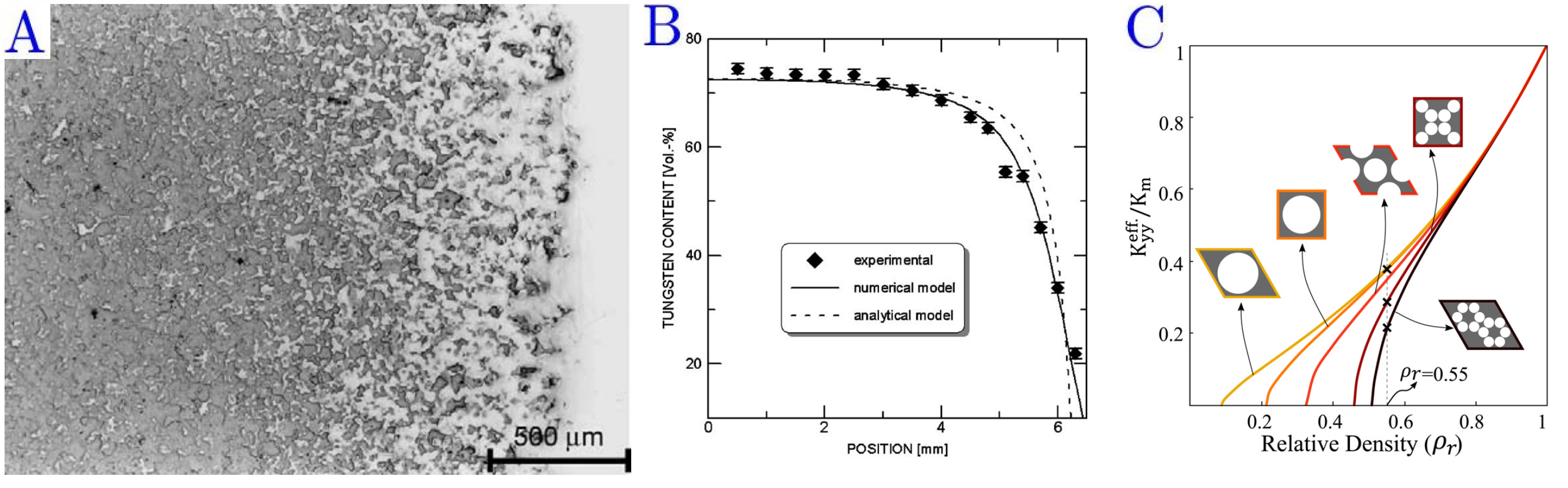}
	\caption{Examples for a functionally graded material and thermal metamaterials. Microstructure of a graded W/Cu composite produced by electrochemical gradation (A) and its tungsten content distribution (B) \cite{KieEtal03}. Figure C) shows the influence of the structure on the effective thermal conductivity as a function of the relative density of heterogeneities \cite{ShirzEtal23}.}
	\label{fig12}
\end{figure}

Regarding the modern state-of-the-art applications of Fourier's law, two outstanding examples should be mentioned: the functionally graded materials (FGM) and thermal metamaterials, even for non-Fourier heat equations \cite{ShirzEtal23}. In an FGM, the material structure varies in space, not necessarily monotonously, it can be periodic, too, thus the thermal properties. For an FGM, the trivial examples are composites and porous materials, usually mechanically optimized. Interesting biomedical applications can be found in \cite{PompeEtal03}, but there are examples for semiconductors, too \cite{MullerEtal03}. Figure \ref{fig12} presents two examples, showing the composition distribution \cite{MiyamotoEtal13b} for an FGM, and Fig.~\ref{fig12}/C presents the change of thermal conductivity with respect to the material structure. The crucial point is about the spatial scale of the material structure, as it essentially influences the homogenization procedure in modeling. That scale can vary in large intervals, from microns to millimeters, which restricts what domain one can substitute the original heterogeneous material with a homogeneous one. Various homogenization procedures can be found in \cite{BoggarapuEtal21} for macroscopic structures. From a thermal point of view, the Voigt-type `rule of mixture' seems physically more adequate, but this question is still open, there are multiple variations depending on the constituents, such as the Reuss-type estimation.
For instance, for the thermal conductivity, one can choose from
\begin{align}
	& \textrm{Voigt-type}: & \lambda_{\textrm{eff}} = V_1 \lambda_1 + V_2 \lambda_2, \\
	&  \textrm{Reuss-type}: & \lambda_{\textrm{eff}} = \frac{1}{\frac{V_1}{\lambda_1}+\frac{V_2}{\lambda_2}}, \\ 
	& \textrm{Markworth et al.}: & \lambda_{\textrm{eff}} = V_1 \lambda_1 + V_2 \lambda_2 + V_1 V_2 \frac{\lambda_1 - \lambda_2}{\frac{3}{\frac{\lambda_2}{\lambda_1}-1}+V_1}, \\
	& \textrm{Wakashima-Tsukamoto}: & \lambda_{\textrm{eff}} = \lambda_1 + \frac{\lambda_1 V_2(\lambda_2 - \lambda_1)}{\lambda_1+ \frac{(\lambda_1-\lambda_2) V_1}{3}},
\end{align}
where $V_1$ and $V_2=1-V_1$ are the corresponding volume fractions of the given materials, and $\lambda_1$ and $\lambda_2$ denote their thermal conductivity. These are only demonstrative examples based on \cite{ChenTong04, MarkEtal95}, showing that the calculation of even one scalar thermophysical property is not a straightforward task despite the knowledge about the constituents or the geometric factors. Moreover, such effective quantities can describe the 'averaged' behavior of a structure, to acquire the more detailed point-wise variations, one needs to use space-dependent thermal parameters. 

Moreover, there could be a threshold in the constituents, at which point a jump appears in the properties. This is analogous with nanofluids (however, the fluid flow and the mixing procedure appear as strong influential conditions, too) \cite{Keb02}. We emphasize again that size effects can significantly influence the outcome in control experiments. Thermal metamaterials appear to be a particular subcase of FGMs, which can be used for thermal cloaking or camouflage \cite{RashedEtal20, YangEtal21} with a thermally optimized material structure about how to control the isotherms.

For space-dependent constituents, the heat equation reads
\begin{align}
	\rho(\mathbf x) c_v(\mathbf x) \partial_t T = \nabla \cdot (\lambda(\mathbf x) \nabla T), \label{fgm1}
\end{align}
in which there is a nonuniform density distribution that differs from its usual meaning, and thus it does not necessarily involve the motion of the conducting medium. Hence, Eq.~\eqref{fgm1} is meaningful without mechanics. Sutradhar et al.~\cite{SutEtal04} suggests a variable transformation,
\begin{align}
	\phi (\mathbf x, t) = \sqrt{\lambda (\mathbf x)} T (\mathbf x, t),
\end{align}
which eases the solution method for non-homogeneous initial and boundary conditions, yielding
\begin{align}
	\frac{\rho(\mathbf x) c_v (\mathbf x)}{\lambda(\mathbf x)}\partial_t \phi = \Delta \phi + \hat \lambda (\mathbf x) \phi, \quad \textrm{with} \quad  \hat \lambda (\mathbf x) = \frac{\nabla \lambda (\mathbf x) \cdot \nabla \lambda (\mathbf x)}{4 \lambda^2 (\mathbf x)} - \frac{\Delta \lambda (\mathbf x)}{2 \lambda (\mathbf x)}. \label{fgm2}
\end{align}
Furthermore, depending on the value of $\hat \lambda$, one can achieve either quadratic or exponential thermal conductivity variations. It is worth noting that the spatial variation is still present for zero $\hat \lambda$, however, simplifying Eq.~\eqref{fgm2}. For further computational details, we refer to \cite{SutEtal04}. 

The thermo-mechanical coupling inherits the spatially-dependent parameters beyond the thermal part \cite{BurEtal17}. Therefore, both Lame constants and the thermal expansion coefficient become space-dependent. The effective Poission's ratio is determined by simply taking $\nu_\textrm{eff} = \nu_1 V_1 + \nu_2 V_2$, nonetheless, the effective Young's modulus and thermal expansion coefficient are found in a more complicated form \cite{FujiNoda01},
\begin{align}
	& \textrm{Young's modulus}: & E_{\textrm{eff}} (\mathbf x) = E_2 \frac{E_2 + (E_1 - E_2)V_1^{2/3}}{E_2 + (E_1 - E_2) (V_1^{2/3} - V_1)}, \\
	& \textrm{Thermal expansion coefficient}: & \chi_{\textrm{eff}}(\mathbf x) = \frac{\chi_1 V_1 E_1/(1-\nu_1) + \chi_2 V_2 E_2/(1-\nu_2)}{V_1 E_1/(1-\nu_1) + V_2 E_2/(1-\nu_2)}.
\end{align}
In order to solve such a model with a finite element technique, a sort of graded method is suggested \cite{BurEtal17}, in which the material properties are numerically integrated for each element.

Although certain tasks require the detailed modeling of the structure (if one knows well the spatial variation of material properties), it might not be necessary for all situations. For instance, a proper non-Fourier equation could catch the overall effects better for a macroscopic problem, especially for a porous or a layered structure, than a computationally intensive simulation on complex geometries \cite{LunEtal22}. This idea started to develop only recently but has outstanding potential and far-reaching consequences. 

However, for a nanoscale problem, the situation becomes more complex as many other aspects emerge. Such nanoscale metamaterials are called superlattices for which Fourier's law (and its kinetic background) must be treated more broadly, including size-dependent properties, various phonon propagation modes, and scattering mechanisms, which altogether point beyond the usual modeling capabilities of continuum frameworks, even for non-Fourier equations. We will discuss the related questions later.

Finally, as a noteworthy analogy, we want to mention the design of mechanical metamaterials, also possessing a designed microstructure to achieve specific mechanical properties such as stiffness, Poisson's ratio, or even damage attributes \cite{SurjadiEtal19, BertoldiEtal17}. The effective behavior, analogously to the non-Fourier equations, can emerge second gradient materials \cite{dIsolaEtal19}. For such material, the energy begins to depend on the density gradient as well \cite{SciarraEtal07}. We refer to \cite{Bertram23b} for a more thorough insight.  

\begin{figure}[]
	\centering
	\includegraphics[width=16cm,height=7cm]{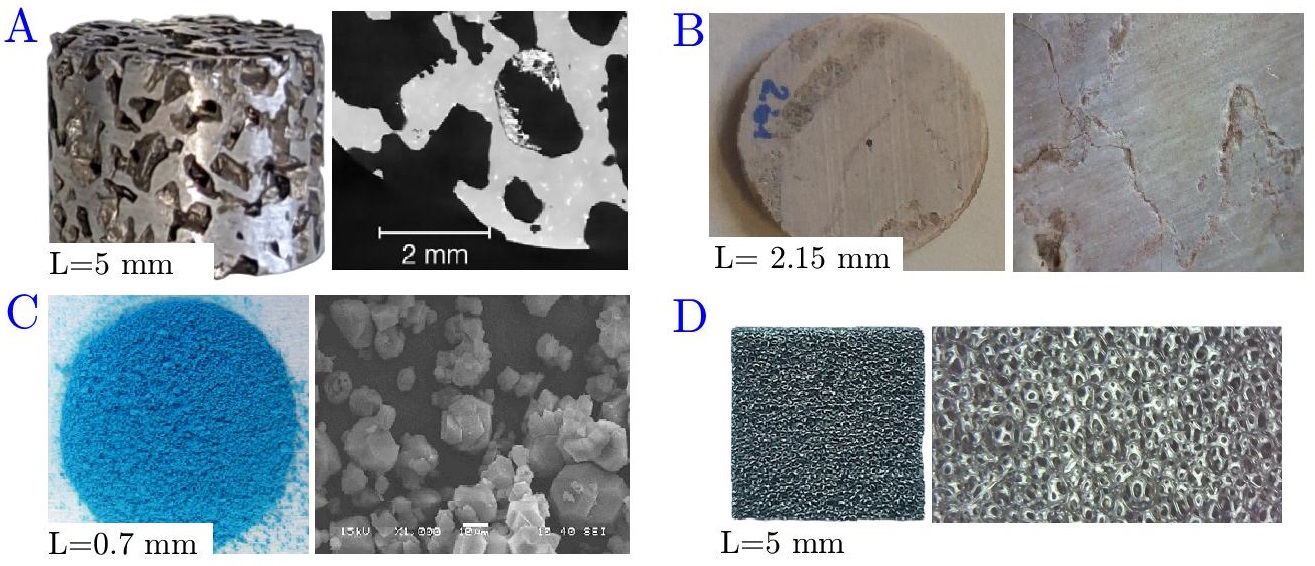}
	\caption{Typical heterogeneous samples for which over-diffusion is observed, with $L$ showing their characteristic thickness. A) Metal foam with large pores \cite{LunEtal22}. B) Szársomlyó limestone \cite{FehEtal21}. C) Metal-organic frameworks. D) Allcomp carbon foam \cite{FehEtal22}. }
	\label{fig2}
\end{figure}

\subsection{Diffusion vs.~over-diffusion.} Let us briefly revisit the background of over-diffusion. Since Fourier's law provides solely one time scale proportional with $L^2/\alpha$ with $L$ being the characteristic sample size, that model cannot properly predict the short time behavior of such material and thus distorts the evaluation, a model with at least two time scales (besides cooling) is necessary. Such alternatives are the two-temperature and Guyer-Krumhansl equations, both providing practically relevant interpretations and insights that could reveal essential attributes for the same material structure. Due to the different levels of modeling, these more advanced models result in different thermal diffusivities, and thus the presence of over-diffusion can significantly affect the thermal parameters we use in practice. Figure \ref{fig2} shows further samples for which over-diffusion is experimentally observed. 

\subsection{Biological heat equations.} That topic is inevitable to include, not simply because it is a hot topic in the literature but also comes with some difficulties in understanding their relations to non-Fourier heat equations. When one models bioheat transfer, it mostly means that particular source terms are introduced into the energy balance \eqref{ebal}, and the constitutive equation -- Fourier's law -- remains valid. In the case when a generalized constitutive model is needed, the dual-phase-lag (DPL, see later) concept enjoys broad interest, also keeping the source terms in the balance equation. Other non-Fourier models are much less prevalent concerning bioheat conduction. For this reason, we find it insightful to provide a brief overview of the relevant source terms and, thus, the basic bioheat models, as those could also have an impact on the research and applications of non-Fourier heat equations.

Thermally modeling living tissues comes with numerous complications. First, the tissue itself is porous and woven through capillaries in which complex coupled chemical, diffusion, and heat transfer processes occur simultaneously. Second, the geometry is also complex, changing from person to person, and the detailed modeling approach would be excessively computationally demanding and problematic in most cases. Therefore only a few aspects can be taken into account. Here, we focus on the heat transfer phenomenon and the related approaches.

The two most common source terms model the metabolic heat generation ($q_{\textrm{met}}$) and blood perfusion,
\begin{align}
	Q_v = q_{\textrm{met}} - \omega_b \rho_b c_b (T_b - T_t),
\end{align}
in which the indices $_t$ and $_b$ stand for the tissue and blood properties, and $\omega_b$ models the blood perfusion rate (volume flow rate of blood), so that term expresses heat convection and entirely neglects the direction-dependence (i.e., isotropic). Pennes was the first who introduced and studied these terms for a human forearm \cite{Penn48}, and thus the name Pennes' heat equation. In most cases, the metabolic term is considered to be uniform and constant. 

Chen and Holmes \cite{ChenHolm80} applied a more detailed approach and distinguished the tissue volume and vascular space, explicitly focusing on blood vessels with a diameter larger than $50$ $\mu$m as those play the most significant part in the heat transfer \cite{ZolfMaer11b, BhowmikEtal13}. Furthermore, for each sub-volume, they included the direction of the upstream blood by its local velocity $\mathbf v$ into the evolution equation for tissue temperature, i.e.,
\begin{align}
	Q_v = q_{\textrm{met}} - \hat \omega_b \rho_b c_b (T_b - T_t) + \rho_b c_b \mathbf v \cdot \nabla T_t + \nabla \cdot (\lambda_p \nabla T_t),
\end{align}
where the blood perfusion rate is characteristic only for the particular sub-volume (viz., not averaged as in Pennes' model). The third term intends to account for the convective effect, and by using $T_t$ here, it must be supposed that blood temperature equilibrates with the tissue; otherwise, neither the blood properties nor the tissue temperature should be present in this term. 
The last term stands for a so-called 'perfusion conductivity', aiming to introduce an effective thermal conductivity for the tissue, $\lambda_{\textrm{eff}} = \lambda_t + \lambda_p$.
We refer to \cite{Charny92} for a more thorough analysis. We note that such a model requires detailed knowledge about the particular vascular geometry and blood velocity. Later, Weinbaum and Jiji \cite{WJ85, WJL84} further developed that model in order to include a particular arrangement of artery and vein pairs working as heat exchangers. This is too complicated and too specific regarding the anatomical information for general use, and thus it has gained much less importance. 

We must mention Wulff's model as well \cite{Wulff74}, as in that approach, although the constitutive equation is modified, this is not a non-Fourier equation. Wulff, along with criticism towards Pennes' heat equation, modeled the blood flow and the energy transfer with
\begin{align}
	\mathbf q = - \lambda_t \nabla T_t + \rho_b h_b \mathbf v_b,
\end{align}
in which $\mathbf v_b$ is the local mean blood flow velocity and $h_b$ represents the specific enthalpy of the blood,
\begin{align}
	h_b = \int\displaylimits_{T_0}^{T_b} c_b(\bar T_b) \textrm{d}\bar T_b + \frac{p_b}{\rho_b} + H_{\textrm{met}}.
\end{align}
Here, $p_b$ is the blood pressure, and $H_{\textrm{met}}$ represents the enthalpy change due to metabolic reactions. 
Directly introducing the velocity in a constitutive equation can easily cause objectivity issues (see in the next section), and it might be more appropriate to use proper material time derivatives wherever necessary. Hence the convective transport of blood appears naturally, similarly to the work of Roetzel and Xuan \cite{XuRoe97, RoeXu98}. 

They introduced a proper two-temperature model \cite{XuRoe97, RoeXu98}, analogously to the Chen-Holmes model. However, instead of supposing any complicated source term, they utilized simple but effective convective coupling terms,
\begin{align}
	\varepsilon \rho_b c_b \left ( \partial_t T_b + \mathbf v_b \cdot \nabla T_b \right) &= \nabla \cdot (\lambda_b \nabla T_b) + h (T_t - T_b), \\
	(1-\varepsilon) \rho_t c_t \partial_t T_t &= \nabla \cdot (\lambda_t \nabla T_t) - h (T_t - T_b) + (1-\varepsilon) q_{\textrm{met}},
\end{align}
where $h$ represents a volumetric heat transfer coefficient between the blood and tissue, and $\varepsilon$ takes into account the porosity of the actual tissue domain. It still requires detailed anatomical data, especially for the blood velocity field. Otherwise, it must be coupled to proper flow equations, probably to the ones which include the rheological properties of blood as well \cite{WajiSan23}.
For further insight into the bioheat models, we refer to \cite{Charny92, ZolfMaer11b, MinSpar09b, ShomEtal22}.

\section{Maxwell-Cattaneo-Vernotte equation}
The MCV model stands as the first hyperbolic generalization of Fourier's constitutive equation, which takes into account the inertial effects (or a phase lag) by introducing the time derivative of the heat flux $\mathbf q$,
\begin{align}
	\tau \dot {\mathbf q} + \mathbf q = - \lambda \nabla T, \label{mcv1}
\end{align}
where $\tau$ is called the relaxation time. The MCV equation is often referred to as a `single-phase-lag' model. Although in parallel with Vernotte \cite{Vernotte58}, Cattaneo questionably obtained this model from a mathematical point of view \cite{Cattaneo58}, it is a thermodynamically compatible model \cite{Gyar77a, VanFul12}, respecting well-posedness and maximum principle, with an asymptotically stable equilibrium. It is crucial to emphasize that thermodynamic compatibility means that Eq.~\eqref{mcv1} can be derived on a thermodynamic basis by exploiting the first and second laws of thermodynamics. In such a way, one clarifies the elements of the state space and how it appears in the potentials of entropy and internal energy. The continuum thermodynamic approaches can greatly differ in that respect, and this question is still not closed completely. 

However, there are less consistent derivation procedures in the literature. One popular approach is taking the Taylor series expansion of $\mathbf q ( \mathbf x, t + \tau) = - \lambda \nabla T$ in time until first order. This is analogous to the famous dual-phase-lag concept. In that way, neither the state space nor the entropy production is determined, therefore the physical interpretation is completely missing. Moreover, it has shortcomings from a mathematical point of view as well, as it is not clear how the Taylor series converges, influencing stability, how the first order is satisfactory for a particular set of initial and boundary conditions, and how the nonlinearities or any anisotropic properties could appear. It is not recommended to follow such procedures due to their weak background and questionable physical and mathematical attributes.

\subsection{Equilibrium or non-equilibrium temperature?} In the 1970s, Taitel's argument \cite{Taitel72} started a debate in the literature about whether the MCV equation was compatible with thermodynamics at a time when modern thermodynamic approaches were not yet really elaborated. Taitel's debate is based on the analytical solution found for boundary conditions in which an immediate temperature jump occurs, and at some instants, it seems that the temperature field is momentarily equilibrated below the boundary temperature. Although it would be indeed nonphysical, such solutions raise questions about the benchmark procedure as it seems that the boundary conditions are not satisfied in every time instant. The immediate temperature jump itself is not physical, i.e., not experimentally feasible. Adding that wave interference can occur in the solution, and accordingly, momentarily, such observation is not necessarily paradoxical. 
Bai and Lavine \cite{BaiLav95} have similar reasoning, arguing that by dropping the temperature down to near absolute zero, the MCV equation can lead to a negative temperature. Although the nonlinearities play an extremely crucial part near absolute zero, the solution is found in the form of a modification of the internal energy following the work of Coleman et al.~\cite{ColEtal86}, and Banerjee and Pao \cite{BanerjeePao74}, discussed soon in Sec.~3.3.

Further studies on compatibility questions rely on the calculation of entropy production, for which it is essential to how the corresponding model is derived. It is further strengthened by the argument that the classical form Eq.~\eqref{f2} shows incompatibility with Eq.~\eqref{mcv1}  \cite{BaiLav95, BarZan97a}. 
In other words, the MCV model cannot fit into the local equilibrium hypothesis. This has led to the idea of non-equilibrium temperature \cite{CasJou94}. Namely, the temperature `close' to (and in) equilibrium differs from the one describing a non-equilibrium process. As it has no experimental background, it remains a theoretical concept only, with multiple variations in the literature \cite{Restuccia16}, and mostly the Extended Irreversible Thermodynamic (EIT) approach exploits that idea \cite{JouRes18}. One possibility is a differential relation between the local equilibrium ($T_{\textrm{loc}}$) and non-equilibrium temperatures ($T$) \cite{CasJou03}:
\begin{align}
	\frac{1}{T} = \frac{1}{T_{\textrm{loc}}} + \frac{1}{2} \frac{\partial \Omega}{\partial T} (\nabla T)^2
	\label{+eq2}
\end{align}
where $\Omega$ is a positive semidefinite function $\Omega=\Omega(\tau, \lambda, T)$, contributing to the entropy density as $s=s_{\textrm{loc}}(e) - \Omega(\tau,\lambda, T) \mathbf q^2$. More specifically, in \cite{CasJou03}, $\Omega = \tau /(\lambda T^2)$, where both $\lambda$ and $T$ are treated as constants while $T$ changes along the process. In general, $\Omega$ can be formulated in numerous ways, also depending on $\rho$ \cite{KovRog20}. For non-constant $\Omega$, Eq.~\eqref{mcv1} is not valid anymore, and further terms appear with modifying the entropy production, which is essential for nonlinear situations \cite{KovRog20}. Both temperatures will be identical with $\tau=0$, and that is how the classical concept of Fourier's law recovered. Despite Eq.~\eqref{+eq2}, it is still an open question how to embed this concept into the usual thermodynamic definition of temperature, viz., $\partial s/\partial e=1/T$. This is even more important for coupled systems for which $e \neq cT$.

This concept is analogous with the semi-empirical temperature (denoted with $\beta$) idea of Cimmelli et al.~\cite{CimKos91, CimKos92, FriCim95, FriCim96}, for which $\beta$ is treated as an additional scalar variable besides $e$, and has an evolution equation $\dot \beta = f(T, \beta)$ \cite{CarloEtal16}. Then Fourier's law is modified as
\begin{align}
	\mathbf q = - \lambda(T) \nabla \beta, \label{seT}
\end{align}
and the relaxation time is defined through the function $f$ as $\tau = (\partial f/\partial T)^{-1}$. Thus the model accounts for the nonlinearities, also including the relaxation time, but not in a way as usually treated in regard to the MCV equation, nevertheless, both approaches can be utilized for the same heat conduction problems. The crucial part is determining the function $f(T, \beta)$. They assume a splitting $f(T, \beta) = f_1(T) + f_2(\beta)$, moreover, $f_2 = - f_1$, which raises questions as these functions depend on different variables. Mathematically, it would mean that both $f_1$ and $f_2$ should be a constant, but that is not true as it would lead to a constant $\beta$ and $T$ as well. That model provides a unique background, being quite flexible with nonlinearities (e.g., with $T$-dependent variables), and can recover the accurate speed of second sound with the cost of the uncertain physical and mathematical aspects. 

From a kinetic point of view, the kinetic temperature is distinguished from the thermodynamic temperature being valid in equilibrium \cite{BarberaEtal99}. This is understood as the kinetic temperature $\vartheta$ measures the kinetic energy of atoms, while the thermodynamic temperature $T$ is interpreted as a factor between the heat and entropy flux, and
\begin{align}
	T= \vartheta \frac{1}{1 + g(\xi, \vartheta, \rho) }, \quad \xi=0 \ \Rightarrow \ g(\xi, \vartheta, \rho)=0,
\end{align}
in which the function $ g(\xi, \vartheta, \rho)$ depends on the actual momentum expansion, and thus on a non-equilibrium variable $\xi$, and $g$ vanishes in equilibrium, and therefore both temperatures become equal.

\subsection{Microstructural effects.} It has also been claimed that microstructural interactions cause finite propagation speed. These are usually modeled with an additional field variable, similar to the semi-empirical temperature concept, but none of them result in the usual MCV equation \eqref{mcv1}. It is essential to mention that these concepts are not strictly based on actual microstructural considerations. Usually, the effects are not determined, only supposed that those might influence the macroscopic observations, thus these are not real microstructural theories in their strict sense \cite{Riha76}.

Although Eringen thoroughly studied the effects of microstructure for both solids and fluids \cite{EriSuh64, Eringen12bI, Eringen12bII}, these works primarily focus on mechanical phenomena and do not on heat conduction. We still feel it necessary to mention Eringen's work because solids with memory (e.g., viscoelastic behavior) can effectively result in a non-Fourier temperature history. Additionally, various couplings are possible, especially for anisotropic solids. Also, the related works \cite{Riha76, Eringen72} acted as motivations for more recent studies. For instance, Grot \cite{Grot69} might be the first who introduced the concept of 'microtemperature' ($\bar{\mathbf T}$) to represent the temperature variation in a microvolume (though it is a vectorial quantity) and treated as a separate thermodynamic variable. In a linear case with negligible deformations, that approach leads to a special two-temperature model (omitting the source terms),
\begin{align}
	\rho c_v \partial_t T &= \kappa_0 \Delta T + \kappa_1 \nabla \cdot \bar{\mathbf T}, \\
	\rho' c_v' \partial_t \bar{\mathbf T} &= (\kappa_4 + \kappa_5) \nabla \nabla \cdot \bar{\mathbf T} + \kappa_6 \Delta \bar{\mathbf T} - \kappa_3 \nabla T - \kappa_2 \bar{\mathbf T},
\end{align} 
in which $\kappa_i$ ($i=0\dots6$) are conduction coefficients, possessing further restrictions \cite{Grot69}. It is possible to eliminate $\bar{\mathbf T}$, and thus obtain
\begin{align}
	k_1 \partial_{tt} T + \partial_t T = k_2 \Delta T + k_3 \partial_t \Delta T + k_4 \Delta \Delta T,
\end{align}
with $k_i>0$ ($i=1\dots4$) are positive coefficients. This structure strongly resembles the usual two-temperature models discussed later in Sec.~4. We want to highlight again that Fourier's law remains valid in such a situation, yet the couplings can introduce seeming deviation, which inspired numerous recent works to pay more attention to such possibilities. 

In the work of Mariano \cite{Mari14, Mariano17}, inspired by Capriz \cite{Capriz85}, it is supposed that a sort of `self-action' takes place, and its dissipative effects contribute to the energy balance, while the Fourier law remains valid. The following constitutive assumptions are made.
First, the heat conductor is supposed to be rigid, but not completely, only 'locally'. Therefore, there is no deformation and velocity field, but the microstructure can still change without macroscopic mechanical effects. This does not found as an outcome of the derivation. Second, that change is induced by introducing a temperature-dependent field, $\nu$, and that $\nu=\nu(T(x,t))$ characterizes the microstructure, claimed to be macroscopically observable. Third, it is also supposed that $e=e(T, \nu, \textrm{D}\nu)$, with $\textrm{D}\nu=(\nabla \nu) \mathbf g$, where $\mathbf g$ is a metric tensor (and $\nabla=D$ when $\mathbf g = \mathbf I$ identity tensor). Lastly, it is assumed that the flux of the inner power performed on the microstructural change has a flux $\mathbf p$ and that $\mathbf p$ directly modifies the heat flux $\mathbf q$, i.e., the balance of internal energy is modified by $\nabla \cdot \mathbf q \rightarrow \nabla \cdot (\mathbf q + \mathbf p)$.
That approach results in a heat equation in $T$-representation (using $\mathbf g = \mathbf I$ for simplicity):
\begin{align}
	a \partial_t T + \mathbf b \cdot \partial_t \nabla T - \lambda \Delta T + \mathbf d \cdot \nabla T + \gamma = 0, \label{+eq3}
\end{align}
where $a, \mathbf b, \mathbf d$ and $\gamma$ are found as
\begin{align}
	a=c_v + \frac{\partial \mathbf p}{\partial \dot \nu} \cdot \nabla  \frac{\textrm{d} \nu}{\textrm{d} T}, \quad \mathbf b = \frac{\partial \mathbf p}{\partial \dot \nu} \frac{\textrm{d} \nu}{\textrm{d} T}, \quad \mathbf d = \frac{\partial \mathbf p}{\partial T}, \quad \gamma = \frac{\partial \mathbf p}{\partial T} \cdot \nabla \nu.
\end{align}
Unfortunately, that model is not yet tested in experiments, and basically, unsolvable in the lack of proper $e=e(T, \nu, \nabla\nu)$ as the coefficients will remain unknown. Recently, an alternative choice $\nu=\nu(x, t,\mathbf q)$ is also studied \cite{CapEtal21}.
Eq.~\eqref{+eq3} is a hyperbolic equation, providing a finite propagation speed, however, the units seemingly have a shortcoming: the characteristic speed is identical to the thermal conductivity \cite{Mariano17}. Hence it is recommended to use that approach with reservations. 

Similarly to Mariano, Berezovski et al.~\cite{BerBer13a, BerEta11a1} also suppose that microstructural effects influence heat conduction in a thermo-mechanical framework. In \cite{BerJurMau11}, these effects are considered with two vectorial internal variables $\boldsymbol \varphi$ and $\boldsymbol \psi$, similarly to \cite{ColGur67}. The first one, $\varphi$, causes micro-stress and internal force, therefore $\boldsymbol \varphi$ is directly connected to mechanics, but it is not a mechanical quantity. It is identified with the micro-temperature \cite{BerJurMau11}, being different from the macroscopically observable temperature. The second internal variable is used as an auxiliary quantity, contributing to the time evolution of $\boldsymbol \varphi$, and helps to achieve a hyperbolic structure for $\varphi$. They obtain a coupled wave equation for both the micro-temperature and displacement ($\mathbf u$), which quantities appear as a source term in the energy balance:
\begin{align}
	\rho c \partial_t T- \nabla \cdot (\lambda \nabla T) = \gamma_1 \partial_t (\nabla \cdot \mathbf u) + \gamma_2 (\partial_t \boldsymbol \varphi)^2, \label{bb1}
\end{align}
in which $\gamma_1$ and $\gamma_2$ are constants. This results in a wave-like propagation for $T$, too. In Eq.~\eqref{bb1}, it may seem that Fourier's law remains valid, nonetheless, that is not true since the classical Fourier equation would not satisfy the corresponding entropy inequality. Therefore they introduce its modification as
\begin{align}
	\mathbf q = - \lambda \nabla T + \boldsymbol \eta \cdot \partial_t \boldsymbol \varphi + \boldsymbol \xi \cdot \partial_t \boldsymbol \psi
\end{align}
where $\boldsymbol \eta$ is the micro-stress and $\boldsymbol \xi$ expresses the change in Helmholtz free energy with respect to $\nabla \boldsymbol \psi$. 
Although no experimental comparison is performed, they also provide a numerical procedure to discretize the coupled system of partial differential equations. Thus it is an open question how this model performs on experimental data and how it could be applicable in engineering practice. Nevertheless, these works highlight the importance of how one extends the state space and how the new variables are implemented.

\subsection{State space, entropy, and internal energy.} To fulfill the local equilibrium hypothesis for a heat conduction phenomenon, using the internal energy $e$ as the only state variable is enough. That choice is surely not enough to obtain Eq.~\eqref{mcv1}, and as previously mentioned, the heat flux $\mathbf q$ is a good candidate. Its continuum thermodynamic implementation originates from Gyarmati \cite{Gyar77a} and is widely applied in EIT \cite{JouVasLeb88ext, JouEtal96b}. In general, the entropy density reads
\begin{align}
	s=s_{\textrm{loc}}(e) - \Omega(\tau,\lambda,\rho,e,Z) \mathbf q^2, \label{s1}
\end{align}
where $\Omega(\tau,\lambda,\rho,e,Z)$ is a positive semi-definite function, the previous $\Omega(\tau,\lambda,T)$ was a particular one. However, depending on the phenomena we aim to model, it can also depend on the mass density $\rho$ and additional variables ($Z$). In \cite{JouEtal96b}, $\Omega=\tau/(\rho \lambda T^2)$ is utilized and based on $\partial^2 s/(\partial e \partial \mathbf q) = \partial^2 s/(\partial \mathbf q \partial e )$, it is claimed that $\Omega$ must be constant. If one accepts that $\Omega=\tau/(\rho \lambda T^2)=\textrm{const.}$, then it becomes a strong consistency condition for a nonlinear (e.g., $T$-dependent) situation. Furthermore, it could be more appropriate to exchange $T$ to $e$; thus, for a thermo-mechanical model, the strain or the thermal expansion coefficient can also influence the non-equilibrium contribution of entropy density \cite{Cimm11}. Eq.~\eqref{s1} is the simplest form that preserves the concavity property of entropy. Additionally, Sobolev and Kudinov \cite{Sobolev18b, SobKud20} investigated $x \ln(x)$-type ($x=x(\mathbf q)$) extensions in the analogy of Gibbs (or Shannon) entropy which reduces to the form of Eq.~\eqref{s1} for small $|\mathbf q|$, and in parallel, also forming a unique non-equilibrium temperature $\theta$, developed only in a one-dimensional setting, and leading to
\begin{align}
	\theta = \frac{T}{1 + \frac{k_\textrm{B} q}{2 c_p v T} \ln\frac{1 + q/(c_p v T)}{1 - q/(c_p v T)}}.
\end{align}

The situation becomes even more complicated for anisotropic materials as the relaxation time $\tau$ becomes a tensor with the thermal conductivity $\lambda$ \cite{ColEtal82, SellCimm19}. This is usual for crystals and significantly impacts the modeling of low-temperature phenomena, including their mechanical properties and dislocation distribution \cite{Mezhov79, Mezhov80, MezMuk11}. Moreover, anisotropy introduces further consequences beyond simply having tensorial coefficients. Let us use the notations of $\mathbf T(T)$ and $\mathbf K(T)$ for the tensorial relaxation time and thermal conductivity with enabling that both depend on the temperature $T$. In that case, Eq.~\eqref{mcv1} reads
\begin{align}
	\mathbf T(T) \partial_t \mathbf q + \mathbf q = - \mathbf K(T) \nabla T \label{mcvAN}
\end{align}
for rigid materials, i.e., mechanics is absent \cite{ColEtal82, ColEtal86}, $\mathbf K(T)$ must be positive definite, and both $\mathbf T(T)$ and $\mathbf K(T)$ must be non-singular. For thermodynamic compatibility, furthermore, the tensor $\mathbf Z (T) = \mathbf K^{-1} \mathbf T$ must be symmetric and the internal energy begins depend on $\mathbf q$ as well:
\begin{align}
	e(T, \mathbf q) = e_0 (T) + \mathbf q \cdot \mathbf A(T) \cdot \mathbf q, \label{INAN}
\end{align}
in which 
\begin{align}
	\mathbf A(T) = \frac{1}{T} \mathbf Z(T) - \frac{1}{2} \frac{\textrm{d}}{\textrm{d}T} \mathbf Z(T), \label{INAN2}
\end{align}
therefore the specific heat is given by $c=\partial_T e(T, \mathbf q)$, thus becoming heat flux-dependent \cite{ColEtal86}. Including that aspect, Bai and Lavine \cite{BaiLav95} call Eq.~\eqref{mcvAN} together with \eqref{INAN} as the modified hyperbolic heat equation, which can lead to more satisfactory behavior for jump-type boundary conditions. It would be an interesting future study to fit together the anisotropic and temperature-dependent properties as Eq.~\eqref{mcvAN} neglects the mechanical contribution appearing by the functional connection between $\mathbf T$ and $\mathbf K$. This would also influence the form and thermodynamic compatibility of \eqref{INAN}.

Concerning internal energy, we mention the work of Rubin \cite{Rubin92}, who argued that the direction of heat flux does not necessarily point from the hotter to the colder, consequently, the MCV equation violates the second law of thermodynamics. Although it seems certainly unusual at first sight, the time derivative term ('heat flux lag') indeed can lead to such situations. Adding that the entropy production of the MCV model is also different, such a lagging phenomenon does not immediately introduce thermodynamic incompatibility. Despite Rubin's unclear (and not necessarily valid) reasoning, it is worth presenting the alternative approach to extend Fourier's law. The core assumption is based on the extension of internal energy, i.e., $e=e_1 + e_2$ in which $e_1$ is the classical one $e_1=cT$, and $e_2$ characterizes the correction for the hyperbolic heat equation and found as $e_2 = \alpha \dot T$ (here $\alpha$ is a constant). Accordingly, the entropy is modified as well, viz., $s=s_1 + s_2$ with $s_1= c \ln(T/T_0)$ and compatibly with $e_2$, $s_2$ reads $s_2 = \alpha(\dot T / T + \beta)$, where $\beta$ is determined from the evolution equation $\dot \beta = (\dot T / T)^2$. A $T$-representation also leads to a hyperbolic, MCV-like equation, though with a completely different background.

For any state variable that appears in the state space beyond $e$, the second law provides a time evolution equation for which we must apply the balances as constraints. In Liu's procedure \cite{Liu72}, for a set of state variables $\mathbf X$, the simplest constitutive state space is given as $(\mathbf X, \nabla \mathbf X)$, and the entropy inequality is constrained by the balances using Lagrange-Farkas multipliers \cite{Farkas18}. It is not simply a mathematically strict derivation procedure through `optimizing' the entropy production but also provides valuable feedback for nonlinearities, e.g., how the transport coefficients can depend on the constitutive state space \cite{Van09b, SzucsKov22}. Even for the classical Navier-Stokes-Fourier equations, it turned out recently that the viscosities and thermal conductivity can depend on $\nabla \mathbf v$, which is not straightforward and missing from the classical literature.

In such case, the Gibbs relation has the form $\textrm{d} e = T \textrm{d}s - \mathbf a \cdot \textrm{d} \mathbf q$, where $\mathbf a$ is the corresponding affinity, $\mathbf a = - \Omega \mathbf q$. This motivates that the internal energy $e$ should take the form $e=e(T, \mathbf q)$, where the specific heat capacity remains $c_V = \partial_T e>0$, and also the contribution of $\mathbf q$ must be positive as described in detail in \cite{CapEtal21, MarEtal22}. Its immediate consequence is that $\partial_{\mathbf q} e \neq \textrm{const.}$ in any case, even when $c$ can be assumed to be constant. Despite that $e=e(T, \mathbf q)$ seems motivated, it rarely appears in the literature, for a recent study, we refer to the work of Mariano \cite{CapEtal21} and Sciacca \cite{SciaccaEtal22}. However, if we think about $e$ as a temporal and $\mathbf q$ as a spatial part of a single internal energy spacetime four-quantity, then it seems inconsistent that the temporal part explicitly depends on the spatial part \cite{SzucsEtal21}, and that spacetime aspect strictly restricts how $e$ can depend on the state space. 

\subsection{Entropy production and nonlinearities.} Let us consider now $s(e,\mathbf q) = s_{\textrm{loc}}(e) - m/2 \ \mathbf q^2$ with $m\geq0$ being constant, and $\mathbf J_s = \mathbf q/T$. Then, the entropy inequality and its linear Onsagerian solution for isotropic materials read
\begin{align}
	\sigma_s = \mathbf q \left (-\rho m \dot{\mathbf q} +\nabla \frac{1}{T}   \right) \geq 0, \quad -\rho m \dot{\mathbf q} +\nabla \frac{1}{T} = l \mathbf q, \quad \tau=\frac{\rho m}{l}, \quad \lambda=\frac{1}{lT^2}, \quad l\in\mathbb R^+. \label{mcv2}
\end{align}
As it is apparent, the thermal conductivity and relaxation time coefficients are not independent. It has crucial consequences in situations with $T$-dependent parameters, when $\lambda(T)$ holds, then $\tau(T)$ holds as well. This is not completely exploited in the study of Frischmuth and Cimmelli \cite{FriCim96}, thus, their work could have further consequences on their modeling strategy, too. Moreover, it might be necessary that the mass density depends on the temperature, which means mechanical interactions are also present \cite{KovRog20}. In such case, the complete derivation procedure must be repeated with including mechanics, i.e., exploiting the momentum balance, kinematic relation, and also with extended internal energy \cite{Lubarda04} such as in a one-dimensional setting,
\begin{align}
	e= cT + \frac{E}{2 \rho} \varepsilon^2 + \frac{E \chi}{\rho} T_0 \varepsilon, \label{etm}
\end{align}
where $E, \chi, \varepsilon$ and $T_0$ are Young's modulus, thermal expansion coefficient, strain, and reference temperature, respectively.

The determination of  $\tau(T)$  highly depends on how it is measured since it is a dynamic quantity and cannot be measured in the same way as $\lambda(T)$, primarily not statically. Usually, the fitting of Eq.~\eqref{mcv1} can supply additional insight when the experiments are performed at different reference temperatures. If so, then the fitting might not be unique as it depends on what nonlinearities we assume. For instance, in \cite{CimFri96}, first $\lambda(T)$ is determined (the particular $\lambda(T)$ function itself is questionable, but the methodology is transparent) and then compared to experiments. Although their results qualitatively differ from the experiments, this is still one possibility and could be helpful for estimations. Examples for $\tau(T)$ can be found in \cite{MasRom17, ColNew88}, and it is still an open question how to take into account the nonlinearities properly. The best way would be to repeat the same experiments with our current understanding.

From a practical point of view, solving the inequality in \eqref{mcv2} should be the starting point for any further simplifications or modifications. In that form, whether the required modification (e.g., the above-mentioned $T$-dependence) has further consequences is visible. Eq.~\eqref{mcv1} alone cannot reflect these properties. The best option would be if any non-Fourier equations are presented with their entropy production and its Onsagerian relations. That would clarify many properties of what we speak about and ease mutual understanding. Moreover, it would immediately highlight whether there is any physical or mathematical contradiction in the model.

For completeness, we also show the fading memory interpretation of the MCV equation. If one supposes that the material has infinite memory, and all the previous time instants influence the future, then, following Gurtin et al.~\cite{GurPip68, ColGur67c}, the heat flux reads
\begin{align}
	\mathbf q (\mathbf x, t) = - \int\displaylimits_{-\infty}^t \kappa(t-s) \nabla T(\mathbf x, s) \textrm{d} s, \quad \kappa(s) = \frac{\lambda}{\tau_q} \exp(-s/\tau_q),
\end{align}
where $\kappa(s)$ is called memory kernel, requiring continuous $T$ and $\kappa$ piecewise continuous functions. The form of $\kappa(s)$ determines how influential the past is. For a profound review of such models, we refer to the books of Amendola et al.~\cite{AmendEtal21b}, Morro and Giorgi \cite{MorroGio23b}.

\subsection{$T$ and $q$-representations.} Considering the $T$-representation of the MCV equation, i.e., combining Eq.~\eqref{mcv1} with the balance of internal energy \eqref{ebal}, one obtains
\begin{align}
	\tau \ddot T + \dot T = \alpha \Delta T + \frac{Q_v}{\rho c_v} + \frac{\tau}{\rho c_v} \dot Q_v, \label{mcv3}
\end{align}
in which the time derivative of the heat source also contributes contrary to the Fourier heat equation. That term is sometimes referred to as `pseudo-source', emerging due to the inertial effect in the heat flux time evolution, and does not independent of the energy balance. Consequently, that `pseudo-source' term vanishes for constant heat sources in time. It is crucial to modeling semi-transparent bodies exposed to (laser) irradiation for which the absorbed energy is often modeled with a source term, e.g., \cite{LamEtal16}. As the $T$-representation is usually treated as a `standard' form for heat equations (inherited the usual convention from Fourier's case), this becomes an essential property. 
Additionally, when $\lambda(T)$ and $\tau(T)$ holds, it is not possible to obtain the $T$-representation, and one must solve as a system  \eqref{ebal}+\eqref{mcv1} without eliminating any of the field variables. 

It is usually argued, e.g., in \cite{Walczak18}, that for a short time ($t\ll\tau$), $\dot T$ becomes negligible, and thus $\tau \ddot T = \alpha \Delta T$ (with $Q_v=0$) is a valid expression; and for $t\gg \tau$, $\dot T = \alpha \Delta T$ holds. While these could provide acceptable approximations, they could possess additional requirements, and more importantly, especially for $\tau \ddot T = \alpha \Delta T$, it cannot be used for compatibility studies as it is not a valid heat equation in general. Moreover, it could lead to false analogies. Despite its wave nature, it is not applicable for ballistic heat conduction as the characteristic wave speed remains identical with the second sound $v=\sqrt{\alpha/\tau}$.
However, these approximations highlight the possibility of separating two different time scales as $\tau$ and $\alpha=\lambda/(\rho c)$ can be independently adjusted in a continuum model. This contradicts the phonon background as $\tau$ explicitly appears in $\lambda$, therefore, while the kinetic-MCV model still describes two propagation phenomena with two distinct time scales, these are not independent of each other, hence such approximations are not possible.

It is interesting to show the $q$-representation as well with a heat source $Q_v$,
\begin{align}
	\tau \ddot{\mathbf q} + \dot{\mathbf q} = \alpha \nabla\nabla\cdot \mathbf q -  \alpha \nabla Q_v \label{mcvq}
\end{align}
for which the source term appears differently, its time derivative does not contribute directly to the evolution, compared to Eq.~\eqref{mcv3}. However, when the $T$-profile is recovered using the balance of internal energy \eqref{ebal} through a time integration, the time dependence of $Q_v$ does contribute to the overall evolution. Furthermore, that form also suggests using the $q$-representation even with source terms as if $Q_v$ is uniform, Eq.~\eqref{mcvq} is much simpler to solve time-dependent $q$-boundaries instead of Eq.~\eqref{mcv3}.

\subsection{Objectivity arguments}
We call the reader's attention again to the evolution equation's objectivity and Galilean covariance properties. As summarized in \cite{Van17gal}, both the balances and constitutive relations must be independent of the reference frame. However, as also pointed out by Christov \cite{ChristovJordan05, Christov09}, it is not straightforward. 
In relation to non-Fourier heat conduction, the formulation of the MCV equation invigorated the debate about which time derivative of the heat flux would be appropriate instead of the partial time derivative, especially for moving observers. According to  Christov and Jordan \cite{ChristovJordan05}, the usage of partial time derivative might lead to paradoxical outcomes if the MCV equation is applied to a body in motion, which is quite straightforward from a space-time point of view. Their claim is that paradox can be eliminated by using the material time derivative,
\begin{align}
	\tau (\partial_t \mathbf q + \mathbf v \cdot \nabla \mathbf q) + \mathbf q = - \lambda \nabla T, \label{+eq5}
\end{align}
with $\mathbf v$ being the velocity field. They also prove its Galilean covariance. Furthermore, that argument is continued in \cite{Christov09} by Christov that in case of Eq.~\eqref{+eq5}, and
\begin{align}
	\rho c(\partial_t T + \mathbf v \cdot \nabla T) + \nabla \cdot \mathbf q = 0,
\end{align}
the heat flux $\mathbf q$ cannot be eliminated from the system, which is described as an undesirable feature that must be fixed. We note that such eliminability of variables cannot be a strict physical requirement. For instance, even for the simplest nonlinearity $\lambda(T)$ and $\tau(T)$, the $T$-representation does not work. Then, Christov modified the material time derivative to 
Oldroyd's upper-convected derivative, i.e.,
\begin{align}
	\tau \Big(\partial_t \mathbf q + \mathbf v \cdot \nabla \mathbf q - \mathbf q \cdot \nabla \mathbf v + (\nabla \cdot \mathbf v) \mathbf q\Big) + \mathbf q = - \lambda \nabla T, \label{+eq6}
\end{align}
proved to be also Galilean covariant, and, additionally, the $T$-representation now exists, and Eq.~\eqref{+eq6} is called Cattaneo-Christov equation. It is worth mentioning the recent work of Angeles \cite{Angeles23}, in which Eq.~\eqref{+eq6} is studied from a different perspective. First, it is argued that Eq.~\eqref{+eq6} supplemented with the conventional balances of mass, energy, and momentum, does not form a hyperbolic set of equations, only a weakly hyperbolic one \cite{Angeles22}. Hence, the criteria to obtain a well-posed problem are different. Second, a more general objective time derivative is parametrically investigated, motivated by Morro \cite{Morro18},
\begin{align}
	\tau \left(\partial_t \mathbf q + \mathbf v \cdot \nabla \mathbf q - \frac{1}{2}(\nabla \mathbf v - \nabla \mathbf v^\textrm{T}) \mathbf q +\frac{\mu}{2}(\nabla \mathbf v - \nabla \mathbf v^\textrm{T}) \mathbf q + \nu (\nabla \cdot \mathbf v) \mathbf q	
	\right) + \mathbf q = - \lambda \nabla T, \label{+eq7}
\end{align}
in which $\mu$ and $\nu$ are the parameters. It is found that $\mu=1$ and $\nu=-1$ make the model objective and hyperbolic, so Eq.~\eqref{+eq7} differs from \eqref{+eq6}. Moreover, Eq.~\eqref{+eq7} is found to be irreducible as well (with $\mu=1$ and $\nu=-1$), so its $T$-representation does not exist; however, as it is a hyperbolic one, it is easier to prove its well-posedness. Notably, Eqs.~\eqref{+eq6} and \eqref{+eq7} coincide in a one-dimensional setting, consequently, the Cattaneo-Christov model can inherit the advantageous properties of Eq.~\eqref{+eq7}, but these do not necessarily hold in a general three-dimensional setting. 
Finally, we want to mention the paper of Tibullo and Zampoli \cite{TibZam11} in which they proved the uniqueness of the solution for \eqref{+eq6} accompanied with incompressible fluids.
It would be worth investigating whether these findings can be transferred to \eqref{+eq7}.

\subsection{Phonon hydrodynamic background I} Let us recall the phonon modeling approach from the previous Section. One must prescribe the possible interaction between phonons and how these quasi-particles behave during the scattering. The two fundamental interactions are resistive and normal collisions. In a resistive collision, the momentum of a particle is not conserved, contrary to the normal collision. In obtaining the MCV equation, one needs to characterize the resistive processes, that is, this model possesses only one characteristic time scale, described by the relaxation time $\tau_R=\tau$. If the number of resistive collisions is prominent in a unit of time, it means small $\tau_R$. However, when, e.g., decreasing the temperature below $20$ K, the number of resistive collisions is significantly decreased as well, resulting in much larger $\tau_R$ values, and the inertial effects take place. In other words, the dominance of normal processes appears only through $\tau_R$. Recalling our previous observation that the continuum-MCV model preserves the two time scales with Eq.~\eqref{mcv3}, it is impossible to do within phonon hydrodynamics since $\tau_R$ directly appears in $\lambda$.
As it is apparent, the thermal conductivity is given and obtained immediately by knowing the relaxation time. Hence that approach decreases the number of parameters to be fitted compared to a continuum model. However, in parallel, the given heat conduction mechanism decreases the region of validity as the phonon approach is restricted only to situations with large Knudsen numbers \cite{DreStr93a}. For a thorough overview of phonon background and the details of collisions, we refer to \cite{GuoWang15, DingEtal18}.

The MCV equation can be obtained through the momentum series expansion of the Boltzmann transport equation, that is,
\begin{align}
	\partial_t f + c \mathbf n \cdot \nabla f = \hat S, \quad \hat S = -\frac{1}{\tau_R} ( f - f_R) - \frac{1}{\tau_N} ( f - f_N), \label{bte00}
\end{align}
in which $\hat S$ represents a production term for phase density $f$, following Callaway's model where the deviation from the corresponding equilibrium (resistive and normal processes) is a special form of the collision integral \cite{Call59}. The momentum expansion results in
\begin{equation}
	\frac{\partial u_{\langle m \rangle}}{\partial t} + \frac{m^2}{4m^2-1}c\frac{\partial u_{\langle m-1 \rangle}}{\partial x}+c\frac{\partial u_{\langle m+1 \rangle}}{\partial x}=\left \{ \begin{array}{ll}
		\displaystyle
		0 \ & \ m=0 \\
		-\frac{1}{\tau_R}u_{\langle 1 \rangle} \ & \ m=1 \\
		- \left( \frac{1}{\tau_R}+\frac{1}{\tau_N} \right )u_{\langle m \rangle} \ & \ 2\leq m\leq M
	\end{array} \right., \label{bal1}
\end{equation}
where $u$ denotes the corresponding momentum quantity with increasing tensorial orders, and $_{\langle \ \rangle}$ denotes the traceless symmetric part of a m$^\textrm{th}$ order tensor. The MCV equation is obtained for $m=1$, without a second order tensor, with truncation closure \cite{DreStr93a}.
The relaxation time of normal processes, $\tau_N$, appears in the successive approximation (for $m=2$) and has a role in ballistic propagation that will be discussed in more detail later. 
Here, as a final remark, we place the thermal conductivity into a more general setting, that is
\begin{align}
	\lambda = \frac{1}{3} c^2 c_v \tau(0,0), \quad \tau(0,0)=\tau(\mathbf k, \omega)|_{\{\mathbf k= \mathbf 0, \omega=0\}}
\end{align}
where $\tau(0,0)$ is found as a projection of the collision operators from normal and resistive processes and corresponds to a steady-state setting for which the wave number $\mathbf k$ and circular frequency $\omega$ are both zero \cite{GK66}. At the level of the MCV equation, and applying the so-called relaxation time approximation, $\tau(0,0)=\tau_R$, but in a more general framework,  $\tau(0,0)=\tau_R$ does not hold.
\subsection{Analytical and numerical solutions.} \label{Analytical and numerical solutions.} 
\begin{figure}[H]
	\centering
	\includegraphics[width=14cm,height=8cm]{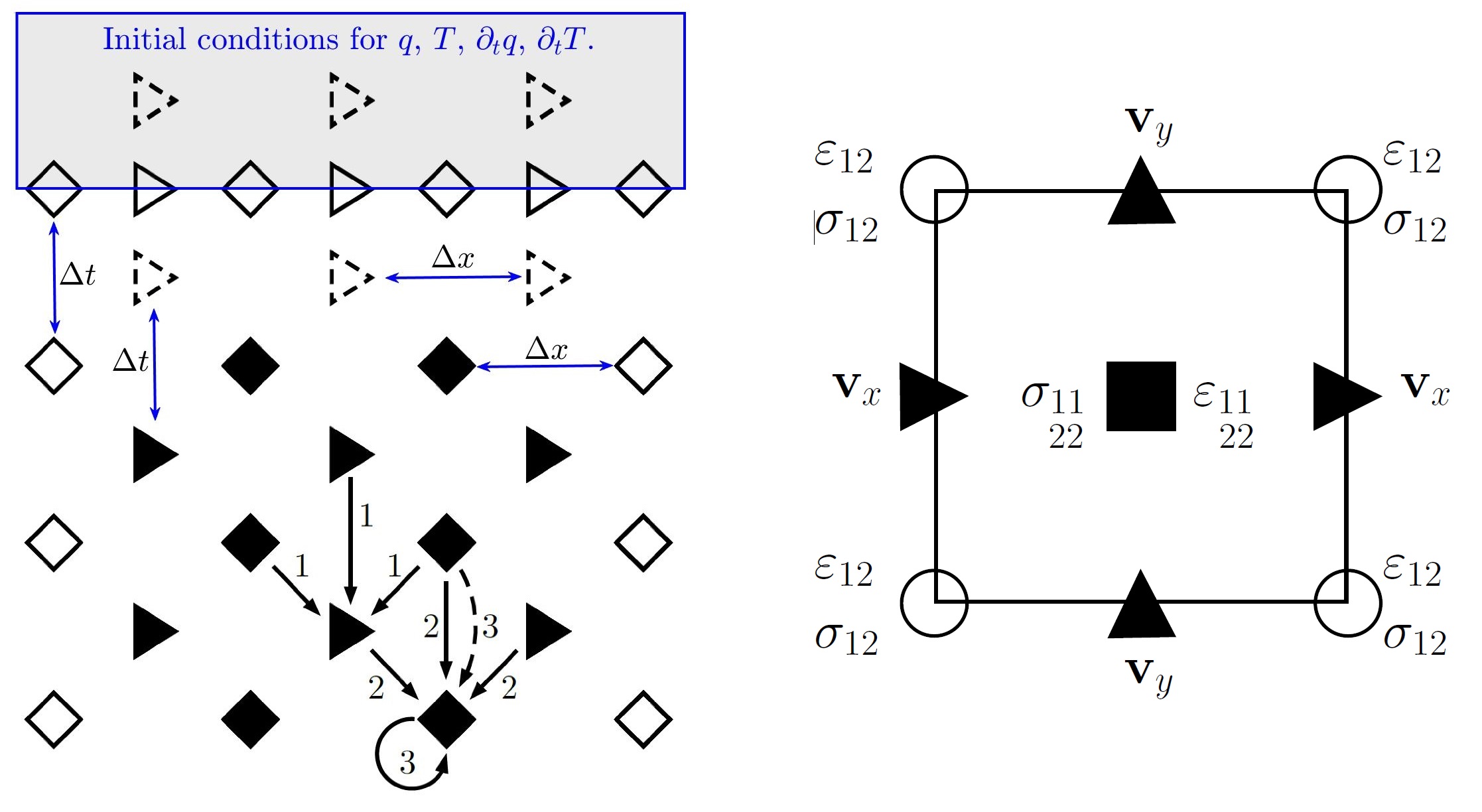}
	\caption{The schematics of the staggered field discretization in a 1D (left) \cite{PozsEtal20} and a 2D (right) \cite{PozsEtal20, JozsKov20b} situation for a finite difference method. Situating the fluxes on the boundary is the most natural allocation of variables as these are `surface-like' quantities. It also applies to tensorial quantities such as the stress ($\boldsymbol{\sigma}$). Following the balance equations, the intensives are `volume-like' quantities and therefore are placed inside. The proper set of initial conditions might need values for more than one time instant. The numbering presents the update order for a symplectic algorithm  \cite{PozsEtal20}.}
	\label{fig3}
\end{figure}

Considering a linear case of \eqref{mcv1} in which both $\lambda$ and $\tau$ are constant, the numerical solution is straightforward compared to the Fourier heat equation, the $T$-representation remains applicable as the treatment of the new time derivative term does not require any special consideration. This does not hold for a nonlinear case for which the $T$-representation cannot be obtained, and one must handle the MCV model as a system, simultaneously solving the balance equation \eqref{ebal} together with the constitutive one \eqref{mcv1}. For such a task, using a staggered field discretization is much more advantageous, and its schematic is visible in Figure \ref{fig3} \cite{PozsEtal20}. This approach eases the discretization and the implementation of initial and boundary conditions and does not require using any $q$ or $T$ representation. Moreover, this becomes even more important later for the Guyer-Krumhansl equation as for such a model, COMSOL can produce false solutions \cite{RietEtal18}. Utilizing a staggered field, one keeps the physics in focus. That method is also helpful for mechanical and thermo-mechanical tasks as well \cite{BallEtal20, FulEtal22}, with a so-called symplectic time stepping \cite{Sanz92, SanzCalv18b, Otti18}. Symplectic methods are particular time-stepping algorithms that preserve the corresponding system's total energy, ensuring the physical admissibility of the obtained numerical solution. 

\begin{figure}[H]
	\centering
	\includegraphics[width=17.5cm,height=4cm]{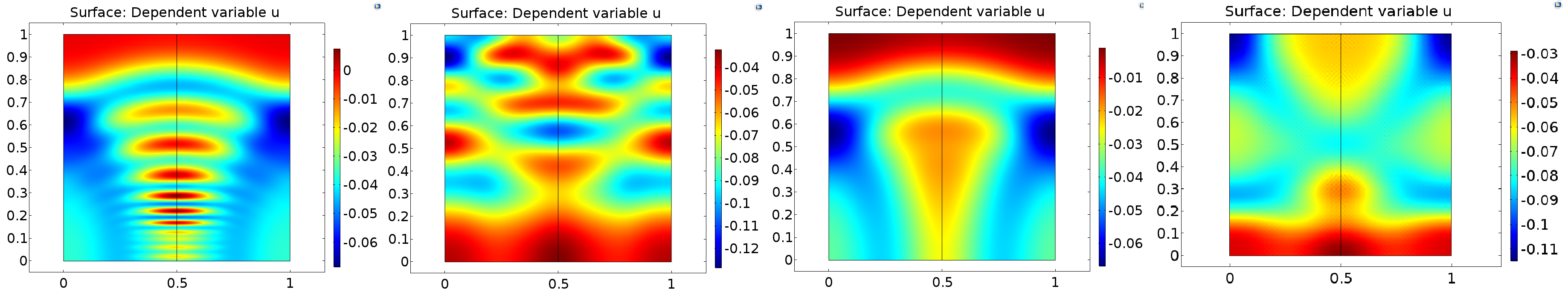}
	\caption{Various COMSOL solutions for a 2D wave equation with a scalar field variable \cite{PozsEtal20}. All simulations are performed for the same set of initial and boundary conditions. The million-dollar question: which one is valid?}
	\label{com}
\end{figure}

That staggered approach can be implemented into a finite element method as well using a so-called mixed formulation in which the field variables are separated and situated at different points of the element, analogously to the one shown on the right side of Fig.~\ref{fig3} \cite{ManMan96, ManMan99}. In other words, eliminating any field variables is not recommended as it will provide a more flexible and helpful approach to implementing any initial and boundary conditions. Following \cite{ManMan96, ManMan99}, Manzari and Manzari applied a Galerkin weighted residual method to transform the evolution equations into an integral form,
\begin{align}
	\int\displaylimits_V w_1 (\rho c_v \partial_t T + \nabla \cdot \mathbf q - Q_v) \textrm{d} V =0,  \quad \Rightarrow \quad  \int\displaylimits_V (w_1 \rho c_v \partial_t T + \nabla w_1 \cdot \mathbf q - w_1 Q_v) \textrm{d} V + \int\displaylimits_{S} \mathbf q \cdot \mathbf n w_1 \textrm{d}S=0, \label{fe1}
\end{align}
\begin{align}
	\int\displaylimits_V w_2 (\tau \partial_t \mathbf q + \mathbf q + \lambda \nabla T) \textrm{d} V =0, \label{fe2}
\end{align}
which formulation likewise works for anisotropic models, and $w_1, w_2$ are weighting functions. Applying partial integration in Eq.~\eqref{fe1} makes the model suitable for a finite element implementation as the surface term ($S$) with the normal vector $\mathbf n$ considers the boundary heat flux conditions. Then each field quantity is interpolated, such as
\begin{align}
	T (\mathbf x, t) = \sum\displaylimits_{j=1}^N \tilde T_j(t) \ ^TN_j(\mathbf x), \label{fe4}
\end{align}
where the interpolation coefficients $\tilde T_j(t)$ are time-dependent for a transient problem, and the interpolation functions $^TN_j(\mathbf x)$ depend only on the spatial coordinate and accordingly on the element type, and the physical quantity for which it is applied to. So that elements of the heat flux fields are interpolated with a different function. All these together form a system of ordinary differential equations, written in a form analogously to mechanics,
\begin{align}
	\mathbf M \dot{\mathbf U} + (\mathbf K_c + \mathbf K_s) \mathbf U = \mathbf F_c + \mathbf F_s, \label{fe3}
\end{align}
in which $\mathbf U$ includes all field variables for all nodes, $\mathbf M$ consists of the heat capacities and relaxation time coefficients, $\mathbf K_c$ constitutes all elements for heat conduction (e.g., thermal conductivities), $\mathbf K_s$ and $\mathbf F_s$ are related to the surface terms (such as prescribing heat convection on the boundary) and $\mathbf F_c$ includes the source terms. Eq.~\eqref{fe3} can be solved in various ways, in \cite{ManMan99}, the time derivative is approximated with
\begin{align}
	\dot{\mathbf U} \approx \Theta \dot{\mathbf U}_{t+\Delta t} + (1-\Theta) \dot{\mathbf U}_t = \frac{{\mathbf U}_{t+\Delta t} - \mathbf U_t}{\Delta t},
\end{align}
i.e., taking the convex combination with $0<\Theta<1$ improves the accuracy and stability properties together. Two well-known choices exist: the Crank-Nicolson ($\Theta=0.5$) and Galerkin ($\Theta=2/3$) schemes. 
Beyond stability and accuracy, these can also notably influence a scheme's dissipative and dispersive properties. Such artificial errors can remarkably distort the solution, hence one must pay attention to what physical problem can be solved reliably with what scheme. One glaring example is published in \cite{PozsEtal20} concerning a two-dimensional (scalar field) wave equation solved with COMSOL: ten different time stepping settings can provide ten distinct solutions (see Fig.~\ref{com} for a few examples). For more details about the properties of various schemes, how to estimate the artificial errors, and how to use entropy as an indicative physical quantity for stability, let us refer to \cite{BerEta14p1, FulEtal20, FulEtal22, Press07b}. Furthermore, we also want to mention the works of Bargmann and Steinmann \cite{BarSte05a, BarSte08} 

Galerkin's technique can be adapted to analytical solutions as well, especially for non-Fourier heat equations, as both the temperature and heat flux have separate time evolution equations \cite{FehKov21}. The fundamental assumption is based on Eq.~\eqref{fe4} that reduces the partial differential equation into a set of ordinary differential equations, depending on the number of field variables. The implementation of time-dependent boundary conditions is straightforward for any variable as these boundary conditions appear as an additional heterogeneous term in the system. Moreover, this method applies to mixed boundary conditions, too. However, contrary to the interpolation function from Eq.~\eqref{fe4}, the spatial functions must form a complete set of orthonormal system. These can be found by determining the corresponding eigensystem of the Laplacian. Interestingly, that system remains the same even for higher-order non-Fourier equations, and seemingly this property is connected to thermodynamic compatibility \cite{FehKov21}. 

Solving a partial differential equation requires both initial and boundary conditions. For most situations, the initial state is supposed to be in equilibrium, i.e., the heat flux field is zero together with its time derivative, and the temperature field is homogeneous. This does not cause any complications, and the usual separation of variables \cite{Farlow93b, Kov18gk}, Laplace transform \cite{CarJae59b}, or operational methods \cite{Zhu16a, Zhu16b} work well for the analytical approach. 
Additionally, the boundary conditions of generalized constitutive equations are not trivial, and thus in this respect, we refer to the excellent overview of Zhou and Yong \cite{ZhouYong21a, ZhouYong21b, YongZhou21}. 

However, what if the initial temperature distribution is not homogeneous, i.e., space-dependent? For such a situation, $\nabla T$ appears as a given inhomogeneous term for Eq.~\eqref{mcv1}, which restricts the initial heat flux field and offers a way to prescribe a compatible set of initial conditions for both fields. Namely, it requires the solution of a partial differential equation \eqref{mcv1} for the initial time instant to obtain a compatible field for $q(x,t=0)$ for a given $T(x,t=0)$. Furthermore, freedom remains on how to constrain the `integration constant', that is, the value of time derivative of $\mathbf q$, which could be space-dependent, too. This needs further knowledge of the history of the system. Interestingly, it is possible to suppose that Fourier's law provides the initial heat flux field, but then the time derivative is zero. If the system is adiabatic, then such a system remains `close' to equilibrium due to the zero initial time derivatives, and therefore the non-Fourier effects remain vague \cite{Kov22a}. This aspect further emphasizes the importance of the thermodynamic background, which is not visible from the $T$-representation. It provides compatibility conditions and helps to avoid nonphysical results, especially negative temperatures \cite{Zhukov16, ZhuSri17}.

\subsection{Experimental background.} 
\begin{figure}[]
	\centering
	\includegraphics[width=15cm,height=6cm]{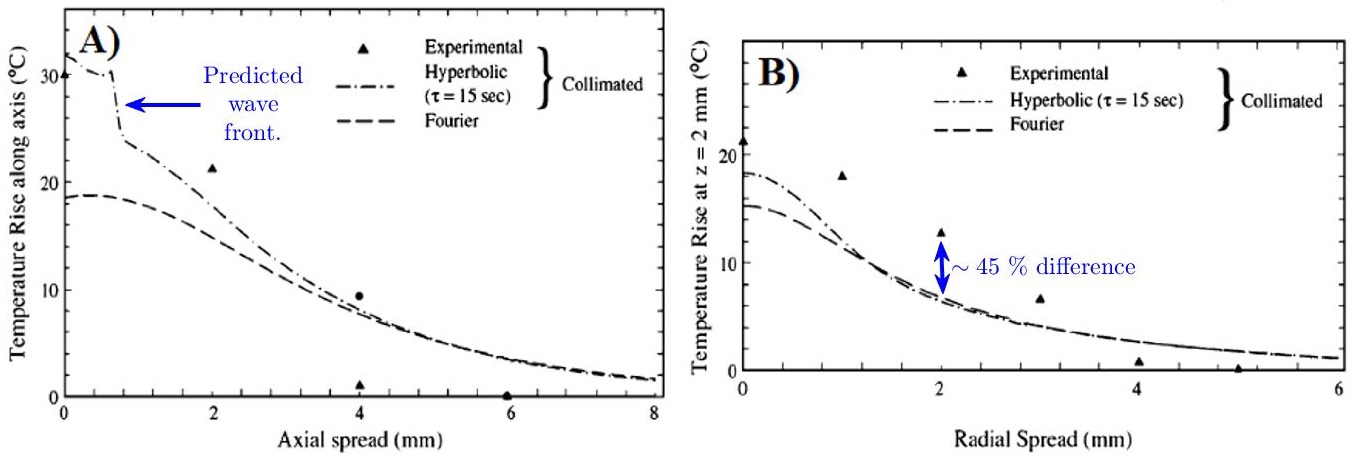}
	\caption{Experimental results of Jaunich et al.~\cite{JauEtal08} on a biological sample. While the MCV equation provides a closer approximation for the measured temperature history with a large relaxation time resulting in sharp wave fronts. Fourier's prediction can be significantly improved with proper heat sources, see \cite{SudEtal21}. }
	\label{fig4}
\end{figure}

First, it must be emphasized that the MCV equation \eqref{mcv3} - in its original sense - is applicable only for the second sound as a low-temperature phenomenon, especially with its phonon background. That argument is further strengthened by \cite{Auriault16}, proving that the MCV equation - interpreted as a phonon hydrodynamic model - is not applicable for macroscale room temperature heat conduction problems, or the additional relaxation term has no notable contribution. However, the continuum model does not require a phonon-based mechanism but still can be helpful in effective modeling, e.g., for heterogeneous materials where all the microstructural effects (or defects) add up and emerge as a delay, some examples are summarized in \cite{Mail2019} when particular time and spatial scales are separable \cite{Auriault91}. On the contrary, there are numerous papers, mainly regarding biological problems \cite{Banetal05, Dharetal15, TanEtal07, ZhouEtal09}, for which the better fit from the MCV (or DPL) model is interpreted as an observed heat wave. For instance, the experiments of Jaunich et al.~\cite{JauEtal08} can be modeled best with Fouries's law with proper heat source terms instead of having a hyperbolic heat conduction model \cite{SudEtal21}. Their results are depicted in Figure \ref{fig4} \cite{JauEtal08}. Despite its misleading statement and modeling approach, the model still can be helpful with strict restrictions, and one has to understand the physical and mathematical framework clearly. Such a misleading approach is spreading in the literature, especially in connection with the DPL model \cite{Zhang09, Hoosetal15, AfrinEtal12, LiuChen10}.

\begin{figure}[]
	\centering
	\includegraphics[width=13cm,height=6cm]{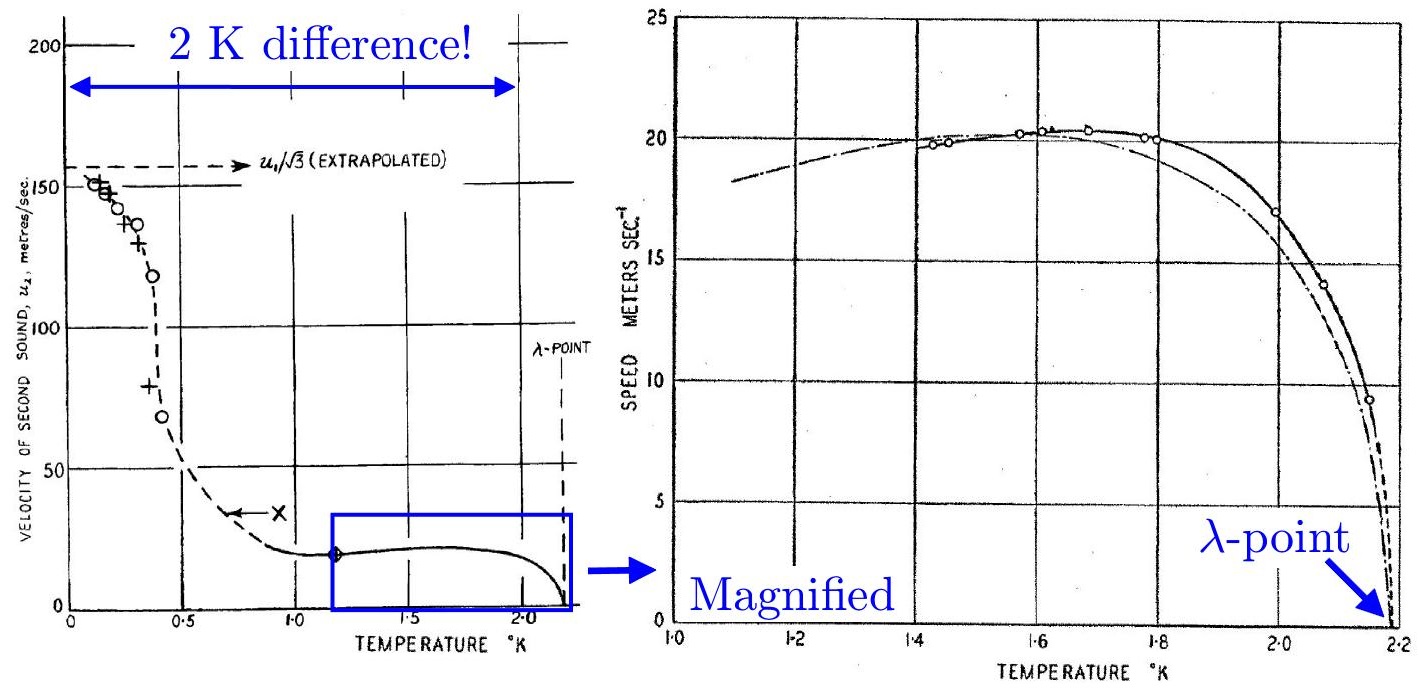}
	\caption{The change in speed of second sound with respect to the temperature, showing a steep increase close to $0$ K \cite{LaneEtal46, AtkOsb50}. }
	\label{fig5}
\end{figure}

\begin{figure}[]
	\centering
	\includegraphics[width=13cm,height=7cm]{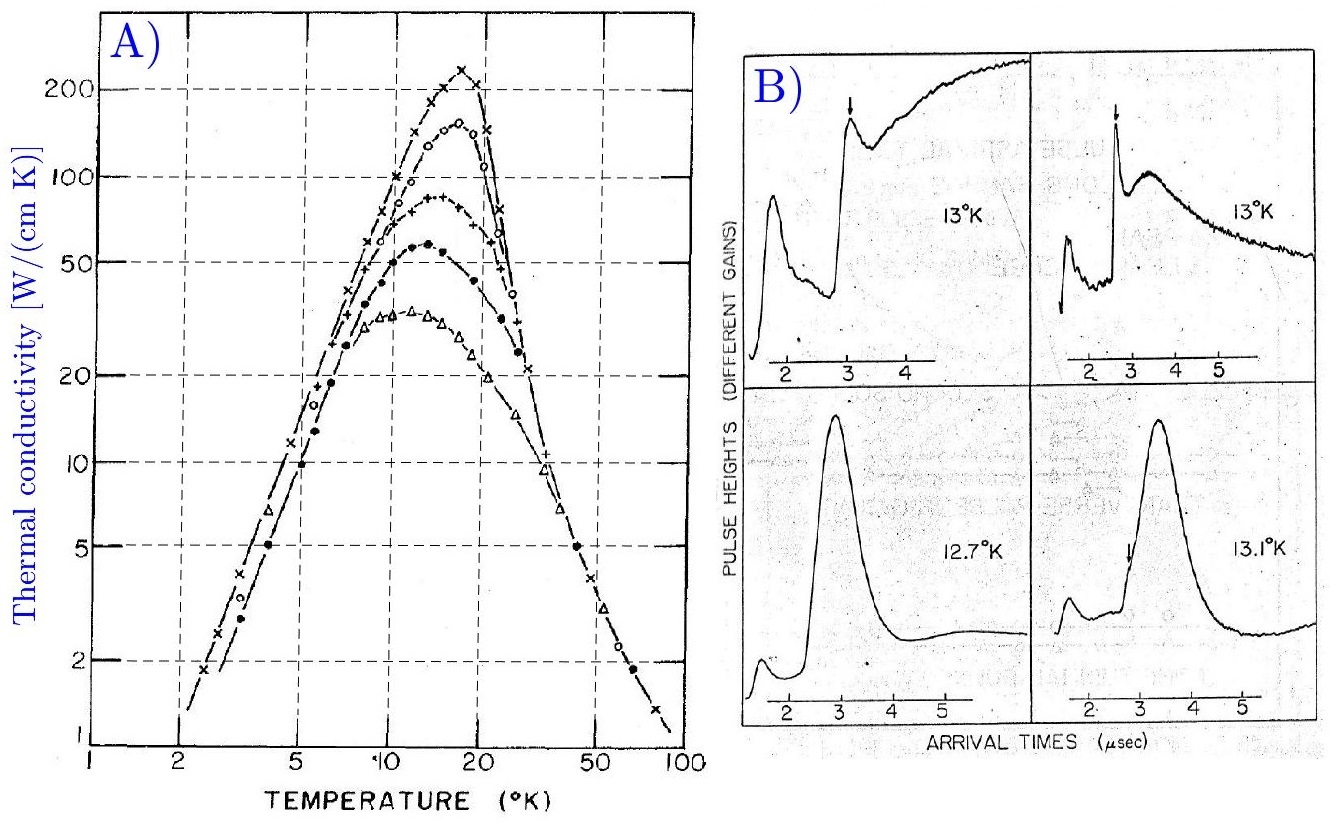}
	\caption{The thermal conductivity of a NaF crystal changes dramatically with purity (A) \cite{JacWalMcN70}, therefore, the occurrence and the shape of the second sound are also highly purity-dependent \cite{McN74t}.}
	\label{fig6}
\end{figure}

For the second sound, the wave propagation speed stands as a central question. Since Eq.~\eqref{mcv3} is hyperbolic, it predicts finite propagation speed $v_T = \sqrt{\alpha/\tau}$, independently of what representation ($T$ or $q$) we use. It was helpful in modeling second sound in superfluid helium \cite{Kapi41, Dres82b, Dres84b}, and has significant $T$-dependence in the low-temperature domain. Figure \ref{fig5} shows the propagation speed of the second sound in helium II \cite{LaneEtal46, AtkOsb50}, from near zero to 2.2 K, as an example. This could also act as a further constraint on what $T$-dependence are physically admissible and how to connect $\lambda(T)$ and $\tau(T)$. Figure \ref{fig6} provides further examples with various NaF crystals about how sensitive the thermal conductivity and second sound are for sample purity \cite{McN74t}. While the MCV equation worked well for the second sound in fluids, it had no further predictive power on how to find the second sound in solids. However, it is not equivalent with that it would not be able to model wave-type heat conduction in solids. 

The predictive power of the MCV equation was helpful in estimating the propagation speed of the second sound but could not provide conditions on how to observe it. This is the point where the so-called Guyer-Krumhansl equation enters the picture. Furthermore, when the ballistic heat propagation is first observed, it immediately turned out that none of these models will be sufficient and further research is necessary.

\section{Guyer-Krumhansl equation}
\subsection{Phonon hydrodynamic background II} Guyer's and Krumhansl's original idea was to develop such a phonon hydrodynamic framework in which they can incorporate both the mechanical (`first sound') and thermal effects (`second sound') coupled through thermal expansion for an isotropic material \cite{GK66}. The famous Guyer-Krumhansl (GK) equation is a sub-case for the decoupled situation. Their starting point is the Boltzmann equation,
\begin{align}
	\partial_t f + \mathbf v \cdot \nabla f = \mathbf C f, \label{bte1}
\end{align}
where $\mathbf C$ is a collision operator (in a more general way than the $\hat S$ that appeared earlier). They apply a particular approximation on the left-hand side of Eq.~\eqref{bte1}, they substitute $f$ with $f_0$, an equilibrium distribution given as $f_0=(\exp(h \omega/(k_B T(\mathbf x)))-1)^{-1}$ (with $h$ being Planck's, and $k_B$ the Boltzmann constant), and they start to seek its solution for a steady-state case first. That is,
\begin{align}
	f=\mathbf C^{-1} \left ( \mathbf v \cdot \nabla T \frac{\partial f_0}{\partial T} \right )
\end{align}
with supposing that $\mathbf C = \mathbf N + \mathbf R$, i.e., $\mathbf C$ is separable into two collision operators, $\mathbf N$ is for the normal processes, and $\mathbf R$ is for the resistive collisions, analogously with the Callaway model ($ \mathbf R \sim 1/\tau_R$ and $ \mathbf N \sim 1/\tau_N$). Their idea is to span the solution in the eigenspace of $\mathbf C$, which is reduced to seek the eigenspace of $\mathbf N$ under low-temperature conditions since the normal processes highly dominate the propagation. The corresponding eigenvectors project $f$ into the internal energy density and heat flux vector, which are also proportional to the corresponding eigenvalues, and the energy conservation appears naturally as a part of their solution.

Within that framework, it is rather natural to have a thermal conductivity as a function of the wave number $\mathbf k$ and frequency $\omega$, i.e., $\lambda(\mathbf k, \omega)$, which is inherited from $\tau(\mathbf k, \omega)$. They obtain expressions for $\tau(0,0)$ for both normal and resistive-dominated processes in terms of the corresponding eigenvectors in a steady situation. For a continuum theory, such $\lambda(\mathbf k, \omega)$ relation must be implemented artificially, as the derivation procedure of evolution equations does not offer any outstanding choice or restrictions for such correlation. Instead, it suggests that thermal conductivity can differ between equilibrium and non-equilibrium situations.

They repeat their analysis for the transient problem, too, separately for $\mathbf R \gg \mathbf N$ and $\mathbf R \ll \mathbf N$, expressing the dominance of each process. Inherently, for the first one, Fourier's law is found for which the $(\mathbf k, \omega)$-dependence appears only as a minor correction. The second case is more interesting, as they take the limit
\begin{align}
	\omega \tau_N \ll 1 \quad \omega \tau_R \gg 1, \label{wc}
\end{align}
and found a $\lambda(\mathbf k, \omega)$ strongly depending on both variables, 
\begin{align}
	\tau_R \dot {\mathbf q} + \mathbf q = - \lambda \nabla T + l^2( \Delta \mathbf q + 2 \nabla\nabla\cdot\mathbf q), \label{gk1}
\end{align}
in which $l^2$ is the square of the mean free path of phonons and can be expressed as $l^2=\frac{3 c^2}{5} \tau_R \tau_N$. Besides, as visible, the characteristic times of both normal and resistive collisions are present, with $\tau_R=\tau$ compared to the Eq.~\eqref{mcv1}. 
We emphasize that Eq.~\eqref{gk1} is valid only in the interval expressed by Eq.~\eqref{wc} and also supposes a linear relationship between the heat flux and the quasi-momentum of phonons (quasi in the sense that phonons are also quasi-particles). Eq.~\eqref{gk1} thus describes the second sound, and as turned out later, Eq.~\eqref{wc} was outstandingly helpful for experiments to find the second sound in solids. This is called window condition. We note that Gurevich and Shkolvskii \cite{GurShk67} also performed a similar study and found exact temperature and frequency criteria for semiconductors.

In order to obtain the thermo-elastic coupling, they introduce the displacement field $\mathbf u$, and $\nabla \cdot \mathbf u$ is called dilatation. That displacement field is coupled to the phonon field through thermal free energy $F_\textrm{th} ( \nabla \cdot \mathbf u, T)$, from which adds a contribution to the phonon pressure $p_\textrm{th} = - \partial F_\textrm{th} / \partial (\nabla \cdot \mathbf u)$. Overall, the coupled system reads
\begin{align}
	\rho_0 \frac{\partial^2 \mathbf u}{\partial t^2} &= \psi(F_m) \Delta \mathbf u - \frac{\partial p_\textrm{th}}{\partial T} \nabla T, \label{bte2} \\
	\frac{\partial e_T}{\partial t} + \nabla \cdot \mathbf q &= - p_\textrm{th} \frac{\partial }{\partial t}(\nabla \cdot \mathbf u), \\
	\frac{\partial \mathbf q}{\partial t} + \frac{c^2}{3} \nabla e_T + \frac{1}{\tau(0,0)} \mathbf q - \frac{\tau_N c}{s} ( \Delta \mathbf q + 2 \nabla \nabla \cdot \mathbf q) &= c^2 p_\textrm{th} \nabla \nabla \cdot \mathbf u, \quad e_T = \frac{\partial e}{\partial T}, \quad s=s(\tau_N, \tau_R), \label{bte3}
\end{align}
whence $\psi(F_m)$ is a coefficient depending on the mechanical free energy $F_m$, and $s$ is $s=\tau_R/\tau_N$ in this framework, but this does not generally hold. For zero thermal expansion coefficient, $p_\textrm{th} = 0$, and these equations become decoupled. In principle, Eqs.~\eqref{bte2}-\eqref{bte3} can be used to model the first and second sounds together, i.e., it could be useful for ballistic propagation. This sometimes misleadingly appears in the literature, for which that property is attributed to Eq.~\eqref{gk1} instead of \eqref{bte2}-\eqref{bte3}. Ballistic propagation is discussed in the next section.

\subsection{Continuum background.} In a continuum approach, the detailed mechanisms are not present, accordingly, the resulting model differs in the coefficients, analogously to the MCV equation. Additionally, the factor $2$ in Eq.~\eqref{gk1} is closely related to the mechanisms and to the applied approximations, and that appears to be a free ($\geq 0$, adjustable in a fitting procedure, but functionally connected) parameter, in a general isotropic case,
\begin{align}
	\tau \dot {\mathbf q} + \mathbf q = - \lambda \nabla T + \eta_1 \Delta \mathbf q + \eta_2 \nabla\nabla\cdot\mathbf q. \label{gk11}
\end{align}
Here, $\eta_1$ is no longer the mean free path but can still be related to some intrinsic length scale, similarly to $\eta_2$ \cite{SzucsEtal21}. The last term with $\eta_2$ can be obtained through a proper isotropic representation of the Onsagerian solution \cite{VanFul12}. The missing detailed propagation mechanism does not disqualify the model from low-temperature applications but extends its applicability region to room-temperature phenomena. While it is not as explanatory as phonon hydrodynamics in that sense, that re-interpretation of the GK equation was necessary and, indeed, it is proved to be useful for over-diffusive problems with its effective modeling capability \cite{FehEtal21}. 

In order to derive Eq.~\eqref{gk11} exploiting the second law as before, one needs the same state space used for the MCV equation, $s=s(e,\mathbf q)$. However, the emphasis is now on the entropy flux $\mathbf J_s$ for which the classical $\mathbf J_s = \mathbf q /T$ definition is insufficient.  
First, Müller proposed a simple extension, called Müller's k-vector, $\mathbf J_s = \mathbf q /T + \mathbf k$ \cite{Muller68}. It is discussed by Verhás \cite{Verhas83} that such extension - together with $\mathbf q$ - must vanish in equilibrium in order to avoid non-zero entropy flux. Additionally, proposed a particular form for $\mathbf k$ as $\mathbf k = \sum \mathbf A \cdot \boldsymbol \xi$, where $ \boldsymbol \xi$ is a set of all dynamic degrees of freedom as additional variables in the state space, and $\mathbf A$ is the set of the corresponding multipliers. Together with $\mathbf k$, any dynamic degree of freedom must also vanish in equilibrium. This is a distinctive property compared to an internal variable approach, as internal variables are not necessarily zero in equilibrium. These variables can represent, e.g., the crack density or other physical attributes characteristic of the material and, in parallel, contribute to the system's time evolution. An even more general setting for  $\mathbf J_s$ is to utilize a current (or Nyíri multiplier \cite{Nyiri91, Van01a}) multiplier $\mathbf B$  as 
\begin{align}
	\mathbf J_s =  \left ( \frac{1}{T} \mathbf I + \mathbf B \right ) \cdot \mathbf q, \label{gkj}
\end{align}
in which $\mathbf B$ is treated as a constitutive second-order tensor and $\mathbf I$ stands for the identity tensor \cite{KovVan15}. Without any prior knowledge of $\mathbf B$, the second law inequality will provide the necessary restrictions,
\begin{align}
	\sigma_s = \mathbf q \cdot \left ( - m \partial_t \mathbf q + \nabla \frac{1}{T} + \nabla \cdot \mathbf B \right ) + \mathbf B \cdot \nabla \mathbf q \geq 0,
\end{align}
with the Onsagerian equations
\begin{align}
	\mathbf q &= \mathbf L_1^{(2)} \left( - m \partial_t \mathbf q + \nabla \frac{1}{T} + \nabla \cdot \mathbf B \right ), \label{+eq4} \\
	\mathbf B &= \mathbf L_2^{(4)} \nabla \mathbf q, \label{gkN2}
\end{align}
for which the second and fourth-order tensors $\mathbf L_1^{(2)}$ and $\mathbf L_2^{(4)}$ offer numerous possibilities for a general, anisotropic situation. Furthermore, Eq.~\eqref{+eq4} possesses a balance form for $\mathbf q$ in which $\mathbf B$ realizes the higher-order flux. Here, let us consider the isotropic case where $L_1^{(2)}$ reduces to a constant $l>0$, and 
\begin{align}
	\mathbf L_2^{(4)} = (\mathbf L_2^{(4)})_{ijkl} = \frac{L_2^{\textrm{sph}} - L_2^{\textrm{dev}}}{3} \delta_{ij}\delta_{kl} + \frac{L_2^{\textrm{dev}} + L_2^{\textrm{A}}}{2} \delta_{ik}\delta_{jl} + \frac{L_2^{\textrm{dev}} - L_2^{\textrm{A}}}{2} \delta_{il}\delta_{jk}
\end{align}
using the index notation with Einstein's summation convention. Additionally, $L_2^{\textrm{sph}}\geq0$, $L_2^{\textrm{dev}}\geq0$ and $L_2^{\textrm{A}}\geq0$ are the spherical, symmetric deviatoric and antisymmetric deviatoric parts of $\mathbf L_2^{(4)}$.
These results were helpful for EIT on how to modify the entropy flux $\mathbf J_s$ as there are no current multipliers in the framework of EIT. It is found that using $\mathbf J_s = \mathbf q/T + \mu \nabla \mathbf q \cdot \mathbf q$ with $\mu\geq0$, and $\mathbf B = \mu \nabla \mathbf q$ makes these approaches to be compatible \cite{JozsKov20b}. However, that prior restriction immediately excludes anisotropic materials for which $\mu$ could be a fourth-order tensor.

Now we can express the coefficients from Eq.~\eqref{gk11} as
\begin{align}
	\tau = l m, \quad \lambda=\frac{l}{T^2}, \quad \eta_1 = l \frac{L_2^{\textrm{dev}} + L_2^{\textrm{A}}}{2}, \quad \eta_2 =l \frac{2 L_2^{\textrm{sph}} + L_2^{\textrm{dev}} -3  L_2^{\textrm{A}}}{6}. \label{GKE}
\end{align}
These relations are necessary for a nonlinear, e.g., $T$-dependent problem. For instance, studying material with a given $\lambda(T)$ (see Fig.~\ref{fig6} for example), then all other coefficients immediately inherit the dependence following from $l(T)=T^2 \lambda(T)$. Thus $\tau=\tau(T), \eta_{1,2}=\eta_{1,2}(T)$, and additional constraints are necessary to obtain the proper $T$-dependence for the other parameters. 

For a demonstrative example \cite{RamosEtal23}, let us consider a one-dimensional setting with $\eta_{1,2}(T)$ reducing to $\hat \eta(T)$ and let both $\lambda$ and $\tau$ be constants, thus $\mathbf L_2^{(4)}$ becomes a scalar $l_2(T)$, and $\hat \eta(T)=l \cdot l_2(T)$. Therefore when substituting the current multiplier from Eq.~\eqref{gkN2} into \eqref{+eq4}, additional terms appear, 
\begin{align}
	\partial_x \Big ( l_2(T) \partial_x q \Big ) = \frac{\textrm{d} l_2 (T)}{\textrm{d} T} \partial_x T \partial_x q + l_2(T) \partial_{xx} q,
\end{align}
from which the first one (from the right hand side) can seemingly modify the thermal conductivity,
\begin{align}
	l m \partial_t q + q = - \left ( \frac{l}{T^2} +  \frac{\textrm{d} l_2 (T)}{\textrm{d} T} \partial_x q \right ) \partial_x T + l l_2(T) \partial_{xx} q.
\end{align}
In other words, the $T$-dependence of $\hat \eta(T)$ introduces a seeming $T$-dependence and even a $\partial_x q$ dependence into $\lambda$, which makes it more difficult to separately determine the proper $T$-dependence of each parameter in an experiment. This is one outstanding non-trivial difference between the Fourier and non-Fourier models, which will require special attention in the future. Furthermore, as the particular appearance of these nonlinear properties can depend on our chosen approach, nonlinearities can be decisive or at least notably helpful to exclude or confirm one approach. 

Such relations are naturally embedded into the phonon hydrodynamic framework through the relaxation times for the propagation mechanisms, and this is not yet discovered using a continuum approach as it provides much more freedom in this regard. Moreover, as observed for the MCV equation, the continuum model is universal and offers more degrees of freedom with adjustable parameters when an effective interpretation is necessary. This emerges to be the cost of extending the model's domain of validity.

Compared to the MCV equation, the crucial difference is in the entropy flux in which the $\nabla \mathbf q$ appears as a consequence of the second law, in accordance with EIT. This is interpreted as a relaxed state variable \cite{SzucsEtal21}, such as $\mathbf Q \sim \nabla \mathbf q$, it is more easily noticed in a spatially one-dimensional setting
\begin{align}
	\tau \partial_t q + q &= -\lambda \partial_x T+\hat \eta \partial_x Q, \label{gk2a} \\
	Q&=\hat \eta \partial_x q, \label{gk2b}
\end{align}
for which $Q$ appears as an independent (relaxed) state variable, contributing to the evolution of heat flux. In fact, in the case of modeling a ballistic propagation, that $Q$ must be included as a non-equilibrium variable and extends Eq.~\eqref{gk2b} with its time derivative, $\tau_Q \partial_t Q$, with an additional relaxation time, representing an additional, third (conduction) time scale. It adds a more substantial physical justification for the current multipliers and is necessary to model ballistic heat conduction. That relaxed state variable does not have a time evolution equation, thus, does not appear in the state space directly. 

Let us add a short remark at that point. If one considers the current multiplier without extending the state space with $\mathbf q$, then the so-called Nyíri equation is obtained \cite{Nyiri91},
\begin{align}
	\mathbf q = - \lambda \nabla T + l^2 \Delta \mathbf q, \label{nyiri1}
\end{align}
that is, only spatial nonlocality is added to Fourier's law but not widely used due to the limited modeling capabilities. 

Overall, the GK equation consists of two time scales regarding the heat conduction mechanism, independent of the framework we use. The interpretation of the time scales themselves depends on the approach. Either way, the coefficients are also not independent, and their thermodynamic origin in Eq.~\eqref{GKE} must be considered for $T$-dependent parameters. 

\subsection{$T$ and $q$-representations.}
These forms are valid only for constant coefficients. Eliminating $\mathbf q$ from Eqs.~\eqref{gk11} and \eqref{ebal}, we obtain
\begin{align}
	\tau \partial_{tt} T + \partial_t T = \alpha \Delta T + (\eta_1+\eta_2) \partial_t \Delta T + \frac{1}{\rho c_v} \left (Q_v + \tau \frac{\partial Q_v}{\partial t} -(\eta_1 + \eta_2) \Delta Q_v \right), \label{gkt}
\end{align}
in which various contributions of the heat source $Q_v$ appear. Considering $Q_v=0$, it is more interesting to observe that both the Fourier heat equation and its time derivative turn up, making it possible to recover Fourier's solution when $\alpha = (\eta_1 + \eta_2)/\tau$, called Fourier resonance \cite{Vanetal17}. Consequently, the antisymmetric deviatoric part $L_2^{\textrm{A}}$ does not contribute to the temperature evolution as well as to the entropy production. When $\alpha > (\eta_1 + \eta_2)/\tau$ holds, it results in an under-damped, wave-like behavior characteristic for the low-temperature phenomena. In the opposite case, $\alpha < (\eta_1 + \eta_2)/\tau$ leads to an over-damped solution, characteristic for the over-diffusive propagation, having outstanding importance for heterogeneous materials discussed soon.
The $q$-representation,
\begin{align}
	\tau \partial_{tt} \mathbf q + \partial_t \mathbf q = \alpha \nabla \nabla \cdot \mathbf q + \eta_1 \partial_t \Delta \mathbf q + \eta_2  \partial_t \nabla \nabla \cdot \mathbf q - \alpha \nabla Q_v.
\end{align}
does not reflect the Fourier resonance immediately, yet that could be achieved with $\eta_1=0$, viz., $L_2^{\textrm{dev}} = - L_2^{\textrm{A}}$, therefore all off-diagonal elements are zero. A more restrictive situation is when both components are zero, thus only the spherical part contributes to the time evolution.

Concerning the MCV equation, staggered discretization is suggested for numerical solutions. It is not different here either, mainly because it does not seem possible to realize $q$-type time-dependent boundary conditions using Eq.~\eqref{gkt}. This statement holds for analytical solutions as well. Otherwise, nonphysical solutions emerge, such as negative temperature, indicating the violation of maximum principle \cite{ProWei12b} or others being even more difficult to realize obtained with COMSOL \cite{RietEtal18}. Here, we note that the situation does not differ from mechanics with non-Hookean (rheological) models, it is recommended to avoid the conventional algorithms \cite{FulEtal20, FulEtal22}. The GK equation (and thus the MCV) is proved to be mathematically well-posed, convergent, and fulfilling the maximum principle \cite{RamosEtal23}.

Additionally, let us note here that a constitutive equation consisting of terms of $q$, $\partial_x q$ and $\partial_{xx} q$ \cite{RogCimm19} (in 1D) is not admissible as $\partial_x q$ is in a different function space than $q$ and $\partial_{xx} q$, and that can best be seen from Galerkin-type analytical solutions \cite{DiErnl11b, FehKov21}.

\subsubsection{Notes on the semi-empirical temperature} Finally, we note that the previously mentioned semi-empirical temperature approach (see Eq.~\eqref{seT}) can lead to an evolution equation being analogous with Eq.~\eqref{gkt} when $\dot \beta = f(T,\beta) = (T-\beta)/\tau$, where $\tau$ is supposed to be the relaxation time, and $\lambda=\lambda(T)$, $e=e(T)$ \cite{FriCim96}. After eliminating $T$, one achieves
\begin{align}
	c_v \tau \partial_{tt} \beta + c_v \partial_t \beta = \lambda \Delta \beta + \nabla \lambda \cdot \nabla \beta,
\end{align}
which model tested for second sound modeling successfully \cite{FriCim96, FriCim95}. However, it failed to model the ballistic effects properly; thus, its thermo-mechanical version is also developed and will be discussed later in more detail.

Alternatively, further modifications of $\dot \beta$ allow to recover the GK equation on a  constitutive level \cite{CimmSellJou10}. That is, if $\dot \beta = (T-\beta)/\tau + 3 l^2/\tau \Delta \beta$ holds, Eq.~\eqref{gk1} is recovered with assuming that the coefficients are inherited from kinetic theory. 
Beyond that, an additional nonlinear term into the time evolution of $\beta$, motivated by phonon hydrodynamics \cite{CimmSellJou10},
\begin{align}
	\dot \beta = (T-\beta)/\tau + 3 l^2/\tau \Delta \beta + f_1(e,\beta) (\nabla \beta)^2,
\end{align}
and that leads to 
\begin{align}
	\tau \dot {\mathbf q} + \mathbf q = - \lambda \nabla T + l^2 (\Delta \mathbf q + 2 \nabla \nabla \cdot \mathbf q) + \frac{2 \tau}{c_v T} \mathbf q \cdot \nabla \mathbf q, \label{eqgktc}
\end{align}
although the exact role of the last term in a constitutive equation is not yet clearly recovered, supposed to have a role in thermal rectification in nanosystems \cite{CarloEtal23}.
We also refer to \cite{CimmSellJou09} for a more thorough discussion of the procedure.

\subsection{Two-temperature models.} The general idea for a two-temperature model is similar to the continuum interpretation of the GK equation. There are two interacting subsystems, each having a temperature $T_1$ and $T_2$, and these subsystems are coupled through a heat transfer term in their energy balance \cite{Sobolev97, Sobolev16}, 
\begin{align}
	\rho_i c_i \partial_t T_i+ \nabla \cdot \mathbf q_i = h (T_j - T_i) + Q_i, \quad \mathbf q_i = - \lambda_i \nabla T_i, \quad (i=\{1,2\}, j=\{1,2 \ | j \neq i \}, \label{ttf1}
\end{align} 
in which $h$ is an intrinsic heat transfer coefficient, each system can likewise possess separate heat sources $Q_i$. Each subsystem obeys Fourier's law, but this is not a requirement. For low-temperature or nanoscale applications, further possibilities appear, such as coupled Fourier and MCV \cite{AndTamm06, VazqRio12, Nosko23}, or coupled MCV and GK equations \cite{Sobolev94, KovFehSob22}. Another possibility is if one of the subsystems has a very high thermal conductivity compared to the other, which makes the subsystem equilibrate much faster, i.e., modeled as a lumped capacitance, similarly to phonon-electron interaction \cite{Tzou95, WhiteEtal14}. It separates the time scales, and the corresponding evolution equation reduces to an ordinary differential equation, reads
\begin{align}
	\rho_1 c_1 \partial_t T_1 + \nabla \cdot \mathbf q_1 &= - h (T_2 - T_1) + Q_1, \quad \mathbf q_1 = - \lambda_1 \nabla T_1,  \label{ttf2} \\
	\rho_2 c_2 \frac{\textrm{d} T_2}{\textrm{d} t} &= h(T_2 - T_1) + Q_2 \label{ttf3}.
\end{align} 
That procedure can easily be extended for larger systems with proper coupling terms. However, at this moment, let us restrict ourselves only to \eqref{ttf1}. While it is neither a non-Fourier model nor directly related to the GK equation, they share one similarity, which motivates why we discuss that model here. That similarity becomes visible in the $T_i$-representation of Eq.~\eqref{ttf1},
\begin{align}
	C_i=\rho_i c_i, \quad \tau \partial_{tt} T_1 + \partial_t T_1 = \Lambda \Delta T_1 + l^2 \partial_t \Delta T_1 - \gamma \Delta \Delta T_1 + Q_1 + Q_2 +  \frac{C_2}{h(C_1 + C_2)} \partial_t Q_1 - \frac{\lambda_2}{h(C_1 + C_2)} \Delta Q_1 \label{ttm0}
\end{align}
with the coefficients
\begin{align}
	\tau= \frac{C_1 C_2}{h (C_1 + C_2)}, \quad \Lambda = \frac{C_1 \alpha_1 + C_2 \alpha_2}{C_1 + C_2}, \quad l^2=\tau (\alpha_1 + \alpha_2), \quad \gamma = \tau \alpha_1 \alpha_2 ,
\end{align}
standing for an analogy with the Eq.~\eqref{gkt}. At first glance, it might seem natural to use that model to interpret the GK-coefficients and might provide some additional insight into how to calculate the parameters in Eq.~\eqref{gkt}. However, despite the analogous $T$-representation (and the additional fourth-order term), Eq.~\eqref{ttf1} describes a different phenomenon, restricted to Fourier heat conduction in systems being easily identifiable and separable. On the contrary, the GK equation fits naturally in the family of non-Fourier heat equations, assuring a transition from low-temperature to room-temperature problems. However, that two-temperature model has an indisputable advantage over the continuum-GK equation. In order to apply Eq.~\eqref{ttf1}, e.g., in evaluating experimental data, one has to decide prior to the solutions which temperature is measured and what are the two dominant components of the system (and with what ratio). Thus, knowledge of the experimental settings significantly reduces the number of fitted parameters, and eventually, only the heat transfer coefficient $h$ remains unknown. This is an ideal case but can occur in exceptional situations. In reality, sadly, this is not that straightforward. For instance, neither the exact components nor their ratio and material parameters are known for a rock sample. On top of this, one also defines the average temperature as $\bar T = (C_1 T_1 + C_2 T_2)/(C_1 + C_2)$, which could be a more reasonable option for mixtures, but without knowing $C_1$ and $C_2$, the fitting procedure can be cumbersome and not necessarily unique \cite{LauererLunev23, KovFehSob22}. Moreover, fitting a $\bar T$ is still viable for a decoupled system with $h=0$.
Furthermore, one can apply Eq.~\eqref{ttf1} for a homogeneous material in which the electrons and phonons have different temperatures \cite{Carlomagno19}, then the classical interpretation of the heat capacities $C_i$ disappears. 
Later, concerning the ballistic propagation on a nanoscale, we will meet similar approaches with coupled non-Fourier heat equations, however, those should not be mixed with two-temperature models.
Finally, we note that Eq.~\eqref{ttf1} might be interpreted as a particular version of the internal variable approach, and the consistent derivation of the two-temperature models through the second law of thermodynamics could result in further structures, e.g., introducing particular couplings between the heat fluxes. An excellent example is shown in \cite{NarEtal22}, in which both theoretical and experimental comparisons are performed for such a coupled model. This could open further discussion in the future in the development of two-temperature models.

\subsection{Over-diffusion and metamaterials.} The GK model was constructive for the detection of second sound in solids due to finding the window condition; see Eq.~\eqref{wc} \cite{GK66}. In this regard, the kinetic background was inevitable. 

Later on, numerous attempts were made to observe the same wave propagation in macroscale objects at room temperature \cite{MitEta95, Kam90}, nevertheless, none of them were reproducible or successful. Sadly, the kinetic theory is not more constructive under such conditions and cannot be utilized to predict the necessary properties for ambient conditions and material parameters. Instead, the continuum background suggested a way, as pointed out in connection with the parallel time scales appearing in the GK equation. The critical point is that not wave propagation is the sole heat conduction model that can show deviation from Fourier's law. Heterogeneous materials also possess multiple heat conduction channels (such as foams); hence the interacting internal parallel time scales are present and are a rough analogy for the normal and resistive processes. Let us recall Figure \ref{fig1}/B in which the over-diffusive propagation is experimentally observed. Independently, Lunev et al.~\cite{LunEtal22} also observed over-diffusion in metal foams, and it turned out that the two-temperature approach is not necessarily capable of properly catching these effects \cite{LauererLunev23}. 
Moreover, as we have seen previously, although a two-temperature model can provide additional insights for the experimental arrangement, it consists of too many parameters that do not simply make the fitting more complicated but can easily result in `over-fitting', or the fit itself is not unique.

This is when the continuum-GK equation shows itself to be useful. It does not simply turn up the idea that parallel heat transfer channels can produce deviation, but Eqs.~\eqref{ebal}+\eqref{gk11} together can characterize various heterogeneous materials with effective parameters, shown in Figure \ref{fig2} for demonstration. Let us recall that the corresponding measurements are heat pulse (flash) experiments wherein the rear side temperature history is recorded (revisit Fig.~\ref{fig1}) and used as a standard method to find the thermal diffusivity. The evaluation with Fourier's law provides a thermal diffusivity valid only after a certain time interval, depending on the matrix material and the particular source of heterogeneity. This Fourier's effective thermal diffusivity ($\alpha_\textrm{F} $) could be interpreted as a `long-time approximation' of the real one for which the GK equation adds a leading order correction. This is more noticeable if we use the structure of the GK equation \eqref{gkt},
\begin{align}
	\alpha_\textrm{F} = \frac{1}{2} \left ( \alpha_\textrm{GK} + \frac{\eta_1+\eta_2}{\tau}\right ), \label{gkexp}
\end{align}
which relation is proved experimentally \cite{FehEtal21}. In other words, Eq.~\eqref{gkexp} expresses that the effects causing the deviation will vanish after a certain time interval, in some sense `averaged'. The GK equation leads to different thermal diffusivity $\alpha_\textrm{GK}$ in the conventional sense, for which all experiments showed $\alpha_\textrm{GK} < \alpha_\textrm{F}$, which stands as a requirement for over-diffusion in the light of Eq.~\eqref{gkexp}. In analogy with the difference between the dynamic and static deformation modulus from mechanics \cite{MortezaEtal20}, this can be interpreted similarly: while the measurement disregards the fast transients, or even closer to the static case, Fourier's thermal conductivity will be measured. Considering such a dynamic measurement for heterogeneous materials, the factor $\lambda_\textrm{F}/\lambda_\textrm{GK} = (1 + (\eta_1+\eta_2)/(\alpha_\textrm{GK} \tau))/2$ characterizes the ratio of static (Fourier) and dynamic (Guyer-Krumhansl) thermal conductivities.

\begin{table}[]
	\centering
	\caption{Typical relaxation times found for various materials, based on \cite{TammaZhou98}.} \label{tab1}
	\begin{tabular}{c|c|c|c}
		Material (metals) & Relaxation time	[$10^{-15}$ s] & Material (heterogeneous) & Relaxation time [s]  \\ \hline
		Na         \cite{HutchEtal64}                                                 & 31                                                                        & Ballotini  ($2--2.5$ mm)   \cite{Kam88}                                         & 13.34                                                          \\
		Cu        \cite{HutchEtal64}                                                  & 27                                                                        & Sand ($0.3--0.5$ mm)        \cite{Kam88}                                       & 3.61                                                           \\
		Ag    \cite{HutchEtal64}                                                     & 41                                                                        & Granule CaCO$_3$ ($1.3--1.6$ mm)   \cite{Kam88}                                & 8.59                                                           \\
		Au        \cite{HutchEtal64}                                                  & 29                                                                        & NaHCO$_3$                   \cite{Kam90}                                       & 28.7                                                           \\
		Ni       \cite{HutchEtal64}                                                   & 10                                                                        & Glass ballotini                    \cite{Kam90}                                & 10.9                                                           \\
		Fe            \cite{HutchEtal64}                                              & 10                                                                        & Szászvár formation                 \cite{FehEtal21}                                & 0.648                                                          \\
		Pt        \cite{HutchEtal64}                                                  & 9                                                                         & Tisza metamorf complex     \cite{FehEtal21}                                         & 0.35                                                           \\
		Cr       \cite{QiuTien93}                                                      & 3                                                                         & Szársomlyó limestone formation  \cite{FehEtal21}                                    & 0.547                                                          \\
		V  \cite{QiuTien93}                                                            & 2                                                                         & Boda claystone formation   \cite{FehEtal21}                                         & 0.4                                                            \\
		Nb      \cite{QiuTien93}                                                       & 4                                                                         & Metal foam ($0.5--1$ mm inclusions)  \cite{FehKov21}                              & 0.3                                                            \\
		W    \cite{QiuTien93}                                                          & 10                                                                        & Capacitor ($3.9$ mm)              \cite{Botetal16}                                 & 0.51                                                           \\
		Pb \cite{QiuTien93}                                                          & 5                                                                         & Carbon foam ($85--90$\% porosity)       \cite{FehEtal22}                           & 0.16                                                          \\ \hline
	\end{tabular}
\end{table}

This leads to the pioneering idea that materials could be designed to enhance over-diffusion artificially or the other way around; if one designed (or produced) a particular structure, then how to predict such effects? That would establish a novel thermal design methodology to create a new class of thermal metamaterials. In parallel, it offers a highly efficient way to characterize heterogeneous materials since there is no need for highly detailed, computationally intensive simulations. Moreover, that would make possible even the real-time monitoring of complex structures. Such thermal design methodology would require knowledge of how each heat transfer mode contributes to the overall thermal behavior. Heat transfer is inescapable, it occurs in any heterogeneous structure; however, the GK parameters probably depend on them in a nonlinear or non-monotonous way. For a relatively simple but not necessarily successful demonstration, one can attempt to use a 3D-printed structure. Here, we note that even manufacturing technologies, especially 3D printing, can be significantly affected by understanding how a porous structure conducts heat. Table \ref{tab1} reflects notable differences between the relaxation times of pure and heterogeneous materials. The models (MCV, GK, or DPL) must be interpreted effectively in the latter case.

\section{Heat equations including ballistic modes}
Contrary to the previous sections, we follow a different pathway to overview models, including ballistic heat conduction. The first reason is that ballistic propagation is usually accompanied by a second sound or diffusion, depending on the particular situation for which their usual treatment considerably differs. The second reason is that ballistic heat conduction is a coupled phenomenon, heat is transported with an elastic wave propagation (first sound) from a continuum point of view. Therefore, it seemed better to present these models with their experimental background.
The most typical experimental conditions are the following:
\begin{itemize}
	\item low-temperature, macroscale objects, requiring extraordinarily pure samples \cite{NarDyn72a, NarDyn75};
	\item room temperature, nanoscale objects such as thin layers and nanotubes \cite{Chen01, WangGuo10a, ZiabEtal18};
	\item rarefied (low-pressure) gases, which needs to generalize Newton's law of fluids, too \cite{Devienne65, Struc05, Shen06b, Sharipov15b}.
\end{itemize}
Their kinetic interpretation helps understand their common property: a significant part of the energy carriers (such as phonons or actual molecules) can propagate without collisions, i.e., the mean free path becomes comparable to the characteristic length of the object. It does not mean that all the energy carriers propagate that way, but the ballistic effects notably contribute to the overall energy transport. 
In the cases mentioned above, the observations of ballistic propagation are performed differently as it is not the temperature history only for which the ballistic effects manifest or the temperature itself is not directly measurable, contrary to the room-temperature heat pulse experiments. 

\subsection{Modeling of low-temperature heat conduction.}
Let us recall Figure \ref{fig1}, which shows the temperature history recorded in a remarkably intriguing heat pulse experiment, called NaF experiments, performed by McNelly et al.~\cite{McNEta70a}. Interestingly, all propagation modes are present in a single experiment: the longitudinal and transversal ballistic propagation modes (which waves are the fastest), the second sound, and diffusion. That remarkable outcome makes this experiment a benchmark problem in this field. 

\subsubsection{Phonon models.} Starting with a kinetic background, we note that, unfortunately, the thermo-mechanical GK system \eqref{bte2}-\eqref{bte3} is not tested; thus, it is unknown whether this coupled model is suitable to model these experiments. On the contrary, the momentum series expansion of the Boltzmann equation, Eq.~\eqref{bal1}, recalled here,
\begin{equation}
	\frac{\partial u_{\langle m \rangle}}{\partial t} + \frac{m^2}{4m^2-1}c\frac{\partial u_{\langle m-1 \rangle}}{\partial x}+c\frac{\partial u_{\langle m+1 \rangle}}{\partial x}=\left \{ \begin{array}{ll}
		\displaystyle
		0 \ & \ m=0 \\
		-\frac{1}{\tau_R}u_{\langle 1 \rangle} \ & \ m=1 \\
		- \left( \frac{1}{\tau_R}+\frac{1}{\tau_N} \right )u_{\langle m \rangle} \ & \ 2\leq m\leq M
	\end{array} \right., \nonumber
\end{equation}
performs well with minor discrepancies, which mostly originate from the properties of that approach. First, one has to utilize at least $M = 30$ momentum equations to obtain a good approximation for the speed of sound (ballistic speed of phonons), that is, for a general three-dimensional problem a 30$^{\textrm{th}}$-order tensor would appear in the model. Moreover, in principle, one would need infinitely many momentum equations for the exact value \cite{MulRug98}. 
Despite these shortcomings, the phonon hydrodynamic equations can be reasonably simplified, and the moments up to $M=3$ provide an acceptable approximation \cite{DreStr93a}, which reads
\begin{align}
	\dot e + c^2 p_x &= 0,  \label{ph1} \\
	\dot p + \frac{1}{3} e_x +N_x &= -\frac{1}{\tau_R} p,  \\
	\dot N + \frac{4}{15} c^2 p_x &= - \left( \frac{1}{\tau_R}+\frac{1}{\tau_N} \right ) N, \label{ph3}
\end{align}
where 
\begin{equation}
	e=hcu, \quad p_i=hu_i, \quad N_{\langle ij \rangle}=hcu_{\langle ij \rangle} ,
\end{equation}
are the energy density, momentum density, and the deviatoric part of the pressure tensor, respectively, have been found for the one-dimensional case in Eq.~\eqref{bal1}. 

Staying with the phonon hydrodynamic background, we have to mention here the work of Ma \cite{Ma13a, Ma13a1} as well, who developed a so-called hybrid phonon gas model based on the papers of Rogers \cite{Rog72a} and Landau \cite{LandauVIeng}. That work was motivated to include both the longitudinal and transversal ballistic modes simultaneously with keeping a one-dimensional model instead of a proper three-dimensional description. To achieve this goal, the internal energy $e$ is divided into two parts, $e=e_l + e_t$, together with the heat flux $ q = q_l +  q_t$, each of them being connected to the corresponding longitudinal and transversal ballistic mode. Furthermore, following Rogers, it is supposed that the heat flux $\mathbf q$ is proportional to the velocity of phonon gas, hence $\mathbf q = e \mathbf v$, and the classical equation of motion, the Navier-Stokes equation,
\begin{align}
	\rho \partial_t \mathbf v + \rho (\mathbf v \cdot \nabla) \mathbf v = - \nabla p + \eta \Delta \mathbf v + \left ( \xi + \frac{1}{3} \eta \right ) \nabla \nabla \cdot \mathbf v,
\end{align}
is modified accordingly and reduced to one dimension. Here, $p$ denotes the scalar pressure, $\eta$ and $\xi$ are for the shear and bulk viscosity. According to Rogers, in such a low-temperature situation, the shear viscosity tends to zero when the normal processes dominate the propagation. Consequently, bulk viscosity plays an essential role in the damping mechanism. For that model, Landau's complex viscosity equation is adopted,
\begin{align}
	\xi = \frac{2 \tau e_t}{3(1 - i \omega \tau)},
\end{align}
in which $\tau^{-1} = \tau_R^{-1} + \tau_N^{-1}$, and $\omega$ stands for the angular frequency of heat waves, and obtains the evolution equation
\begin{align}
	\partial_t q + \frac{1}{\tau_R} q =- \frac{1}{3} \partial_x e+ \frac{2 \tau}{3(1 - i \omega \tau)} \partial_{xx} q. \label{hpg1}
\end{align}
Eventually, Eq.~\eqref{hpg1} stands for both $q_t$ and $q_l$, with separate balances for the internal energies $e_l$ and $e_t$, and the relation $e=e_l + e_t$ realizes the coupling to obtain the temperature evolution. This appears to be a particular extension of a two-temperature model in which two constitutive equations of GK-type are coupled, thus it is neither a pure phonon nor a continuum model. 

\subsubsection{Continuum models.} 
\begin{figure}[]
	\centering
	\includegraphics[width=13cm,height=8cm]{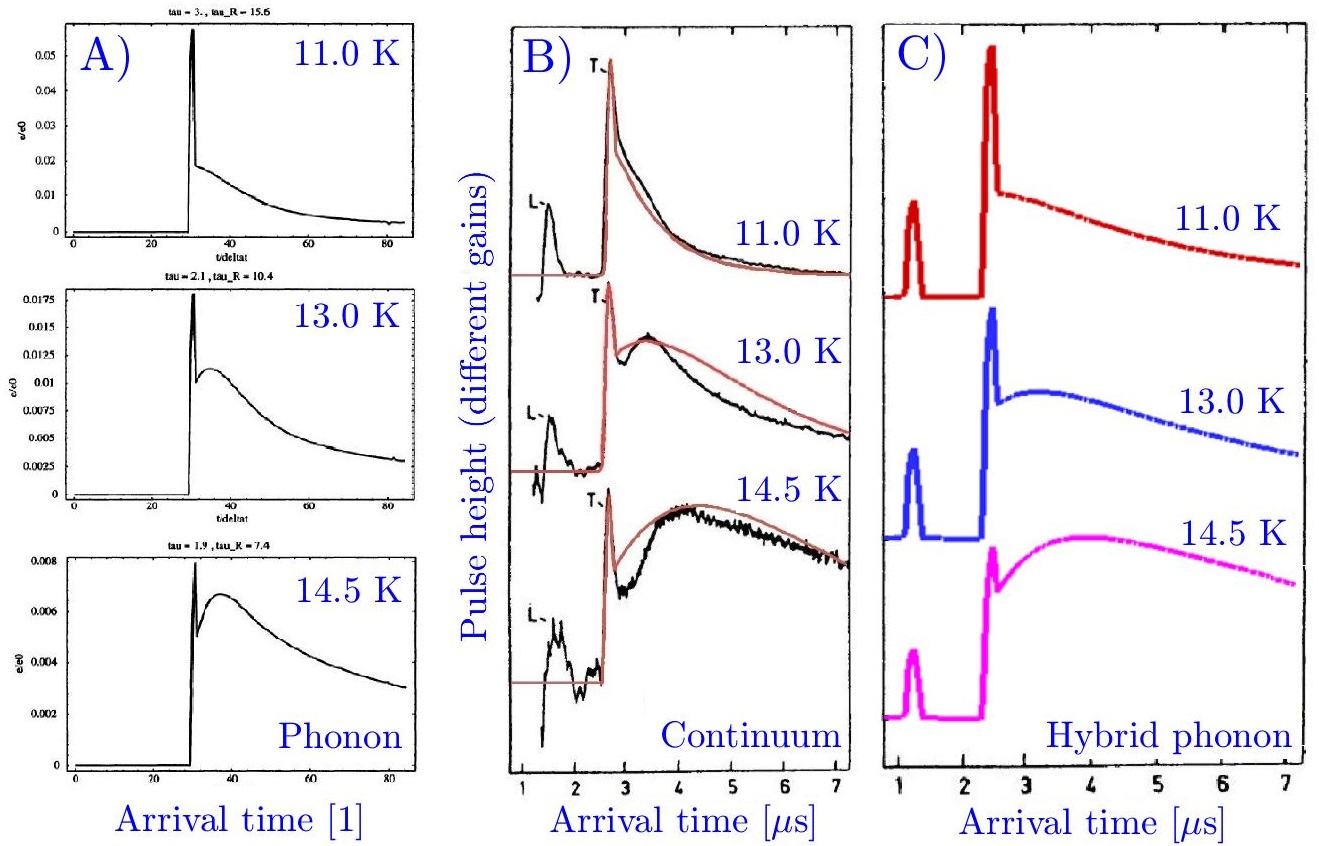}
	\caption{McNelly's NaF experiments are evaluated with (A) phonon \cite{MulRug98, KovVan18}, (B) continuum \cite{KovVan18}, and (C) hybrid phonon models \cite{Ma13a1}. Part (B) also shows the originally recorded data for comparison.}
	\label{fig7}
\end{figure}

In the analogy of the MCV and GK equations, it is also possible to achieve analogy with the phonon hydrodynamic approach \eqref{ph1}-\eqref{ph3}, in the same way as the GK equation is derived \cite{KovVan15}, alternatively, in a GENERIC framework \cite{SzucsEtal21}. However, as Eq.~\eqref{ph3} suggests, we need to include a second-order tensor in the state space as well, viz., $s(e,\mathbf q, \mathbf Q) = s_{\textrm{le}}(e) - m_1/2 \mathbf q^2- m_2/2 \mathbf Q: \mathbf Q$. Concerning the entropy flux, Eq.~\eqref{gkj} remains suitable, and the current multiplier $\mathbf B$ is helpful to realize the coupling between $\mathbf q$ and $\mathbf Q$.
It becomes more transparent in the corresponding entropy production,
\begin{align}
	\sigma_s = \mathbf q \cdot \left (- \rho m_1 \partial_t \mathbf q + \nabla \frac{1}{T} \mathbf I + \nabla \cdot \mathbf B \right ) - \rho m_2 \mathbf Q : \partial_t \mathbf Q + \mathbf B : \nabla \mathbf q, \label{spbal}
\end{align}
resulting in the Onsagerian relations
\begin{align}
	- \rho m_1 \partial_t \mathbf q + \nabla \frac{1}{T} + \nabla \cdot \mathbf B &= l \mathbf q, \label{Qbal1} \\
	- \rho m_2 \partial_t \mathbf Q &= L_{11} \mathbf Q + L_{12} \nabla \mathbf q, \label{Qbal2}\\
	\mathbf B = L_{21} \mathbf Q + L_{22} \nabla \mathbf q, \label{Qbal3}
\end{align}
with $l\geq0$, $L_{11}\geq0$, $L_{22}\geq0$, and $L_{11} L_{22} - L_{12} L_{21} \geq 0$, expressing the positive semi-definiteness of the inequality \eqref{spbal}. For the compatibility, we take $L_{22}=0$, thus identifying $\mathbf B = L_{21} \mathbf Q$. In that approach, $\mathbf Q$ is no longer directly proportional with $\nabla \mathbf q$ on the contrary to the continuum GK model, Eq.~\eqref{Qbal2} describes its time evolution, The removal of the non-equilibrium contribution of $\mathbf Q$ from the entropy density naturally relaxes $\mathbf Q$ to $\nabla \mathbf q$. That picture is also helpful in the interpretation of the current multiplier. Consequently, Eq.~\eqref{gk2b} is extended with the time derivative of $\mathbf Q$, with a relaxation time denoted by $\tau_Q$. Reducing the system \eqref{Qbal1}-\eqref{Qbal3} to the one spatial dimension leads to 
\begin{align}
	\tau_q \partial_t q + q &= -\lambda \partial_x T+l \partial_x Q, \label{bal2a} \\
	\tau_Q \partial_t Q + Q&=l \partial_x q, \label{bal2b}
\end{align}
where $\tau_q$ and $\tau_Q$ are the corresponding relaxation times. Eqs.~\eqref{bal2a}-\eqref{bal2b} found as a hyperbolic approximation of a more general parabolic set of equations \cite{KovVan15}. Its complete isotropic representation is presented in \cite{FamaEtal21}, showing the detailed relationships between the coefficients for various situations.
Here, we emphasize that these relaxation times are independent of each other, and the compatibility with the phonon model \eqref{ph1}-\eqref{ph3} is formal but still comparable \cite{KovVan15, KovEtal18rg}. One can apply the kinetic interpretation, which is a matter of choice, but not the only option. When one attempts to model the NaF experiments, for instance, then the coefficient $l$ can be adjusted to obtain the exact speed of sound of ballistic mode, together with the relaxation times; overall, having the same number of parameters to be fit \cite{KovVan16, KovVan18}. Figure \ref{fig7} summarizes the modeling capabilities on experimental data of different models.
That difference becomes more crucial for rarefied (real) gases, and what flexibility appears to be an advantage here will become a disadvantage there. Therefore, no such `universal' model can be suggested for any heat conduction problems, but there are alternatives from which one can choose the best one for a particular task.

Evaluation of the NaF experiments with both the phonon and continuum models showed the temperature dependence of relaxation times, besides the thermal conductivity \cite{KovVan18}. Interestingly, as mentioned in the case of the MCV equation, the direct implementation of such nonlinearities would open new perspectives toward modeling low-temperature phenomena. 

It also stands as further motivation to investigate thermo-mechanical models, sharing that perspective with Frischmuth and Cimmelli \cite{FriCim95, FriCim96, FriCim98}. Here, we restrict ourselves to the small deformation regime. They attempted to improve the modeling capabilities of the semi-empirical temperature approach by introducing mechanics through thermal expansion. While the mechanical subsystem remains completely classical, they modify the thermal part by exchanging the heat flux $\mathbf q$ to $\mathbf g = \nabla \beta$, postulated as a new state variable, for which an MCV-like equation is supposed to be valid such as
\begin{align}
	C \partial_t T &= \lambda \nabla \cdot \mathbf g + r, \label{ebals}\\
	\tau \partial_t \mathbf g &= \nabla T - \mathbf g, \label{eg}
\end{align}
where $r$ stands for a source term, including the mechanical contribution $r \sim T_0 \nabla \cdot \mathbf v$, and $C$ is specific heat capacity. We note that the energy balance \eqref{ebals} is also arbitrarily modified, its thermodynamic compatibility is unclear. However, the modification was necessary to match the units in Eq.~\eqref{eg}, otherwise, the thermal conductivity disappears, and eliminating $\mathbf g$ still can recover the $T$-representation of the MCV equation in the form
\begin{align}
	\tau \partial_{tt} T + \partial_t T = \alpha \Delta T + \frac{r}{C} + \tau \frac{\dot r }{C}, \label{mcvs}
\end{align}
but Eq.~\eqref{mcvs} cannot reflect the underlying assumptions and model details. More importantly, it cannot justify the thermodynamic compatibility of the system \eqref{ebals}-\eqref{eg}. Eq.~\eqref{eg} must be supplemented with the dynamic equation $\dot \beta = f(T, \beta)$, adding further flexibility to the model. Sadly, they did not manage to compare its solutions directly to experimental data, but in a test solution, both the ballistic and second sound modes are present.

\subsubsection{Notes on the thermo-mechanical couplings.} Despite the previous model's questionable points, it is reasonable to couple thermal expansion to the MCV equation \cite{BallEtal20}. Keeping thermodynamic compatibility in mind, we start with the internal energy $e$ (in 1D) by recalling Eq.~\eqref{etm}
\begin{align}
	e = c T + \frac{E}{2 \rho} \varepsilon^2 + \frac{E \chi}{\rho} T_0 \varepsilon, \label{emech}
\end{align}
where, we recall that, $\chi$ stands for the thermal expansion coefficient, and $E$ is Young's modulus. We note that for a low-temperature situation, this could be valid for minimal temperature excitation for a narrow temperature interval due to the emerging nonlinearities. This is avoided here. Furthermore, in a general three-dimensional treatment, separating the spherical and deviatoric terms of the strain tensor $\boldsymbol \varepsilon$ would be expedient. The mechanical contribution in the energy balance appears as a $\sigma \partial_t \varepsilon$ source term
\begin{align}
	\rho \partial_t e + \partial_x q = \sigma \partial_t \varepsilon,
\end{align}
in which the stress $\sigma = E \varepsilon - E \chi (T - T_0)$, and that energy balance is coupled to the system
\begin{align}
	\rho \partial_t v - \partial_x \sigma &=0, \label{mcvte1} \\
	\partial_t \varepsilon &= \partial_x v, \\
	\tau \partial_t q + q &= - \lambda \partial_x T. \label{mcvte2}
\end{align}
In summary, this is a classical thermo-mechanical model including isotropic Hooke's elasticity for stress, and exchanging Fourier's law to the MCV equation, referred to as MCV-TE model in \cite{BallEtal20}. It is worth investigating the temperature representation for a simplified ($\tau=0$, i.e., Fourier), but a three-dimensional system, derived in \cite{FulEtal18e},
\begin{align}
	\frac{1}{c_{||}^2} \partial_{tt} (\gamma_1 \partial_t T - \lambda \Delta T) = \Delta \left [ \left ( \rho c - \frac{6 E^{\textrm{dev}} E^{\textrm{sph}} \chi^2 T_0}{E^{\textrm{sph}} + 2 E^{\textrm{dev}}} \right ) \partial_t T - \lambda \Delta T \right ], \label{ful1} 
\end{align} 
with parameters
\begin{align}
	c^2_{||} = \frac{E^{\textrm{sph}} + 2E^{\textrm{dev}}}{3 \rho}, \quad \gamma_1 = \rho c - 3 E^{\textrm{sph}} \chi^2 T_0, \quad \gamma_2 = \rho c - \frac{6 E^{\textrm{dev}} E^{\textrm{sph}} \chi^2 T_0}{E^{\textrm{sph}} + 2 E^{\textrm{dev}}}, \quad E^{\textrm{sph}}=3K, \quad E^{\textrm{dev}}=2G
\end{align}
in which $c_{||}$ is the longitudinal elastic wave propagation speed, $K$ and $G$ are the bulk and shear moduli. Interestingly, the classical thermal diffusivity $\alpha=\lambda/(\rho c)$ is modified by thermal expansion as $\hat \alpha = \alpha (\rho c / \gamma_2)$; if the thermal expansion coefficient $\chi=0$, $\gamma_1=\gamma_2=\rho c$ follows, and the correction factor $(\rho c / \gamma_2)=1$ holds. In other words, neglecting thermal expansion effects in the evaluation of the experimental data detunes the thermal diffusivity, e.g., by 6\% for aluminum. 

Considering the NaF experiments, the coefficients $\rho(T)$ and $\chi(T)$ are unknown, and therefore it seems reasonable to find $\rho$ and $\chi$ for each reference temperature, but this has not been done so far. Similarly, for Eqs.~\eqref{bal2a}-\eqref{bal2b}, one must apply the measured speeds of first and second sounds as constraints for the fitted parameters, and that again reduces the number of free parameters. The analytical and numerical solutions of ballistic equations are analogous to the previous models, and the staggered spatial discretization is more advantageous from a numerical point of view as it eases the implementation of boundary conditions. 

In fact, the MCV-TE model \eqref{mcvte1}-\eqref{mcvte2} is not a completely novel idea. Lord and Shulman \cite{LorShu67} were the first who proposed to couple the MCV equation with thermal expansion. At that time, much before the work of Coleman et al.~\cite{ColEtal86}, they formulated the anisotropic MCV model as 
\begin{align}
	\mathbf A \cdot \dot {\mathbf q} + a \dot {\mathbf q} + \mathbf q = b \nabla T + \mathbf B \cdot \nabla T,
\end{align}
where $a, b, \mathbf A$, and $\mathbf B$ are all considered as material parameters so that the requirements for such model provided later by \cite{ColEtal86} are missing (see Eqs.~\eqref{INAN} and \eqref{INAN2}). Furthermore, these parameters are not necessarily positive, e.g., Fourier's law is recovered with $b=-\lambda$, ($\lambda>0$). However, only the simplest isotropic form is used later with constant scalar coefficients ($\mathbf A= \mathbf 0$, $a=\tau$, $b=-\lambda$, $\mathbf B= \mathbf 0$). In the state space, they used the small strain tensor  $\boldsymbol \varepsilon$ and $T$, so that $e=e(T, \boldsymbol \varepsilon)$, and $s=s(T,\boldsymbol \varepsilon)$. The energy balance is formulated with the help of the Helmholtz free energy $\varphi (T,\boldsymbol \varepsilon)=e(T, \boldsymbol \varepsilon) - T s(T, \boldsymbol \varepsilon)$,
\begin{align}
	\nabla \cdot \mathbf q = \rho T \left ( \frac{\partial^2 \varphi}{\partial T^2} \dot T + \frac{\partial^2 \varphi}{\partial \boldsymbol \varepsilon \partial T} \dot{\boldsymbol \varepsilon} \right ). \label{ebal2}
\end{align}
Eliminating $\mathbf q$ using Eqs.~\eqref{mcv1} and \eqref{ebal2} leads to a complex nonlinear model in which terms $\dot T^2$, $\dot{\boldsymbol \varepsilon} \dot T$, and $\dot{\boldsymbol \varepsilon}^2$ are neglected, and yields
\begin{align}
	\lambda \Delta T = - \rho T \left[ \frac{\partial^2 \varphi}{\partial T^2} ( \dot T + \tau \ddot T) + \frac{\partial^2 \varphi}{\partial {\boldsymbol \varepsilon} \partial T} (\dot{\boldsymbol \varepsilon} + \tau \ddot{\boldsymbol \varepsilon}) \right ], \label{ls1}
\end{align}
that can be solved together with the equation of motion and with a particular free energy function, resulting in a more complicated structure for the time evolution than Eq.~\eqref{ful1}. Dhaliwal and Sherief utilize the same concept \cite{DhaliSher80} with anisotropic thermal conductivity, though with scalar relaxation time, for which uniqueness is proved for homogeneous boundary conditions. That property is inherited to \eqref{ls1}, too.
Compared to the MCV-TE model, the one-dimensional versions can be identical; thus, the Lord-Shulman theory is helpful in modeling ballistic heat conduction experiments. 
Since the heat flux evolution equation is arbitrarily chosen, further, more general thermoelastic models can be derived on the same basis. 

The next class of thermoelastic models is called temperature-rate dependent equations, where the time derivative of the temperature $\dot T$ also influences the stress and heat flux evolutions. This is developed by Green and Lindsay \cite{GreLin72a}. Considering the material to be anisotropic with constant coefficients, the heat flux, and stress are
\begin{align}
	\mathbf q = -T_0 (\mathbf b \dot T + \mathbf K \nabla T), \quad \boldsymbol{\sigma} = \mathbf C^{(4)} \boldsymbol{\varepsilon} + \mathbf A (T-T_0 + a\dot T),
\end{align}
where $a, \mathbf b$ and $\mathbf A$ are material parameters, $a$ might be related to the relaxation time, but basically, these parameters have an undiscovered physical interpretation.

The concept of elastic heat flow vector field by Ignaczak \cite{Ign90} is also worth mentioning. This is based on the assumption that an additional vector field $\boldsymbol{\beta}$ influences the resulting heat flux (similarly to an internal variable),
\begin{align}
	\mathbf q = - \lambda \nabla T + \boldsymbol{\beta}, \quad \dot{\boldsymbol{\beta}} = - a \frac{\nabla T}{T}
\end{align}
with $a>0$ parameter. Consequently, both the internal energy and entropy depend on $\boldsymbol{\beta}$, having importance in modeling low-temperature wave propagation. This inevitably results in a nonlinear evolution for the temperature field, specifically developed to find thermal soliton-like solutions. In a one-dimensional case, two characteristic wave speeds are propagating in the same direction, therefore describing different modes of heat conduction. However, that theory is not yet directly tested on ballistic experimental data. 

For a more thorough review and for further details with mathematical proofs, let us refer to \cite{Chand98, IgnOst09b, HetIgn99, HetIgn00}, and \cite{HetEtal09b}. Furthermore, we also want to highlight the book of Berezovski and Ván \cite{BerVan17b} for a broader discussion on the role of internal variables and thermodynamics concerning thermoelasticity, the corresponding numerical techniques are presented, too. In regard to thermomechanics, it is worth mentioning that from an effective point of view, although on significantly longer time scales, rheology can also stand as a factor for non-Fourier heat conduction \cite{FulEtal22, FulEta14m1}. For a profound mathematical treatment of even more complex situations related to thermopiezoelecricity, requiring anisotropy and further couplings with the electric field, let us refer to the recent works of Tibullo et al.~\cite{CiarlettaEtal22, TibuNunzi23}. These topics still have immense research potential. From a theoretical point of view, the development (or improvement) of the thermodynamic background is crucial, as it is helpful to consistently introduce the necessary couplings, revealing the functional dependencies and, in parallel, ensuring compatibility with the second law. From a practical point of view, such advanced models have various application possibilities, and their presence additionally motivates the improvement of analytical and numerical solution techniques, and therefore the referred mathematical analysis is inevitable.

\subsection{Micro and nanoscale heat conduction.}

\begin{figure}[H]
	\centering
	\includegraphics[width=13cm,height=7cm]{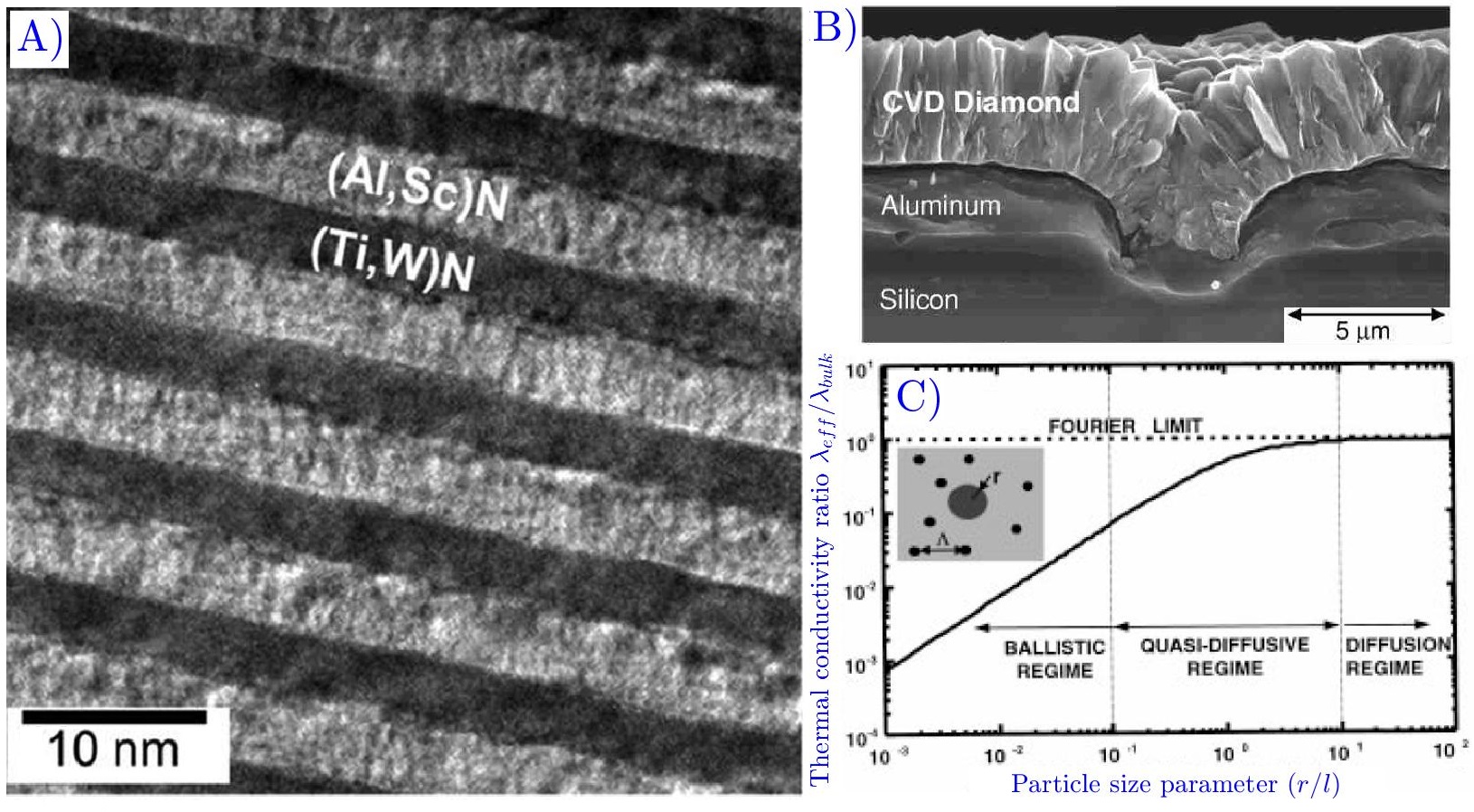}
	\caption{A) Magnified image of a diamond film applied to enhance heat removal \cite{SahaEtal16}. B) TEM micrograph of a superlattice of Ti$_{0.7}$W$_{0.3}$N and Al$_{0.72}$Sc$_{0.28}$N layers \cite{SchShiGood05}. C) The change of effective thermal conductivity of nanostructures with respect to the ratio of the nanoparticle size ($r$) and mean free path ($l$) \cite{Chen00b}.}
	\label{fig8}
\end{figure}

While the Knudsen number ($\textrm{Kn} = l/L$) serves as a natural connection between the macroscopic low-temperature and microscopic room-temperature heat conduction problems, there are some essential differences. First, the experimental background, e.g., the realization of experiments and how to measure the temperature, can greatly differ. In this respect, we refer to the paper of Goodson and Ju \cite{GoodJu99} for further details. Second, the low-temperature conditions can enhance couplings and nonlinearities, which are not necessarily present in a room-temperature environment despite the nanometer length scale. In contrast, the size dependence of thermal conductivity stands as a central question since even the steady-state heat transport is influenced by ballistic contributions \cite{Chen01}. In other words, that means $\lambda=\lambda(\textrm{Kn})$ in which $\lambda \rightarrow \lambda_0$ takes the macroscopic bulk limit when $\textrm{Kn} \ll 1$. However, the other direction ($\textrm{Kn} \sim 1$) is still unclear, and there are multiple approaches in this respect.
First, according to \cite{Maj93}, thermal conductivity cannot be defined when only ballistic propagation is present in the material, and thus radiative transfer occurs between the boundaries instead of conduction, standing as an alternative approach. Consequently, it is worth noting that the following models are valid only when the diffusive mode is also present, which provides a lower bound for the length scale. Second, these ballistic boundary scattering contributions add further thermal resistance to the structure and thus reduce the thermal conductivity \cite{Maj91, Chen00}. From a practical point of view, the superlattices, thin layers, and thermoelectric devices stand as outstanding examples \cite{Chen99, MinnichEtal09}. From an experimental point of view, the so-called time-domain thermoreflectance (TDTR) method is a widely used standard procedure to detect the thermal conductivity of such material structure as the reflected optical signal is sensitive to any temperature change \cite{RegnerEtal13, BeardoEtal20}. For a detailed experimental overview, we refer to the papers of Jiang et al.~\cite{JiangEtal18} and Saha et al.~\cite{SahaEtal11, SahaEtal16}. Since thin layers (and therefore superlattices) are used as a periodic structure in a microelectronic device, these naturally show anisotropic properties with strong temperature dependence \cite{Yao87, YuEtal95}. Here, the size dependence is meant to be related to the cross-plane thermal conductivity, depending on the number of layers (or period thickness) \cite{SahaEtal16, GuEtal18, VazVanKov20}. 
Figure \ref{fig8}/(A and B) present an example of thin films and superlattices, and (C) demonstrates the appearance of ballistic effects when the mean free path changes.

\subsubsection{Size-dependent thermal conductivity.} 
\begin{figure}[]
	\centering
	\includegraphics[width=14cm,height=5.5cm]{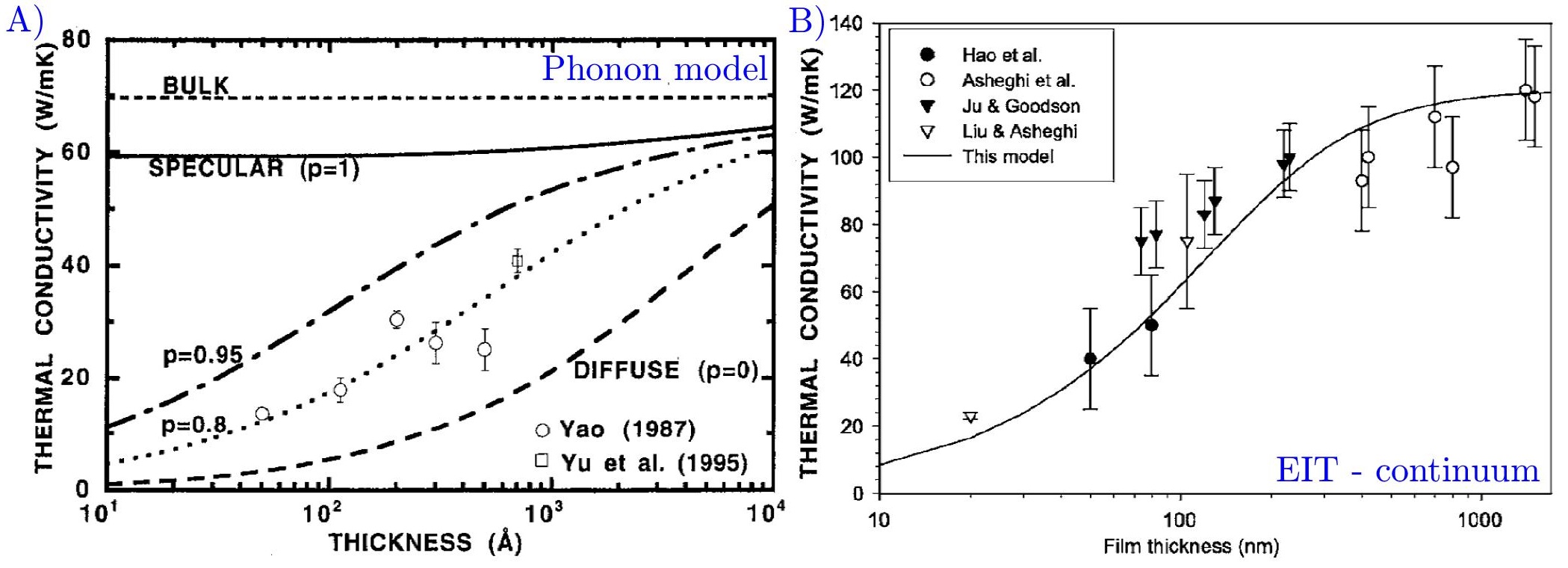}
	\caption{Comparing model predictions to experimental data. A) Phonon model with adjustable interface scattering parameter \cite{Chen00}. B) EIT continuum model utilizing Eq.~\eqref{eit3} with assuming $\omega=0$ and all $l_i$ mean free paths are equal \cite{AlvJou08}.}
	\label{fig9}
\end{figure}

A continuum model does not build the governing equations employing detailed transport mechanisms, we again begin with the kinetic theory background. It would be more accurate to use `structure-dependent' thermal conductivity instead of size dependence since it is not only the size that matters. For instance, the interface roughness, the material type, and the temperature conditions influence the effective thermal conductivity. Also, it is apparent now that Fourier's law using Debye's definition for thermal conductivity \eqref{macrok} cannot be valid at such scales, and one must utilize a more general treatment. This is more visible from the general definition of the bulk thermal conductivity from kinetic theory \cite{Chen97}, called the dispersion model,
\begin{align}
	\lambda = \frac{1}{3} \sum_p \int c(\omega)_p v(\omega)_p l(\omega)_p \textrm{d} \omega, \label{nanok}
\end{align}
in which the summation denotes different phonon branches, such as optical and acoustic, and the latter can be further separated into transverse and longitudinal modes. The integration considers the whole spectrum for each branch. Furthermore, $v(\omega)$ considers each branch's characteristic group velocity, and $c(\omega)$ stands for the frequency-dependent specific heat capacity; hence, the internal energy is also $\omega$-dependent. Eq.~\eqref{nanok} can be simplified with the so-called gray medium approximation for which the mean free path $l$ becomes independent of $\omega$. For instance, consider a superlattice with two components, one can estimate its bulk thermal conductivity by knowing the component properties
\begin{align}
	\lambda = \sum_{i=1}^2 \chi_i \lambda_i \big( 1- 1.5 p(1- \alpha_j/\alpha_i) A_{1}/\xi_i - 1.5(1-p) A_2/\xi_i\big), \quad p \approx \exp(-16 \pi^3 \delta^2/\omega^2),
\end{align}
where $\chi$, $\xi$, and $\delta$ are the relative layer thickness and film thickness normalized to the mean free path and mean interface surface roughness, respectively \cite{Chen97}. Here, $p$ stands for the interface scattering parameter suggested by Ziman \cite{Ziman01}, ranging from $0$ (diffuse) to $1$ (specular), see Figure \ref{fig9}/A for its influence. For a more flexible treatment, one can consider $p$ independent of the frequency $\omega$ and used as a fitting parameter as $p$ can significantly influence how the thermal conductivity decreases with the layer thickness. Additionally, $A_1$ and $A_2$ express the direction and spectral-dependent transmissivity, and their exact expression can be found in \cite{Chen97}.

More recent studies by Saha et al.~\cite{SahaEtal16, SahaEtal18} provides two methods to estimate the interface thermal resistance. Their starting point is that the overall thermal resistance of a superlattice is expressed as the sum of each component, i.e., $R_{\textrm{total}} = R_{\textrm{layer 1}} + R_{\textrm{layer 2}} + R_{\textrm{interface}}$, yielding
\begin{align}
	\frac{N R_{\textrm{interface}}}{L} = \frac{1}{\lambda_{\textrm{total}}} - \frac{1}{2} \left (\frac{1}{\lambda_1} + \frac{1}{\lambda_2} \right ),
\end{align}
for which the two components have the same thickness, and $L$ denotes the total thickness of the superlattice in which there are $N$ number of interfaces. In their first scenario, they suppose that $\lambda_1$ and $\lambda_2$ thermal conductivities are the same as the measured $240$ nm thin film, which implies a diffusive phonon scattering mechanism for thicknesses larger than the mean free path. The other scenario assumes the ballistic propagation of phonons; therefore, the thermal conductivities $\lambda_1$ and $\lambda_2$ do not contribute to the overall resistance, which could be reasonable for layer thicknesses being much smaller than the mean free path. Hence, after measuring the effective total thermal conductivity $\lambda_{\textrm{total}}$ for multiple situations, one can estimate how the interfaces contribute to the overall thermal behavior in a superlattice.

The above-discussed results are further extended with the work of Alvarez et al.~\cite{AlvEtal12, AlvEtal11}, in which the compatibility with kinetic theory is considered less rigorously but obtained multiple helpful $\lambda(\textrm{Kn})$ expressions depending on the boundary contributions of phonons and investigated their thermodynamic compatibility in the framework of EIT. They focus only on longitudinal flow in nanowires with radius $R$. Their starting point is the GK equation \eqref{gk1} under stationary conditions, viz. there is no change in time ($\nabla \cdot \mathbf q = 0$). Moreover, $\mathbf q$ is considered to be much smaller than $l^2 \Delta \mathbf q$ (but $\mathbf q \not\equiv 0$, otherwise the solution would be trivial), thus solving
\begin{align}
	\Delta \mathbf q = \frac{\lambda_0}{l^2} \nabla T. \label{alvgk1}
\end{align}
They also assume that the heat flux $\mathbf q$ can be divided into a bulk $\mathbf q_b$ and a wall contribution part $\mathbf q_w$, that is, $\mathbf q=\mathbf q_b + \mathbf q_w$. They solve Eq.~\eqref{alvgk1} first with $q_b = 0$ boundary condition at both ends ($r=\pm R$) in a one-dimensional setting (along the radius $r$), and supposing that a slip boundary condition is valid for the wall contribution,
\begin{align}
	q_w = C l \left (\frac{\partial q_b}{\partial r} \right)_{r=R},
\end{align}
the constant $C$ expresses the diffusive and specular boundary scattering (similarly to the parameter $p$ before). Let us realize that $C=0$ is excluded, otherwise, the system could not be in a stationary state and would result in $\mathbf q=0$ and $\nabla T=0$. Then, the effective, size-dependent thermal conductivity is found as
\begin{align}
	\lambda (\textrm{Kn}) = \frac{1}{\pi R^2} \frac{L}{\delta T} Q_{\textrm{tot}} = \frac{\lambda_0}{8 \textrm{Kn}^2}(1 + 4 C \textrm{Kn}), \quad Q_{\textrm{tot}} = \int\displaylimits_0^R 2 \pi \big(q_b(r) + q_w (r)\big) \textrm{d}r, \label{alvgk2}
\end{align}
where $\delta T$ expresses the temperature difference, and, more interestingly, this $\lambda (\textrm{Kn})$ formula cannot provide the bulk limit $\lambda_0$ for very small Knudsen numbers. They test the thermodynamic compatibility of Eq.~\eqref{alvgk2} by substituting it into the classical entropy production \eqref{f2} with local equilibrium assumption for which $\sigma_s \geq 0$ is satisfied. They continue their analysis by extending the boundary contribution with a second-order term as
\begin{align}
	q_w = C l \left (\frac{\partial q_b}{\partial r} \right)_{r=R} - \gamma l^2 \left (\frac{\partial^2 q_b}{\partial r^2} \right)_{r=R}
\end{align}
for which the parameter $\gamma>0$ can be chosen to recover various slip boundary conditions. Repeating the previous approach,
\begin{align}
	\lambda (\textrm{Kn}) = \frac{\lambda_0}{8 \textrm{Kn}^2}(1 + 4 C \textrm{Kn} - 4 \gamma \textrm{Kn}^2), \label{alvgk3}
\end{align}
is found and investigated later with more detail by Vázquez and Márkus \cite{VazqMar09, MarkGamb13, VazqEtal17}. Here, the situation seems more complicated since, in the bulk limit, it would lead to negative thermal conductivity, thus, there exists a critical $\textrm{Kn}$, which acts as an upper bound. When first testing its thermodynamic compatibility, using Eq.~\eqref{f2} (the local equilibrium version) again, they found that such second-order extension for the slip boundary condition is not thermodynamically compatible, however, exchanging it to a generalized one with extended state space (for which both $\mathbf q$ and $\mathbf Q \sim \nabla \mathbf q$ are included as before), then it restores the compatibility.

At this point, we want to place some necessary criticism regarding the thermodynamic treatment to be clear and avoid misunderstandings. First, while it could be reasonable under some special conditions to neglect $\mathbf q$, such a term can significantly modify the solutions of the equation \eqref{alvgk1} and put the thermodynamic compatibility into a different view. Second, utilizing a local equilibrium entropy production for the (simplified) GK equation is not physically reasonable since the GK equation itself is derived from a weakly nonlocal framework. Third, it seems unnatural that these $\lambda(\textrm{Kn})$ expressions have various validity limits, basically depending on an arbitrary choice of boundary conditions (e.g., how to choose $\gamma$), not explicitly on the rarefiedness of the medium what an intuition would suggest. However, despite these flaws, Eq.~\eqref{alvgk3} still can be helpful in modeling experiments \cite{AlvEtal10}, but one has to keep in mind the restrictions together with all the related assumptions.

Alvarez and Jou \cite{AlvJou07KN, AlvJou08} also investigated a different approach, exploiting the hierarchical structure of the EIT evolution equations, having the same characteristic hierarchical structure as the system \eqref{bal1}, with one notable difference: in EIT, there is a freedom to employ the coefficients from kinetic theory. Briefly, the EIT system reads,
\begin{align}
	a_1 \partial_t \mathbf q^{(1)} - \nabla \frac{1}{T}  - b_1 \nabla \cdot \mathbf q^{(2)} &=- c_1 \mathbf q^{(1)} \label{eit1} \\
	a_n \partial_t \mathbf q^{(n)} - b_{n-1} \nabla \mathbf q^{(n-1)}  - b_n \nabla \cdot \mathbf q^{(n+1)} &=-c_n \mathbf q^{(n)}, \label{eit2}
\end{align}
where the coefficients $a_i$, $b_i$, and $c_i$ are now phenomenological, and the upper indices $(n)$ denote the tensorial order. This is also characteristic of the internal variable approach \cite{BerVan17b, JozsKov20b}. Nevertheless, EIT requires a hierarchical structure similar to the momentum series expansion to have a hyperbolic system, but this is not a requirement in the internal variable approach. Interestingly, when a current multiplier is supposed to contribute to the entropy flux, it always leads to a parabolic set of evolution equations. 
The system \eqref{eit1}-\eqref{eit2} is closed by truncation, i.e., for the $n$$^{\textrm{th}}$ variable, the $(n+1)$$^{\textrm{th}}$ flux is considered to be zero. This is the most straightforward closure, the framework of RET is more sophisticated, requiring Galilean covariance and involving the maximum entropy principle to close the hierarchy \cite{RugSug15}. We refer to \cite{CimmEtal14} for a more comprehensive comparative study.

After taking the Fourier transform of \eqref{eit1}-\eqref{eit2} (interestingly, without using the energy balance, thus this system is neither mathematically nor physically closed), they express the thermal conductivity from the resulting dispersion relation and substitute it into the Fourier law,
\begin{align}
	\mathbf q(\omega, k) = - i k \lambda (\omega, k) \hat T(\omega, k),
\end{align}
with $i$ is the imaginary unit, and $\lambda (\omega, k)$ is found to be a continued-fraction, depending on which level one truncates the hierarchy,
\begin{align}
	\lambda (\omega, k) = \frac{\lambda_0}{1 + i \omega \tau_1 + \frac{k^2 l^2_1}{1 + i \omega \tau_2 + \frac{k^2 l_2^2}{1 + i \omega \tau_3 + \dots}}}. \label{eit3}
\end{align}
Here, the relaxation times $\tau_i$ are formed from $a_i/c_i$, and the intrinsic length scales $l_i$ originate from $b_i/(c_i c_{i+1})$. While the bulk term $\lambda_0$ can depend on temperature, all the other coefficients must be constant.
The characteristic length of the system $L$ is introduced through the wave number $k=2 \pi /L$; thus, the expression \eqref{eit3} differs from the usual dispersion relations in this respect. For a stationary case, they assume $\omega=0$, and the continuation requires further knowledge on the relaxation times $\tau_i$ and the intrinsic length scales $l_i$. Figure \ref{fig9}/B compares its outcome for numerous experimental results \cite{AlvJou08}. For further possibilities, we refer to \cite{AlvJou07KN, AlvJou08}, and we add that such size-dependent thermal conductivity is applied within the transient theory of EIT \cite{AlvEtal12}. These aspects are not restricted to superlattices and also prevail concerning nanotubes \cite{CamRom14, FujiEtal05, RomanoEtal15, ShiMar06}.

\subsubsection{Short note on nanofluids.} We close this part by clarifying the role of nanofluids as it might be misleading that, e.g., in \cite{Lebon14}, nanofluids are mentioned concerning nanoscale heat conduction. Nanofluids are ``ordinary'' fluids (such as water) mixed with a small portion of nanoparticles in order to enhance the fluid thermal properties and to improve the heat transfer capabilities \cite{Keb02, EapenEtal10}, e.g., in solar panels \cite{KhanVaf18, MuneEtal21}. Hence, indeed, the presence of nanoparticles induces nanoscale phenomena, but usually, these are effectively modeled in a macroscale approach, for instance, by determining the effective thermal conductivity for a particular composition, including particle size effects as well \cite{KumarEtal04}. Nevertheless, the determination of the effective thermal conductivity of a nanofluid stands as a challenging and still open question, and there are no general approaches as $\lambda$ might depend on the particle shape, volume and mass fractions, and sedimentation attributes, see for instance \cite{EastmanEtal04, WanMuj07, AjeeEtal22, ModiEtal23}. Despite the seeming analogy, non-Fourier effects are neither expected nor observed in nanofluids on macroscale. We refer to \cite{SobPet08b} for a detailed overview of nanofluids and their technological background.

\subsubsection{Transient heat conduction.}
Similarly to the steady-state situation, the central question remains about how the ballistic and diffusive phonon propagation modes contribute to the overall heat transport. As a particular extremum of these modes, Majumdar et al.~\cite{Maj93, JosMaj93} suppose only ballistic modes, and in such case, phonons become analogous to photons from many aspects, and thus the heat flux becomes 
\begin{align}
	\mathbf q = \sigma_{\textrm{phonon}} (T_1^4 - T_2^4), \label{maj1}
\end{align}
between two parallel plates (e.g., in thin films), with $\sigma_{\textrm{phonon}}$ being the phonon Stefan-Boltzmann constant depending on the particular phonon modes and their corresponding propagation speeds. It must be emphasized that the relation \eqref{maj1} is valid only in the Casimir limit \cite{Casimir38} and when the temperature is much below the Debye temperature \cite{Garai07} (i.e., only the acoustic modes are present). Otherwise, the heat transfer is not purely ballistic, the phonon-phonon scattering becomes significant. 
Due to the analogy with photons, it is suitable to introduce the phonon intensity $I_\omega$ as
\begin{align}
	I_\omega (\mathbf r, \mathbf v, t) = \frac{\hbar \omega}{4 \pi} D(\omega) |\mathbf v | f(\mathbf r, \mathbf v, t), \label{maj2}
\end{align}
where $\mathbf v$ is the group velocity of heat carriers and $D(\omega)$ describes the phonon density of states per unit volume. In a one-dimensional case, the Boltzmann transport equation is written as
\begin{align}
	\frac{\partial f_\omega}{\partial t} + v_x \frac{\partial f_\omega}{\partial x} = \frac{f_\omega^0 - f_\omega}{\tau} \Rightarrow \big(\textrm{Eq.~}\eqref{maj2}\big) \Rightarrow \frac{1}{v} \frac{ \partial I_\omega}{\partial t} + \mu \frac{\partial I_\omega}{\partial x} = \frac{I_\omega^0(T(x)) - I_\omega}{v \tau(\omega, T)}, \label{maj3}
\end{align}
with $v_x = \mu v$ with $-1<\mu<1$ being a geometrical factor ($\mu>0$ means forward, and $\mu<0$ means backward directions), and also, the relaxation time approximation is applied with the equilibrium intensity corresponding to a black body at a temperature below the Debye temperature, and following the Bose-Einstein statistics. For gray bodies, as before, the relaxation time becomes independent of the frequency $\omega$, and the equilibrium intensity distribution can be approximated with 
\begin{align}
	I_\omega^0(T)= \frac{1}{2} \int\displaylimits^1_{-1} I_\omega \textrm{d}\mu, \quad \textrm{and} \quad \sigma_{\textrm{phonon}} T^4 = \pi \int \displaylimits_0^{\omega_D} I_\omega^0(T) \textrm{d}\omega \label{maj4}
\end{align}
with $\omega_D$ being the cut-off frequency related to the Debye temperature. Equations \eqref{maj3} and \eqref{maj4} together form the so-called `Equations of Phonon Radiative Transfer' (EPRT) model \cite{JosMaj93}, valid below the Debye temperature.

Chen's approach \cite{Chen01}, however, explicitly separates the ballistic and diffusive contributions as $I_\omega (\mathbf r, t) = I_{\omega, b} (\mathbf r, t) + I_{\omega, d} (\mathbf r, t)$, consequently, $\mathbf q = \mathbf q_b + \mathbf q _d$ and $e = e_b + e_d=C T$, but it is not a two-temperature model since the temperature $T$ is associated to the internal energy $e$ only, not separately to $e_b$ and $e_d$, and $C$ is the specific heat capacity, not assigned to each propagation mode either. The ballistic contribution $ I_{\omega, b} (\mathbf r, t)$ is calculated using Eq.~\eqref{maj3} with one essential difference, for ballistic phonons, there is no associated equilibrium state.
For the diffusive part, $I_{\omega, d} (\mathbf r, t)$, however, the equilibrium $I_\omega^0(T)$ remains meaningful, and the time evolution is again given in the form of Eq.~\eqref{maj3}. It significantly eases the solution procedure to apply the diffusion approximation for thermal radiation, being valid only for optically thick media for which the radiative heat flux is obtained in the form of Fourier's law. The model is further simplified by considering the spherical harmonic expansion of  $I_{\omega, d} (\mathbf r, t)$ as $I_{\omega, d} (\mathbf r, t)=J_{\omega,0}(\mathbf r, t) + \mathbf n \cdot \mathbf J_{\omega,1}(\mathbf r, t)$, which breaks down the original integral-partial differential evolution equation into a set of coupled partial differential equations, similar to the momentum expansion. However, here the orthogonality of spherical harmonics can be exploited. The coupled system reads
\begin{align}
	\frac{4 \pi}{| \mathbf v |} \frac{\partial J_{\omega,0}}{\partial t} + \frac{4 \pi}{3} \nabla \cdot \mathbf J_{\omega,1} &= -\frac{4 \pi}{| \mathbf v | \tau_\omega } ( J_{\omega,0} - I_\omega^0), \label{chen1} \\
	\frac{1}{| \mathbf v |}  \frac{\partial \mathbf J_{\omega,1}}{\partial t}+ \nabla J_{\omega,0} &=-\frac{1}{| \mathbf v | \tau_\omega } \mathbf J_{\omega,1}, \label{chen2}
\end{align}
where Eq.~\eqref{chen1} is of balance type for the (diffusive) internal energy, and Eq.~\eqref{chen2} recalls the MCV-type constitutive equation, indeed,
\begin{align}
	u_d = \int \frac{4 \pi}{| \mathbf v |} J_{\omega,0} \textrm{d}\omega, \quad u_b = \int \int \frac{1}{| \mathbf v |} I_{\omega, b} \textrm{d} \hat{ \boldsymbol \Omega} \textrm{d}\omega, \quad  \mathbf q_d = \frac{4 \pi}{3} \int \mathbf J_{\omega,1} \textrm{d} \omega, \quad \mathbf q_b = \int \int  I_{\omega, b} \cos \theta \textrm{d} \hat{\boldsymbol \Omega} \textrm{d} \omega, \label{chen3}
\end{align}
where the integration of $\textrm{d} \hat{ \boldsymbol \Omega}$ expresses the summation over the entire solid angle.
Overall, these equations yield 
\begin{align}
	\frac{\partial e_d}{\partial t} + \frac{\partial e_b}{\partial t} &= - \nabla \cdot \mathbf q_d - \nabla \cdot \mathbf q_b, \label{chen4} \\
	\tau \frac{\partial e_b}{\partial t} + \nabla \cdot \mathbf q_b &= - e_b, \label{chen5} \\
	\tau \frac{\partial \mathbf q_d}{\partial t} + \mathbf q_d &= - \frac{\lambda_d}{C} \nabla e_d, \label{chen6}
\end{align}
and the ballistic heat flux contribution $\mathbf q_b$ can directly be expressed using Eqs.~\eqref{maj3} and \eqref{chen3}. This model is called the `ballistic-diffusive' equation. The balance of $e_d$ can be obtained by subtracting Eqs.~\eqref{chen5} from \eqref{chen4}, this is only a matter of choice, but only one of them can be used to avoid the over-determination of the model. 

However, according to Lebon et al., one could approximate $\mathbf q_b$ with an MCV-type equation, too, with different relaxation time and thermal conductivity, and thus the compatibility with EIT would be easily accessible \cite{Lebon14, LebonEtal11}. Therefore, this is the moment when we also introduce their EIT approach. This is purely a macroscopic approach for which it remains arbitrary whether to imply any assumptions from kinetic theory. The EIT model also applies the separation of ballistic and diffusive contributions such as $e=e_b + e_d$ and $\mathbf q = \mathbf q_b + \mathbf q_d$, and renders a sort of `quasi-temperature' $T_b$ and $T_d$ to the internal energies with heat capacities $c_b$ and $c_d$, choosing $c_d=c_b$, therefore $T = T_b + T_d$. We must keep in mind the intensive attribute of $T$, thus $T_b$ and $T_d$ are not real thermodynamics quantities, but both field variables are driven by their balance equations,
\begin{align}
	\frac{\partial e_i}{\partial t} + \nabla \cdot \mathbf q_i = r_j, \quad \textrm{with} \quad i,j=\{b, \ d, | \ i\neq j\}
\end{align}
and the total internal energy $e$ fulfills the first law of thermodynamics with a source term $r=r_b + r_d$, and using Chen's model, $r_b = - e_b/\tau_b$, and $r_d$ is constrained through the given $r$. Then, based on the earlier results from kinetic theory, it is supposed that the diffusive part of phonons can properly be modeled with the MCV equation, and the ballistic part, however, is assumed to be driven by the GK equation, 
\begin{align}
	\tau_d \frac{\partial \mathbf q_d}{\partial t} + \mathbf q_d &= - \lambda_d \nabla T_d, \label{neit1} \\
	\tau_b \frac{\partial \mathbf q_b}{\partial t} + \mathbf q_b & = - \lambda_b \nabla T_b + l_b^2 ( \Delta \mathbf q_b + 2 \nabla \nabla \cdot \mathbf q_b), \label{neit2}
\end{align}
with the coefficients supposed to follow the kinetic theory,
\begin{align}
	\lambda_d = \frac{1}{3} c_d v_d^2 \tau_d, \quad \lambda_b = \frac{1}{3} c_b v_b^2 \tau_b, \quad v_d = \frac{l_d}{\tau_d}, \quad v_b = \frac{l_b}{\tau_b}. \label{neit3}
\end{align}
At this point, we must note some further comments about the system \eqref{neit1}-\eqref{neit3}. First, sadly, this system is not derived in the same way as before, i.e., defining the elements of the state space and how the entropy flux constitutes the heat fluxes. It might be necessary since the Onsagerian solution of the entropy production would naturally offer a coupling between $\mathbf q_b$ and $\mathbf q_d$, and these parts are omitted in \eqref{neit1}-\eqref{neit3}, although could have outstanding importance even in more straightforward situations. Second, although \cite{Lebon14} states that the system \eqref{neit1}-\eqref{neit3} provides a macroscopic description, it is not entirely true since that approach is only applicable for nanosystems considering the size dependence of $\lambda$, together with the kinetic assumptions. Despite these hidden attributes, the EIT model is a valuable approach \cite{JouEtal96b}, however, with further undiscovered potential, which would be helpful to recognize the limits of continuum approaches and discover the modeling capabilities. For a more detailed overview of the microscale heat transfer and further coupled heat equations, let us refer to \cite{JouCimm16, Chen21}.

\subsection{Comments on the quantum concepts} So far, we reviewed one branch of quantum transport focusing on the phonon-phonon interaction without introducing (and exploiting) the concepts of wave functions, quantum statistics, and density matrices. Moreover, the phonon transport models have counterparts from continuum frameworks, recalling the exact structure of evolution equations. This does not hold for electron transport, and even
properly formulating Fourier's law based on quantum mechanics has been an old problem, especially when the heat flux needs further generalizations. While the utilization of phonons is also a valid quantum approach that has been discussed so far, a more general framework should include, e.g., the electrons and electron-phonon interactions. However, these particles possess a notably different quantum nature, thus influencing the corresponding set of wave functions and their symmetry attributes, summarized in \cite{Walczak18}. 
The introduction of probability distributions is inevitable. The existence of fluctuations stands as further challenging properties of quantum systems and might result in negative entropy production \cite{SaitoDhar07}. 

Let us imagine a stationary problem with two heat reservoirs characterized by the temperatures $T_1$ and $T_2\neq T_1$, and connected via a conductor. The electron transport simultaneously induces charge ($I_c$) and heat currents ($I_h$) between the reservoirs. Instead of the temperature difference, the resulting currents are proportional to the difference between the distribution functions ($\vartheta_{1,2}$), following \cite{PekKar21},
\begin{align}
	I_c &= \tilde q \sum_p \int\displaylimits_{0}^\infty \frac{1}{2 \pi} \mathbf v_p(\mathbf k) (\vartheta_1(\mathbf k) - \vartheta_2(\mathbf k)) \mathcal{T}_p (\mathbf k) \textrm{d} \mathbf k, \quad &\Rightarrow \quad &I_c = \frac{\tilde q}{h} \sum_p \int\displaylimits_{\epsilon(0)}^\infty   (\vartheta_1(\epsilon) - \vartheta_2(\epsilon)) \mathcal{T}_p (\epsilon) \textrm{d} \epsilon \label{qu1} \\
	I_h &=  \sum_p \int\displaylimits_{0}^\infty  \frac{1}{2 \pi} \epsilon_p (\mathbf k) \mathbf v_p(\mathbf k) (\vartheta_1(\mathbf k) - \vartheta_2(\mathbf k)) \mathcal{T}_p (\mathbf k) \textrm{d} \mathbf k, \quad &\Rightarrow \quad &I_h = \frac{1}{h} \sum_p \int\displaylimits_{\epsilon(0)}^\infty  \epsilon (\vartheta_1(\epsilon) - \vartheta_2(\epsilon)) \mathcal{T}_p (\epsilon) \textrm{d} \epsilon \label{qu2}
\end{align}
where $\tilde q$ is the particle charge, and $\epsilon=\epsilon_p (\mathbf k=\mathbf 0)$ indicates the energy of the given particle as a function of the wave number $\mathbf k$, for all independent modes $p$. Furthermore, $\mathcal T_p$ stands for the particle transmission probability, for ballistic transport $\mathcal T_p (\mathbf k)\equiv1$. Eqs.~\eqref{qu1} and \eqref{qu2} are applicable for both bosons and fermions with their corresponding distribution function and charge, thus, the resulting electron transport can also be characterized. 
Assuming a single ideal ballistic channel between the reservoirs leads to the observation that the thermal conductivity becomes quantized,
\begin{align}
	\lambda_0 = \frac{\pi^2 k_B^2}{3h} T,
\end{align}
similarly to the electric conductance, being the same for both bosons and fermions and first experimentally observed by Scwab et al.~\cite{SchwabEtal00}. For further experimental and theoretical details, let us refer to \cite{MeschEtal06, PekKar21}.
Although such an approach provides a quantitative estimate for the energy (and charge) transfer through a given conductor (not necessarily in ballistic mode), being helpful in quantum electronics design, it is limited in a sense to offer a field equation to model the transients of the studied domain. 

The paper of Greiner et al.~\cite{GreinerEtal97} provides the closest formalism that we have used so far for a one-dimensional conductor (valid for, e.g., a nanowire) of length $\ell$, considering electrons to be heat carriers. The solution for the entropy inequality is provided in the form of
\begin{align}
	\mathbf j_\mu = \sum_\nu L_{\mu \nu} \mathbf X_{\nu}, \quad \mu, \nu=1,2
\end{align}
in which $L_{\mu \nu}$ are the Onsagerian coefficients, $\mathbf j_\mu$ represent the electrical current and heat flux densities, and $\mathbf X_{\nu}$ express the corresponding gradients. Exploiting the fluctuation-dissipation theorem, the linear response is found using a correlation function $C_{\mu \nu}(t)$. In that particular case, it reads
\begin{align}
	C_{\mu \nu}(t) = \frac{1}{\pi} \left(\frac{e \hbar }{m L}\right)^2  \left(\frac{\hbar^2 }{2 m  e}\right)^{\mu + \nu - 2} e^{-t/\tau_c} \int\displaylimits_{-L/2}^{L/2} \textrm{d}z \int\displaylimits_{k^-}^{k^+} k^{2(\mu + \nu -1 )} f_0(1- f_0) \textrm{d}k, \quad k^{\pm} = \frac{m}{\hbar t} \left(\pm \frac{L}{2} - z\right), \label{+eq8}
\end{align}
where $e$ is the electron charge, $m$ is the carrier's effective mass, $\hbar$ stands for the reduced Planck's constant, $k$ is the wave vector, and $f_0$ is the Fermi distribution. The exponential term $e^{-t/\tau_c}$ primarily determines the transport regime through the average scattering time (scattering on the lattice imperfections). That is, the transport is called ballistic if  $e^{-t/\tau_c}=1$, implying that $\tau_c \rightarrow \infty$, and also providing a continuous transition towards the diffusive transport when $C_{\mu \nu}(t) \sim e^{-t/\tau_c}$. The system's response is determined on the spectral properties of $C_{\mu \nu}(t)$; therefore, they apply a Fourier-Laplace transform on Eq.~\eqref{+eq8}, finding $S_{\mu \nu}(\omega)$, hence the conduction coefficients $L_{\mu \nu}$  and the thermal conductivity $\lambda$ are
\begin{align}
	L_{\mu \nu}(\omega) = \frac{L S_{\mu \nu}(\omega)}{k_\textrm{B} T} \quad \Rightarrow \quad \lambda (\omega) = \frac{L_{11}(\omega) L_{22}(\omega) - L_{12}(\omega)L_{21}(\omega)}{L_{11}(\omega) T }.
\end{align}
That approach naturally includes the size dependence in $\lambda$. Moreover, it also returns the universal thermal conductance (reciprocal of the thermal resistance) of $\lambda_0=\pi^2 k_\textrm{B}^2 T/(3 h)$ being valid for perfect (or ideal) channels. That ideal thermal conductance can be transferred to non-ideal situations \cite{GotsEtal22}. 

Interestingly, Michel et al.~\cite{MichelEtal05, MichelEtal06} simulated heat conduction in a one-dimensional chain, consisting of $N$ identical and weakly coupled subsystems, analogously to the Fermi-Pasta-Ulam problem \cite{FePaUl55}. The system is described by the Hamiltonian
\begin{align}
	H = \sum_{\mu=1} ^N H_{\textrm{loc}}(\mu) + g \sum_{\mu=1}^{N-1} V(\mu, \mu+1),
\end{align}
in which the weakness of the coupling is expressed by $g \ll \Delta E$, i.e., the interaction $ V(\mu, \mu+1)$ between each subsystem is re-scaled with a parameter $g$, being much smaller than the local energy gap $\Delta E$ between the subsystems. Each subsystem has $n$ numbers of equally distributed levels plus their ground level. Additionally, the local Hamiltonian describes the local energy using $\Psi(t)  H_{\textrm{loc}}  \Psi(t) = \Delta E P_\mu$, where $\Psi(t)$ is a wave function and $P_\mu$ is the probability of finding the $\mu^\textrm{th}$ subsystem in an excited state. With introducing a time evolution operator $D(\hat \tau)$ in a way $\Psi(t+\hat \tau) =D(\hat \tau)  \Psi(t)$ (the notation $\hat \tau$ is introduced for the transition time to distinguish it from the relaxation time $\tau$). Consequently, the $P_\mu$ probabilities inherit that time dependence and thus form a dynamics, a set of ordinary differential equations for all $P_\mu$ for small values of $\hat \tau$. It will reduce to the Fourier heat equation only if the conditions
\begin{align}
	2 g \frac{n}{\delta \epsilon} \geq 1, \quad g^2 \frac{n}{\delta \epsilon^2} \ll 1
\end{align}
are satisfied, and $\delta \epsilon \ll \Delta E$ is the bandwidth. 

Manzano et al.~\cite{ManzanoEtal12} follow an analogous procedure, imaging a $N$-element chain of a two-level quantum system with the Hamiltonian
\begin{align}
	H = \sum_{j=1} ^N \frac{\hbar \omega}{2} \sigma_j^z + \sum_{j=1} ^{N-1} \hbar g(\sigma_j^+\sigma_{j+1}^- + \sigma_j^- \sigma_{j+1}^+),
\end{align}
where $\sigma_j^z$, $\sigma_j^+$, and $\sigma_j^-$ are Pauli's $z$, raising and lowering operator, and $g$ is again a coupling constant. The master equation of Lindblad form describes the time evolution of the quantum system,
\begin{align}
	\dot \rho = - \frac{i}{\hbar} [H, \rho] + \mathcal{L}_1 \rho + \mathcal{L}_2, \label{qu3}
\end{align}
in which $\rho$ denotes the density matrix, and $\mathcal{L}_{1,2}$ models the interactions with the bosonic heat reservoirs. The resulting heat current can be found using $\textrm{Tr}(H\dot \rho)$. However, one must add a so-called dephasing term to Eq.~\eqref{qu3} to obtain a heat flux compatible with Fourier's law. The dephasing term introduces a coherence decay into the system with
\begin{align}
	\mathcal{L}_{\textrm{deph}} \rho = \gamma \sum_{j=1} ^N \left( \sigma_j^+\sigma_j^-\rho \sigma_j^+\sigma_j^- - \frac{1}{2}\{ \sigma_j^+\sigma_j^-,\rho\}    \right),
\end{align}
with an adjustable parameter $\gamma$ to investigate the sensitivity of the resulting heat current to dephasing. Although the two-level chain itself strictly restricts the heat capacity, and thus the dependence on the reservoir's temperature, a proper size-dependent behavior can be modeled \cite{ManzanoEtal12}. 

\begin{figure}[H]
	\centering
	\includegraphics[width=9cm,height=6cm]{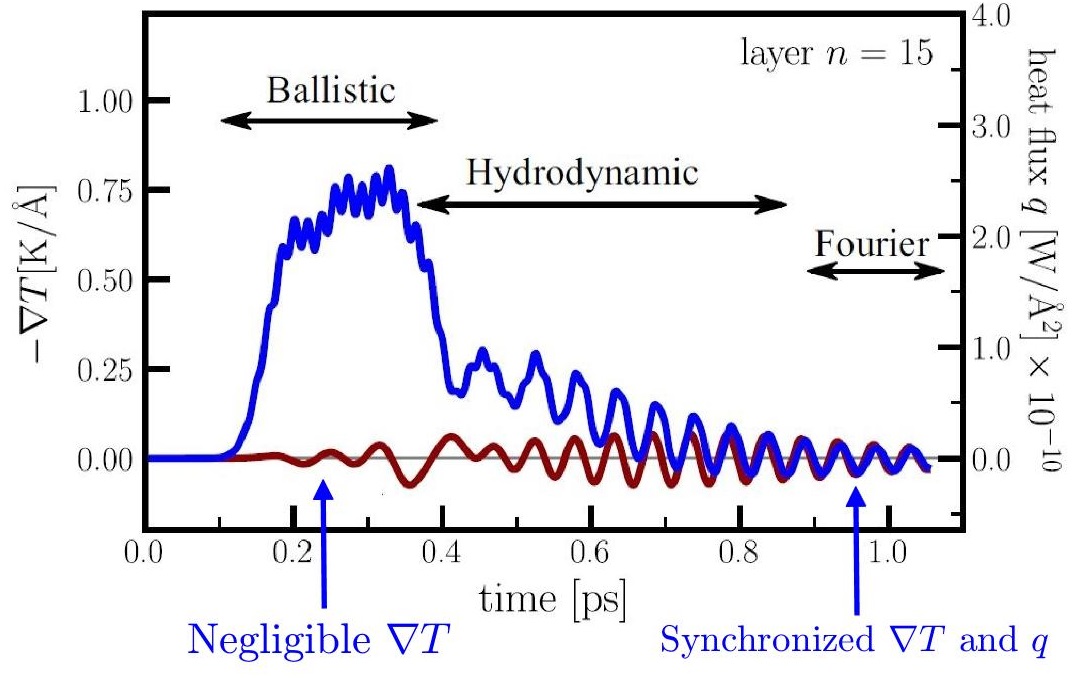}
	\caption{The time evolution of heat flux and temperature gradient, presenting the characteristics for each propagation mode \cite{MazzaEtal21}.}
	\label{fig14}
\end{figure}

These approaches can reconstruct the diffusive energy transport, but the non-diffusive modes are missing, and one can define multiple Hamiltonians to create quantum chains with diffusive behavior and proper scaling. From that point of view, the work of Mazza et al.~\cite{MazzaEtal21} could be a notable progress. They investigate the dynamics of a so-called layered single-band Hubbard model, considering the sub-picosecond time scale for electronic interactions. The topmost layer is excited with an ultra-fast light pulse, creating hot electrons. The interaction between the cold and hot electrons introduces different time scales, which are implemented into the time evolution of the density matrix.
The results include the presence of all propagation modes. The heat flux shows a strong correlation with the density of hot electrons. During ballistic propagation, the temperature gradient is negligible. It starts to be significant in the hydrodynamic regime, and lastly, in the diffusive regime, the temperature gradient and heat flux are synchronized. These relationships between $\nabla T$ and $q$ are often disregarded and would be essential to study in the future for different systems, whether these are more general or not. For further details, see Figure \ref{fig14} and \cite{MazzaEtal21}.

We find it insightful to mention the recent work of Ván \cite{Van23}, about investigating the relations between capillary fluids, superfluids, and quantum mechanics. Interestingly, these seemingly different models (with different physical meanings) are connected due to a strong common property called classical holographic property, meaning that the pressure of perfect fluids ($\mathbf P$) is connected to a scalar potential ($\phi$) via the relation
\begin{align}
	\nabla \cdot \mathbf P = \rho \nabla \phi.
\end{align}
This happens to be a thermodynamic consequence under zero dissipation and thus can be helpful to find a quantized form of the heat conduction field equation directly without assuming particular chains and interactions among them. 

\subsubsection{Thermal rectification} Let us imagine a situation in which we have a one-dimensional conductor between two thermal reservoirs. In a steady-state, $Q_{1\rightarrow 2}$ current occurs. Now let us exchange the temperature of reservoirs, i.e., reversing the direction of the temperature gradient. The resulting $Q_{2\rightarrow 1}$ current is not necessarily the same if certain conditions are satisfied. This thermal rectification can be characterized in various ways, such as the ratio of these heat currents. Although that phenomenon does not demand either the presence of quantum effects or particular length, time, or energy scales, it is still strongly connected to quantum systems due to its practical applications. Thermal rectification can be exploited in electronic systems as thermal diodes (on the analogy of electronic diodes). 

Asymmetry stands as the most critical requirement to achieve such a thermal rectification effect. A straightforward macroscopic example is a functionally graded material in which the material properties depend on space, consequently also possessing different thermal conductivity. Moreover, their temperature dependence is also different, hence nonlinearities can induce asymmetry as well \cite{Dames09}. 
The additional, macroscopically feasible cause is the presence of distinct interface attributes. The different surface and mechanical properties can also induce thermal rectification, which is called thermal strain or warping \cite{RobWalk11}.

However, such an effect can be realized in practice in particular heterogeneous structures (recall the aim of thermal metamaterials), for which the material structure with heterogeneities is manufactured in an intended way. For very insightful examples, we refer to the paper of Zhao et al.~\cite{ZhaoEtal22}, in which paper they collected and reviewed a vast number of experimental works with discussing the possible causes. Interestingly, shape is also a factor, Yang et al.~\cite{YangEtal21b} designed and then experimentally tested several trapezoid-shaped microelectrodes. 

From a quantum mechanical point of view, thermal rectification can be realized with a graded system suited to a proper long-range interaction \cite{PerAvil13, Pereira19}. The corresponding Hamiltonian reads
\begin{align}
	H= \sum_{j=1} ^N \left( \frac{p_j^2}{2 m_j} + \frac{u_j^4}{4} \right ) + \sum_{j,k} \frac{(u_j - u_k)^2}{2(1+|j-k|^\beta)},
\end{align}
where $p_j$ denotes the momentum of the j$^\textrm{th}$ particle with mass $m_j$, and $u_j$ is its displacement. The graded system means that the mass distribution of the particles satisfies $m_j> m_{j-1}$, nevertheless, at the ends $m_{0}=m_1$ and $m_N=m_{N-1}$. The last term represents the interaction potential, where $\beta$ characterizes the polynomial decay. Both the mass asymmetry and the long-range interaction are in favor of the rectification phenomenon and remain significant even for long chains $N>200$. We stress that asymmetry remains crucial, e.g., a chain consisting of equal masses can still present rectification if an additional, space-dependent external field can act on the system \cite{ChenEtal15}. 
Let us recall that Eq.~\eqref{eqgktc} is motivated by the same reasons, and a particular nonlocal (on the analogy of long-range interaction) is implemented into a continuum framework, leading to further modifications of the Guyer-Krumhansl equation \cite{CarloEtal23}.
We refer to \cite{Pereira19, BalEtal18, SeniorEtal20} for further quantum mechanical aspects and insight. Thermal transistors are constructed analogously, in that regard, we only refer to \cite{JoulEtal16, GhoshEtal21} as that topic points beyond the limits of the present paper.

\subsection{Rarefied gases.} 
\begin{figure}[]
	\centering
	\includegraphics[width=15cm,height=5.5cm]{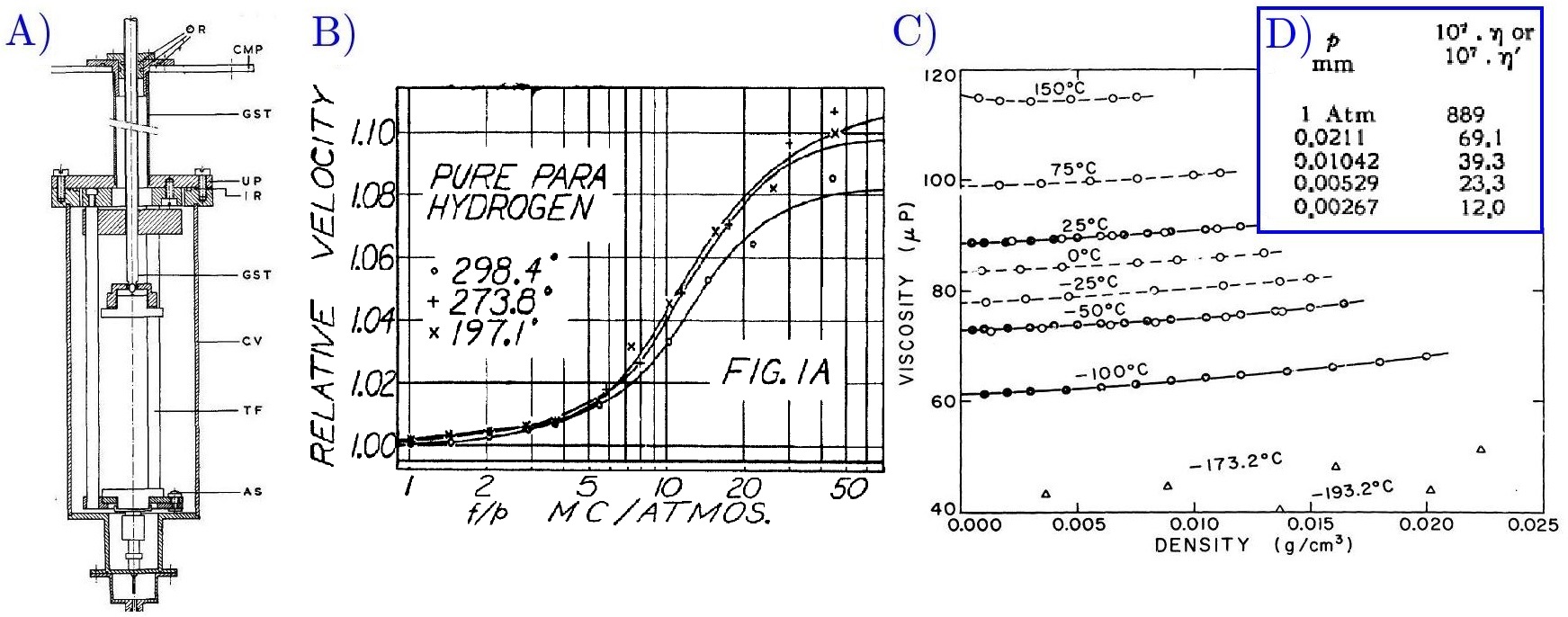}
	\caption{Rarefied gases from an experimental point of view. A) Typical measurement arrangement \cite{SluiEtal64}. B) Change in the speed of sound with respect to mass density \cite{Rhod46}. C) Mass density dependence of shear viscosity towards the dense states \cite{GrackiEtal69a, GrackiEtal69b}. D) Mass density dependence of shear viscosity towards the rarefied states \cite{IttPae40} (different from part C)).}
	\label{fig10}
\end{figure}

It is worth starting this topic again with kinetic theory as there is a clear analogy between rarefied phonon gas and rarefied real gaseous materials, the Knudsen number is similarly large in both cases, but the carrier is different. The modeling of that difference over the energy carriers distinguishes between the models based on kinetic theory. For instance, it does matter whether the gas is monatomic or polyatomic and hence how the source term in the Boltzmann equation is formulated. Such details are missing from the continuum models, but they appear again to be more flexible, especially when nonlinearities enter the picture. From an experimental point of view, the rarefied state is achieved by decreasing the system's pressure while keeping its volume constant. There should be a transition from the classical Navier-Stokes-Fourier (NSF) equations to the more advanced, generalized systems by having the mass density dependence of the transport coefficients such as viscosities, thermal conductivity, and, therefore, the new coefficients (relaxation times and various coupling parameters). Such transition is observed as the change in the speed of sound, starting from the dense (normal) state to the rarefied region.
However, modeling that transition is not straightforward, and thus various models can be derived on kinetic and continuum bases. To be more precise, Figure \ref{fig10} presents a typical experimental arrangement and data recorded for the change of speed of sound as a function of frequency over pressure ($\omega/ p$), which scaling property will be discussed soon. The classical NSF system cannot explain these observations, and these experiments act as a benchmark for extended theories.

\subsubsection{Kinetic approach.} The starting point is a suitable approximation for the collision integral in the Boltzmann transport equation \eqref{bte00}, such as the Bhatnagar-Gross-Krook (BGK) or ellipsoidal-statistical (ES)-BGK model \cite{BGK54}, in the form of $S = - \nu ( f - f_{\textrm{eq}})$, where $\nu$ is the collision frequency ($\sim 1/\tau$), and $f_{\textrm{eq}}$ expresses the corresponding equilibrium distribution for a monatomic gas. These models differ in how the collision frequency and the equilibrium distribution are utilized (e.g., on what quantities they depend on), what molecule type with what potential is included \cite{Cercignani00b, Struc05}. Moreover, it is also a matter of choice at which level we want to approximate the Boltzmann equation together with its collision integral. Consequently, here we restrict ourselves to the conceptual presentation only due to the large number of modeling possibilities.
Since the transport coefficients are given as an outcome, these approximations also restrict their ratio (like the Prandtl number) and hence how realistic the model is for a particular medium. For instance, for the BGK model, the Prandtl number is 1, which could be quite inaccurate, but in parallel, it offers a simpler approach than the ES-BGK approximation. However, the latter model allows for adjustment of the Prandtl number through the equilibrium distribution and also consists of the BGK model as a particular choice. 

Both the kinetic and continuum approaches require the basic conservation laws for mass, momentum, and energy. In the kinetic theory, these balances are found through the collision invariants $\psi=\psi(\mathbf r, \mathbf v, t)$, such as 
\begin{align}
	\psi=1: \ \rho= m\int f_{\textrm{eq}} \textrm{d} \mathbf c, \quad \psi=\mathbf c: \ \rho \mathbf v = m \int \mathbf c  f_{\textrm{eq}} \textrm{d} \mathbf c, \quad \psi=\frac{1}{2} c^2: \ \rho e = \frac{3}{2} \rho T = \frac{m}{2} \int (\mathbf c - \mathbf v)^2  f_{\textrm{eq}} \textrm{d} \mathbf c, \label{compcond}
\end{align}
for which $\mathbf v$ stands as the barycentric velocity and $\mathbf c$ is the phase velocity, and the corresponding balances are
\begin{align}
	&\textrm{mass:} &  \dot \rho + \rho \nabla \cdot \mathbf v &=0, \label{mbal} \\
	&\textrm{momentum:} &\rho \dot {\mathbf v} + \mathbf P \cdot \nabla &= \rho \mathbf f, \label{mombal} \\
	&\textrm{total energy:} &\rho \dotr{\left (e + \frac{1}{2} v^2 \right)} + (\mathbf v \cdot \mathbf P  + \mathbf q ) \cdot \nabla &= \rho \mathbf f \cdot \mathbf v, \\
	&\textrm{internal energy:} &\rho \dot e + \nabla \cdot \mathbf q &= - \mathbf P : \nabla \mathbf v, \label{enebal}
\end{align}
with $\mathbf P$ being the pressure tensor with a suitable decomposition of $\mathbf P = p \mathbf I + \boldsymbol \Pi$ in which $p$ is the hydrostatic pressure (not identical with the spherical part), and $\boldsymbol \Pi = \boldsymbol \Pi^{\textrm{dev}} + \Pi \ \mathbf I$ is the viscous pressure, where $\boldsymbol \Pi^{\textrm{dev}}$ is the traceless deviatoric part, and $\Pi$ is called dynamic pressure. Furthermore, $\mathbf f$ is a force density (such as gravitation). 
These balances are the starting point in continuum theories. In a kinetic approach, these follow from the Boltzmann equation. Furthermore, any approximations of the collision integral $S$ must satisfy these balances as well as the $h$-theorem. Here, we note that similarly to the balance equations \eqref{mbal}-\eqref{enebal}, the entropy balance can also be derived with a collision invariant $\psi = - k_{\textrm{B}} \ln (f/y)$ with $k_{\textrm{B}}$ being the Boltzmann constant, resulting in the kinetic counterpart of Eq.~\eqref{slaw}, using the same notations as Eq.~\eqref{slaw},
\begin{align}
	s=- k_{\textrm{B}} \int f \ln\frac{f}{y} \textrm{d} \mathbf c, \quad \mathbf J_s = - k_{\textrm{B}} \int \mathbf c f \ln\frac{f}{y} \textrm{d} \mathbf c, \quad \sigma_s=- k_{\textrm{B}} \int S f \ln\frac{f}{y} \textrm{d} \mathbf c. \label{slaw2}
\end{align}
Again, we emphasize that it becomes a matter of choice on what level we want to approximate the Boltzmann equation, viz., how to close these set of balances. The calculations and the resulting models can greatly vary in this respect, and therefore we refer to these works for a more detailed overview \cite{Cercignani00b, Struc05}. 

Conceptually, the aim is to find a suitable and manageable approximation of the Boltzmann equation, with a given set of potentials, and molecule types, for a given state space, and the calculability strongly depends on the chosen setting. Each procedure results in a unique distribution function $f$, which is used to close the hierarchical system with constitutive equations and construct the corresponding entropy function. We emphasize again that due to the numerous possibilities and technical details, we keep focusing on the conceptual questions.

Let us begin with the Chapman-Enskog (CE) expansion for which the main idea is to expand the phase density $f$ into a series $f=f^{(0)}+ \varepsilon f^{(1)} + \varepsilon^2 f^{(2)} + \dots$, where $\varepsilon$ stands for a small parameter. In the CE expansion, this is identified with the Knudsen number, however, this is not necessary and could depend on the particular molecular model. Furthermore, any expansion must satisfy the compatibility conditions, i.e., when $f_{\textrm{eq}}$ substituted with $f^{(0)}$ in Eq.~\eqref{compcond}, hence $f^{(0)}$ is found as the local Maxwellian distribution; and these are supplemented with 
\begin{align}
	0 = m \int f^{(\alpha)} \textrm{d} \mathbf c, \quad \mathbf 0 =  m \int \mathbf c f^{(\alpha)} \textrm{d} \mathbf c, \quad 0 = \frac{m}{2} \int (\mathbf c - \mathbf v)^2 f^{(\alpha)} \textrm{d} \mathbf c, \quad \alpha \geq 1.
\end{align}
The CE expansion includes five fields ($\rho$, $\mathbf v$ and $T$), and closes the balances with
\begin{align}
	q^{(\alpha)}_{i} = \frac{m}{2} \int c^2 c_i f^{(\alpha)} \textrm{d} \mathbf c, \quad \sigma^{(\alpha)}_{ij} = m \int c_{\langle i} c_{j \rangle} f^{(\alpha)} \textrm{d} \mathbf c, 
\end{align}
which can be used to construct the series of heat flux $q_i = \varepsilon q_i^{(1)} + \varepsilon^2 q_i^{(2)} + \dots$, and pressure $\sigma_{ij} = \varepsilon \sigma_{ij}^{(1)} + \varepsilon^2 \sigma_{ij}^{(2)} + \dots$ up to an arbitrary order, and $\sigma_{ij}$ stands for the symmetric part of $\boldsymbol \Pi^{\textrm{dev}}$, and thus it implies that the dynamic pressure $\Pi$ is assumed to be zero together with the antisymmetric part of $\boldsymbol \Pi^{\textrm{dev}}$. The notations are following Eq.~\eqref{bal1}. Since $f^{(0)}$ is given as a local Maxwellian, hence  $P_{ij}^{(0)}=0$, and $q_i^{(0)}=0$, and the equations of Euler fluid is given. In the first order, the classical NSF equations are found with 
\begin{align}
	q_i^{(1)} = - \frac{5}{2}  \frac{p}{\nu} \frac{\partial T}{\partial x_i}, \quad \sigma_{ij}^{(1)} = -\frac{2}{1-b} \frac{p}{\nu} \frac{\partial v_{\langle i}}{\partial x_{j\rangle}}, \quad \lambda = \frac{5}{2}  \frac{p}{\nu}, \quad \mu = \frac{1}{1-b} \frac{p}{\nu} ,
\end{align}
where the parameter $-1/2 \leq b< 1$ is used to adjust the Prandtl number (Pr $\sim \mu/\lambda$) in the ES-BGK approximation (where $\mu$ is shear viscosity). Higher-order expansions result in the Burnett (second-order) and super-Burnett (third-order) equations. However, these are found to be unstable under particular conditions but might be used only for steady-state problems \cite{StrTah11, Struc12}. Further orders are extremely cumbersome to determine, and other methods are more suitable. This highlights an issue that the sequential approximations and series expansions are not necessarily convergent and might result in nonphysical solutions. Despite these drawbacks, the kinetic approach could be undoubtedly valuable when the transport coefficients must be determined and measurements are unavailable, especially for the coupled thermo-diffusion problems \cite{Yang49, Ham60}, or heat transfer coefficient in a rarefied medium.  

Interestingly, one might find the thermal conductivity and viscosity to depend only on the temperature, not on the mass density \cite{MichEtal10, IttPae40, GulSel02, Haynes73}. The literature can be divisive in that sense as some argue that these quantities remain constant for a given temperature, but other measurements show its opposite \cite{Kov18rg}. Furthermore, the mentioned $\omega/p$ scaling is strongly related to this property. Such scaling means that by calculating the dispersion relation of the transport equations, one finds the phase velocity as a function of $\omega/p$ only, however, that is true only for an ideal gas with constant transport properties, and the mass density dependence would violate that scaling. This also holds for higher-order approximations \cite{Arietal12, Arietal15}, and it has a crucial role in experimental modeling. To be clear, in numerous experiments \cite{Rhod46, SluiEtal64, MeySess57}, only the temperature is kept constant, and either the $\omega$ or $p$ are varied, sometimes only the pressure is changed in a considerable interval; thus only the mass density dependence could cause the deviation from the NSF equations. We may add that such theoretical result leads to nonzero transport coefficients in the zero density limit \cite{RugSug15} contrary to the experiments \cite{ItterCla38}, we refer to Fig.~\ref{fig10} for the related experimental background. Kinetic theory is relatively inflexible from this perspective compared to a continuum approach. Besides, numerous experiments show the mass density dependence of viscosity, for instance, for both high and low-pressure intervals \cite{GulTrap86, GrackiEtal69a}, which require further corrections, e.g., as a function of the Knudsen number \cite{GulSel02, Dym87, UmVes14}.

Furthermore, it turned out that the CE expansion can lead to negative phase densities \cite{Struc05}. This property is shared with the so-called Grad's method \cite{Grad49, Grad58}, for which the set of variables is extended with the heat flux $\mathbf q$ and deviatoric pressure $\mathbf \sigma$ (13-moment equations), or even further, higher tensorial order moments can be included (26 or more moment equations). Similarly to the previous theories, such extension of state space leads to the appearance of time derivatives of these quantities. The constitutive equations are found in a balance form, generalizing the classical NSF equations in a particular way. The resulting system of equations respects the symmetric hyperbolic structure, together with the balances \eqref{mbal}-\eqref{enebal}, the 13-moment equations reads
\begin{align}
	\dot q_i + \frac{5}{2}\rho T \frac{\partial T}{\partial x_i} + \frac{5}{2} \sigma_{ik} \frac{\partial T}{\partial x_k} - \sigma_{ik} T \frac{\partial \ln \rho}{\partial x_k} - \frac{\sigma_{ik}}{\rho} \frac{\partial \sigma_{kl}}{\partial x_l} + T \frac{\partial \sigma_{ik}}{\partial x_k} + \frac{7}{5} q_i \frac{\partial v_k}{\partial x_k} + \frac{7}{5} q_k \frac{\partial v_i}{\partial x_k}+ \frac{2}{5} q_k \frac{\partial v_k}{\partial x_i}&=-\frac{2}{3} \frac{p}{\mu} q_i, \label{grad1} \\
	\dot \sigma_{ij} + \frac{4}{5} \frac{\partial q_{\langle i}}{\partial x_{j \rangle}} + 2 \sigma_{k \langle i} \frac{\partial v_{j \rangle}}{\partial x_k} + \sigma_{ij} \frac{\partial v_k}{\partial x_k} + 2 \rho T \frac{\partial v_{\langle i}}{\partial x_{j\rangle}} &= -\frac{p}{\mu} \sigma_{ij},\label{grad2}
\end{align}
being valid only for Maxwell molecules (for which the potential $V(r)\sim r^{-4}$, and that represents a borderline between hard and soft potentials, too \cite{Liboff03b}). It is also claimed that the convergence with moments is rather slow, nonetheless, it still needs to model fast phenomena or large gradients (such as shocks). It means that the Guyer-Krumhansl-type equations are excluded, and non-local terms (such as $\Delta \mathbf q$) are not allowed. Furthermore, a critical Mach number exists, depending on the number of moments for which sub-shocks and discontinuities occur. These can be contradictory with experiments and molecular-level simulations, and their appearance depends on the number of moments \cite{Struc05, TaniRugg18}. 

Since the phase density can be negative, even with small amplitudes, it is impossible to construct the corresponding entropy function, which is feasible only for linearized equations. Moreover, besides the second law of thermodynamics, even hyperbolicity can be broken \cite{Torrilhon16}, and thus this approach is accompanied by additional conditions on the hyperbolicity radius around equilibrium \cite{MulRug98}.
This shortcoming can partially be solved using the so-called Maximum Entropy (abbreviated as 'MaxEnt' \cite{Dreyer87, Levermore96}) closure, being valid only for ideal gases \cite{OttStruc20}. Here, we seek an appropriate $f$, which maximizes the entropy density from Eq.~\eqref{slaw2}, and the resulting phase density is
\begin{align}
	f_{\textrm{max}} = y \exp\big( - \sum\displaylimits_A \Lambda_A \phi_A\big), \quad A=\{1,2,\dots,N\},
\end{align} 
where $\Lambda_A$ are the Lagrange multipliers for the $N$-moment system with moments $F_A$, and $\phi_A$ are even-order polynomials of $\mathbf c$, otherwise, the resulting system will not be thermodynamically compatible (hence unstable). That requirement excludes the 13-moment equations but supports the 14-moment system. Additionally, above the 10-moment equations (13-moment equations without $\mathbf q$), this closure becomes feasible only numerically. The symmetry $F_A(\Lambda_A) \longleftrightarrow \Lambda_A(F_A)$ shows the symmetric property of the hyperbolic system, and it becomes visible only when the system is represented in with the variables $\Lambda_A$; therefore the variables $\Lambda_A$ are together called main field \cite{Struc05}.

Another approach to obtain a more reliable set of equations is a so-called regularization procedure, analogous to the current multipliers seen earlier regarding Eq.~\eqref{gkj}. In other words, one introduces additional relaxed variables (i.e., without time dependence). Here, they appear as auxiliary (relaxed) moments with a much faster time scale than the main field. For the detailed derivation, see \cite{Struc05}. The resulting model becomes parabolic (such as the GK equation). In the example of 13-moment equations, the regularization yields,
\begin{align}
	\dot q_i + \frac{5}{2}\rho T \frac{\partial T}{\partial x_i} + \frac{5}{2} \sigma_{ik} \frac{\partial T}{\partial x_k} - \sigma_{ik} T \frac{\partial \ln \rho}{\partial x_k} - \frac{\sigma_{ik}}{\rho} \frac{\partial \sigma_{kl}}{\partial x_l} + T \frac{\partial \sigma_{ik}}{\partial x_k} +& \nonumber\\ + \frac{7}{5} q_i \frac{\partial v_k}{\partial x_k} + \frac{7}{5} q_k \frac{\partial v_i}{\partial x_k}+ \frac{2}{5} q_k \frac{\partial v_k}{\partial x_i}+&\frac{1}{2} \frac{\partial R_{ik}}{\partial x_k} + \frac{1}{6} \frac{\partial \Delta}{\partial x_i} + m_{ikl} \frac{\partial v_k}{\partial x_l} =-\frac{2}{3} \frac{p}{\mu} q_i, \label{grad3} \\
	\dot \sigma_{ij} + \frac{4}{5} \frac{\partial q_{\langle i}}{\partial x_{\rangle j}} + 2 \sigma_{k \langle i} \frac{\partial v_{j \rangle}}{\partial x_k} + \sigma_{ij} \frac{\partial v_k}{\partial x_k} +& 2 \rho T \frac{\partial v_{\langle i}}{\partial x_{j\rangle}} +\frac{\partial m_{ijk}}{\partial x_k}= -\frac{p}{\mu} \sigma_{ij},\label{grad4} 
\end{align}
\begin{align} 
	m_{ijk} = \frac{4}{3} \frac{\sigma_{\langle i j} q_{k \rangle}}{p} - 2 \mu T \frac{\partial}{\partial x_{\langle i}} \left (\frac{\sigma_{jk \rangle}}{p}\right ), \quad R_{ij} = \frac{20}{7} \frac{\sigma_{k \langle i}\sigma_{j \rangle k}}{\rho} + \frac{64}{25} \frac{q_{\langle i} q_{j\rangle}}{p} - \frac{24}{5} \mu T \frac{\partial}{\partial x_{\langle i}} \left ( \frac{q_{j\rangle}}{p}\right), \nonumber
\end{align} \vspace{-0.15cm}
\begin{align} 
	\Delta = 5 \frac{\sigma_{kl} \sigma_{kl}}{\rho} + \frac{56}{5} \frac{q_k q_k}{p} - 12 \mu T \frac{\partial}{\partial x_k} \left ( \frac{q_k}{p} \right ), \label{grad5}
\end{align}
for which we have kept the original notations for the additional moments following for the sake of comparability and valid only for Maxwell molecules. The set of equations \eqref{grad3}-\eqref{grad5} are denoted with R13, and taking the additional moments ($m_{ijk}$, $R_{ij}$, and $\Delta$) to be zero, it reduces to Eqs.~\eqref{grad1}-\eqref{grad2}. These relaxed variables appear as a combination of the lower-order moments contrary to the current multipliers, and the coefficients strongly depend on the molecule model. For a more exhaustive systematic overview of the kinetic models, we refer to Struchtrup \cite{Struc05, RhaStruc16, StrTah11} and Cercignani \cite{Cercignani00b}. Furthermore, it is also possible to derive the 13-moment equations in the framework of GENERIC, providing further insights into the proper thermodynamic formulation and, in parallel, avoiding the original shortcomings (e.g., loss of hyperbolicity) \cite{StrucOtt22}. An alternative approach to stabilize the momentum equations and find a stable version of the Burnett equation is presented in \cite{SinghEtal17}.

For monatomic gases, Rational Extended Thermodynamics inherits the hierarchical structure of moment expansion, therefore, it becomes possible to obtain identical evolution equations. However, the difference occurs in the closure as RET does not attempt to rely on any particular distribution function or its approximations but is based on Galilean covariance and entropy principle with Liu's procedure. Concerning Grad's 13-moment model, Eqs.~\eqref{grad1}-\eqref{grad2}, these approaches can result in a completely identical outcome \cite{RugSug15}.

The situation changes for polyatomic gases due to the additional degrees of freedom for which the existing single-hierarchy ($F$-series) is replaced with a double-hierarchy ($F$ and $G$-series),
\begin{alignat}{2}\label{ret1}
	\partial_{t} F + \partial_{k} F_{k} & = 0, & \quad & \quad
	\nonumber \\
	\partial_{t} F_{i} + \partial_{k} F_{ik} & = 0, & \quad & \quad
	\nonumber \\
	\partial_{t} F_{ij} + \partial_{k} F_{ijk} & = S_{ij}, & \quad \partial_{t} G_{ll} + \partial_{k} G_{llk} & = 0, \nonumber \\
	& \quad & \quad \partial_{t} G_{lli} + \partial_{k} G_{llik} & = Q_{lli}, 
\end{alignat}
where the densities can be expressed in terms of physical variables as follows:
\begin{alignat}{2}\label{ret4}
	F & = \rho, & \quad & \quad  \nonumber\\
	F_{i} & = \rho v_{i}, & \quad & \quad  \nonumber\\
	F_{ij} & = \rho v_{i} v_{j} + (p + \Pi) \delta_{ij} + \Pi_{\langle ij \rangle}, & \quad
	G_{ll} & = \rho v^{2} + 2 \rho e, \nonumber \\
	& \quad & \quad G_{lli} & = \rho v^{2} v_{i} + 2 (\rho e + p + \Pi) v_{i} + 2 \Pi_{\langle ki \rangle} v_{k} + 2 q_{i},   
\end{alignat}
thus the $G$-hierarchy becomes independent implying that $G_{ll} \neq F_{ll}$, viz., $2\rho e \neq 3p$, moreover, these elements of the $G$-hierarchy represent Galilean transformation rules for energy and heat flux \cite{Van17gal}. Consequently, that structure also enables the treatment of the dynamic pressure $\Pi$ as a new field variable. We note that the size of the $G$-hierarchy must always be smaller by $1$ than the $F$-hierarchy to respect Galilean covariance. It has particular importance in modeling rarefied gases, especially when $\omega ~\sim 1/\tau$, hence $\Pi$ has its evolution equation together with a relaxation time. Its simplest version is called Meixner's theory \cite{Meix43a}. In RET, Meixner's theory is called ET6 ($6$-field) theory, in which the Euler fluid model is extended with a relaxation equation in the rarefied gas limit, it reads
\begin{align}
	\dot \rho + \rho \frac{\partial v_k}{\partial x_k} &=0, \label{ret2} \\
	\rho \dot v_i + \frac{\partial p}{\partial x_i} + \frac{\partial \Pi}{\partial x_i} &=0, \\
	\dot T + \frac{2 m}{D k_{\textrm{B}} \rho} \Big ( p + \Pi \Big )\frac{\partial v_k}{\partial x_k} &=0, \\
	\tau_{\Pi} \dot \Pi + \Pi + \eta \frac{\partial v_k}{\partial x_k} + \frac{5 D -6}{3D}\Pi \tau_\Pi \frac{\partial v_k}{\partial x_k} &=0, \label{ret3}
\end{align}
where the relaxation time $\tau_\Pi $ and the bulk viscosity $\eta$ are given as a function of degrees of freedom $D$,
\begin{align}
	\tau_\Pi = \frac{2 p T (D-3)}{3 D \zeta}, \quad \eta = \frac{2 (D-3)}{3 D}p \tau_\Pi, \quad p = \frac{k_{\textrm{B}}}{m} \rho T, \quad e = \frac{D}{2}  \frac{k_{\textrm{B}}}{m} T, \quad \zeta = \zeta(\rho,T) >0 \ \ \textrm{for any}\ \  \{\rho,T\}.
\end{align}
It is visible that Meixner's theory reduces to the Euler fluid for monatomic gases as $D=3 \Rightarrow \tau_\Pi=0, \eta=0 \Rightarrow \Pi=0$ \cite{AriEtal12c, Arietal15}. The system can only be obtained from kinetic theory if the phase space of $f$ is extended with a variable denoted by $I$ and that is related to (but not identical to) the internal degrees of freedom \cite{Verhas97}, thus $f=f(t, \mathbf x, \mathbf c, I)$, together with a non-negative measure $\varphi(I) \textrm{d} I$, and $\varphi(I) = I^n$, $n=(D-5)/2$, called state density. It yields
\begin{align}
	2 \rho e = \int\displaylimits_{-\infty}^\infty \int\displaylimits_0^\infty (m C^2 + 2 I) f(t, \mathbf x, \mathbf c, I) \varphi(I) \textrm{d} I \textrm{d} \mathbf c, \quad
	3(p + \Pi) =  \int\displaylimits_{-\infty}^\infty \int\displaylimits_0^\infty m C^2 f(t, \mathbf x, \mathbf c, I)  \varphi(I) \textrm{d} I \textrm{d} \mathbf c.
\end{align}
Sadly, it is proved that the Eqs.~\eqref{ret2}-\eqref{ret3} cannot adequately model rarefied gases, thus further extensions are necessary. Since the $13$-moment equation proved to have stability and convergence issues, it is reasonable to consider the ET14 theory (14-momentum equations) as the next candidate.
One has to consider the $F$ and $G$-hierarchies plotted in Eqs.~\eqref{ret1} and \eqref{ret4}, and thus the balances are
\begin{align} 
	\dot{\rho} + \rho \partial_{k} v_{k} &= 0, \label{ret00} \\
	\rho \dot{v}_{i} + \partial_{j} \left[ (p + \Pi) \delta_{ij} + \Pi_{\langle ij \rangle} \right] &= 0_{i}, \\
	\rho \dot{e} + \left[ (p + \Pi) \delta_{ij} + \Pi_{\langle ij \rangle} \right] \partial_{i} v_{j} + \partial_{i} q_{i} &= 0,
\end{align}
and their closure results 
\begin{align} \label{ret5}
	\dot{\Pi} + \left( \frac{2 \hat{c}_{v} - 3}{3 \hat{c}_{v}} p  + \frac{5 \hat{c}_{v} - 3}{3 \hat{c}_{v}} \Pi \right) \partial_{k} v_{k} + \frac{2 \hat{c}_{v} - 3}{3 \hat{c}_{v}} \Pi_{\langle ik \rangle} \partial_{\langle i} v_{k \rangle}
	- \frac{5}{3} \frac{1}{(1 +\hat{c}_{v})^{2}} \frac{\textrm{d}\hat{c}_{v}}{\textrm{d}T} q_{k} \partial_{k} T + \frac{2 \hat{c}_{v} - 3}{3 \hat{c}_{v}(1 + \hat{c}_{v})} \partial_{k} q_{k}  &= - \frac{1}{\tau_{\Pi}} \Pi, \\
	\dot{\Pi}_{\langle ij \rangle} + \Pi_{\langle ij \rangle} \partial_{k} v_{k} + 2 \partial_{k} v_{\langle i} \Pi_{\langle j \rangle k \rangle}+ 2 (p + \Pi) \partial_{\langle i} v_{j \rangle} - \frac{2}{(1 + \hat{c}_{v})^{2}} \frac{\textrm{d}\hat{c}_{v}}{\textrm{d}T}
	\partial_{k}T q_{\langle i} \delta_{j \rangle k}+ \frac{2}{1 + \hat{c}_{v}} \partial_{\langle j} q_{i \rangle}    &= - \frac{1}{\tau_{S}} \Pi_{\langle ij \rangle}, \label{ret6}
\end{align}
for the viscous and dynamic pressure with $\hat c_v = c_v m /k_{\textrm{B}}$, and for the heat flux,
\begin{align}\label{ret7}
	&  {\dot{q}_{i} + \frac{2 + \hat{c}_{v}}{1 + \hat{c}_{v}} q_{i} \partial_{k} v_{k}
		+ \frac{1}{1 + \hat{c}_{v}} q_{k} \partial_{i} v_{k}
		+ \frac{2 + \hat{c}_{v}}{1 + \hat{c}_{v}} q_{k} \partial_{k} v_{i}
	}  +\frac{k_{\mathrm{B}}}{m} \left\{
	(1 + \hat{c}_{v}) p \delta_{ki} + (2 + \hat{c}_{v})
	(\Pi \delta_{ki} + \Pi_{\langle ki \rangle}) \right\} \partial_{k}T \nonumber
	\\
	&  {\quad  - \frac{k_{\mathrm{B}}}{m} T \partial_{i}p + \frac{1}{\rho} \left\{ (p - \Pi)\delta_{ki} - \Pi_{\langle ki \rangle} \right\}
		\partial_{l} \left\{ (p + \Pi)\delta_{kl} + \Pi_{\langle kl \rangle} \right\}
		= - \frac{1}{\tau_{q}} q_{i}.
	}
\end{align}
The linearized form of Eqs.~\eqref{ret5}-\eqref{ret7} could have significant practical relevance, as it turned out from the evaluations of acoustic experimental data \cite{Arietal13}. ET14 has its counterpart from a continuum point of view \cite{KovEtal18rg}, however, their compatibility is not yet fully discovered in the nonlinear regime. Overall, the linearized-ET14 equations are
\begin{align} \label{ret8}
	\dot{\Pi} + \frac{2 \hat{c}_{v}^{(0)} - 3}{3 \hat{c}_{v}^{(0)}} p^{(0)} \partial_{k} v_{k}
	+ \frac{2 \hat{c}_{v}^{(0)} - 3}{3 \hat{c}_{v}^{(0)} (1 + \hat{c}_{v}^{(0)})} \partial_{k} q_{k}
	&= - \frac{1}{\tau_{\Pi}} \Pi, \\ \label{ret9}
	\dot{\Pi}_{\langle ij \rangle} + 2 p^{(0)} \partial_{\langle j} v_{i \rangle}
	+ \frac{2}{1 + \hat{c}_{v}^{(0)}} \partial_{\langle j} q_{i \rangle} 
	& = - \frac{1}{\tau_{S}} \Pi_{\langle ij \rangle}, \\ \label{ret10}
	\dot{q}_{i} - \frac{k_{\mathrm{B}}}{m} T^{(0)} \partial_{i}p 
	+ \frac{k_{\mathrm{B}}}{m} (1 + \hat{c}_{v}^{(0)}) p^{(0)} \partial_{i}T + \frac{p^{(0)}}{\rho^{(0)}} \partial_{k} \left\{ (p + \Pi) \delta_{ik}  \Pi_{\langle ik \rangle} \right\} &= - \frac{1}{\tau_{q}} q_{i},
\end{align}
for which the linearization is performed around a set of reference values $(\rho^{(0)}, v_{i}^{(0)}, T^{(0)}, \Pi=0, \Pi_{\langle ij \rangle}=0_{ij}, q_i=0_{i}) = \mathrm{const}$. Eqs.~\eqref{ret8}-\eqref{ret10} can exactly be reproduced with both EIT and NET-IV approaches \cite{KovEtal18rg}. The closure of the nonlinear model \eqref{ret00}-\eqref{ret7} yields the following (approximate) entropy density and entropy flux,
\begin{align}
	s(e, \rho, q_i, \Pi, \Pi_{\langle ij \rangle}) &= s_{\textrm{eq}}(e, \rho) - \frac{3 \hat c_v}{2(2 \hat c_v- 3) p T} \Pi^2 - \frac{1}{4 \rho T} \Pi_{\langle ij \rangle} \Pi_{\langle ij \rangle} - \frac{\rho}{2 p^2 T (1 + \hat c_v)} q_i q_i + \mathcal{O}(3), \label{ret11} \\
	J_k &= \frac{1}{T} q_k - \frac{1}{p T (1 + \hat c_v)} \Pi q_k - \frac{1}{p T (1 + \hat c_v)} q_i \Pi_{\langle ik \rangle} + \mathcal{O}(3), \quad 2 \hat c_v - 3 >0, \label{ret12}
\end{align}
in which the third-order non-equilibrium terms are neglected. Interestingly, while these entropy relations are found in the closure procedure as an outcome, EIT and NET-IV thermodynamic approaches use them as a starting point, constraining the evolution equations from the beginning. Hence Eqs.~\eqref{ret11}-\eqref{ret12} serve as a strong connection between various thermodynamic approaches and provide insight into the compatibility conditions. Additionally, from Eq.~\eqref{ret12}, the appearance of a current multiplier, the viscous pressure is clear, however, it must be separated into spherical and deviatoric parts in order to implement the proper coupling to the heat flux.

Further refining on the ET14 model is possible by splitting the variable $I$ into rotational $I^{\textrm{r}}$ and vibrational modes $I^{\textrm{v}}$, and that leads to a triple ($F$, $G$ and $H$) hierarchies, using $f=f(t, \mathbf x, \mathbf c, I^{\textrm{r}}, I^{\textrm{v}})$ with the corresponding state densities. It does not influence the derivation procedure, and the resulting equations again fulfill the second law of thermodynamics and Galilean covariance. For further details, we refer to the works of Ruggeri et al.~\cite{Arietal12, RuggSug21b}.

\subsubsection{Continuum models.} Although, in the light of knowing the relations \eqref{ret11}-\eqref{ret12}, it would be straightforward to achieve compatibility, at least in the linear regime, this is not the primary aim for a continuum approach. Kinetic theory and RET models are specific, valid only for ideal gases, and implement various molecule properties. On the contrary, the continuum approach can be more adaptable, facilitate the adaptation of measured state-dependent transport coefficients, and extend the region of validity towards the dense states. Moreover, in the analogy of the effective modeling we have seen concerning heterogeneous materials, the generalized NSF equations could be helpful in modeling two-phase flows for which the droplets in the gas flow behave like molecules in the rarefied flow, but these analogies are not yet experimentally tested. Thus, compatibility is a secondary but still important attribute in comparing and benchmarking different approaches, as it is a reasonable expectation that all (acceptable) theories should reproduce the modeling of certain phenomena. In that case, the compatibility is tested on the rarefied gas experiments mentioned earlier \cite{Kov18rg, KovEtal18rg, KovRogJou20}. 

The EIT approach for rarefied gas modeling results in a model more similar to the R13 equations than the system of ET14 since the dynamic pressure $\Pi$ is not among the state variables, but the current multipliers in the entropy flux are different \cite{CarMorr72a, CarMorr72b, LebCloo89},
\begin{align}
	\mathbf J_s = \frac{1}{T} \mathbf q + \alpha_1 \boldsymbol \Pi^{\textrm{dev}} \cdot \mathbf q + \alpha_2 \nabla \mathbf q \cdot \mathbf q + \alpha_3 \nabla  \boldsymbol \Pi^{\textrm{dev}} :  \boldsymbol \Pi^{\textrm{dev}}, \label{eit4}
\end{align}
and the non-local $\nabla \mathbf q$ and $\nabla  \boldsymbol \Pi^{\textrm{dev}}$ terms make the resulting evolution equations to be parabolic. Together with the $s=s(e, \rho, q_i, \Pi_{\langle ij \rangle})$ entropy density, the evolution equations are
\begin{align}
	\tau_q \dot {\mathbf q} + \mathbf q &= - \lambda \nabla T - \lambda_2  \boldsymbol \Pi^{\textrm{dev}} \cdot \nabla - \lambda_3 \Delta \mathbf q, \\
	\tau_S \dot{ \boldsymbol \Pi}^{\textrm{dev}} +  \boldsymbol \Pi^{\textrm{dev}} &= - \mu (\nabla \mathbf v)^{\textrm{dev}} - \mu_2 (\nabla \mathbf q)^{\textrm{dev}} - \mu_3 \Delta  \boldsymbol \Pi^{\textrm{dev}},
\end{align}
with the coefficients
\begin{align}
	\tau_q = \frac{3}{2} \tau_S, \quad \lambda = \frac{15}{4} p \tau_S, \quad \lambda_2 = \frac{3}{2} R T \tau_S, \quad \mu=2 p \tau_S, \quad \mu_2=\frac{4}{5} \tau_S, \quad \mu_3 = - RT \tau_S^2
\end{align}
inheriting the coefficients from the R13-moment equations \eqref{grad3}-\eqref{grad4}, except $\lambda_3$, which could also be found from the Eq.~\eqref{grad5}, and the nonlinear terms are not present here. Overall, both $\lambda_3$ and $\tau_S$ are to be fitted to experimental data. We note that such correspondence between the coefficients remains arbitrary, but for a rarefied gas model, it seems reasonable and advantageous to adopt the kinetic relations to reduce the number of free parameters from $6$ to $2$ (or $1$, if suitable).

\begin{figure}[]
	\centering
	\includegraphics[width=16.5cm,height=9.5cm]{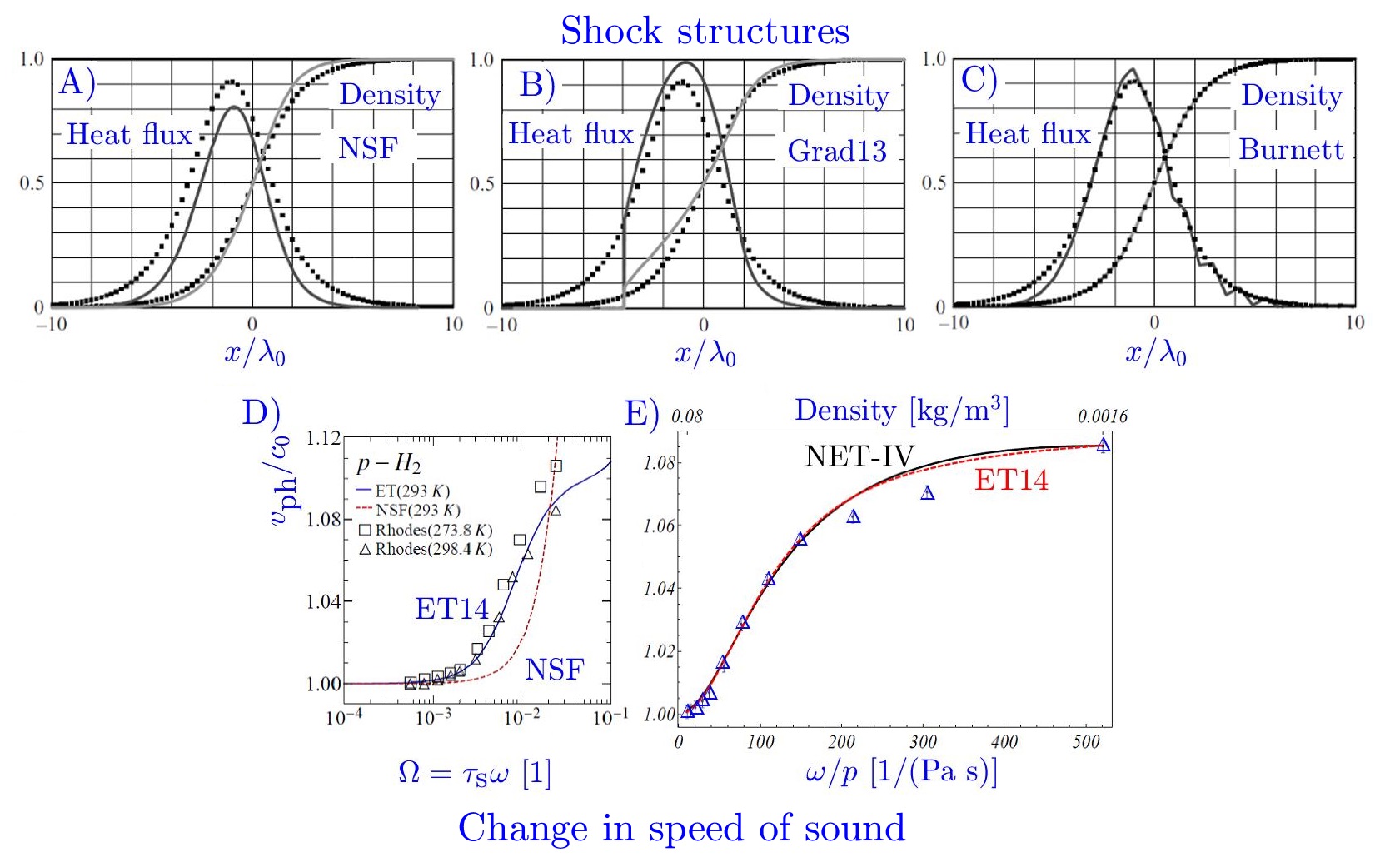}
	\caption{Model benchmarking for two different problems. First row: shock structure studies, in which the NSF (A), Grad13 (B), and Burnett (C) equations are compared to direct-simulation Monte Carlo data \cite{TorrStruc04}, $\lambda_0$ denotes the mean free path. Second row: velocity measurements compared to NSF, ET14 (D) \cite{Arietal13}, and NET-IV (E) theories \cite{Kov18rg}. Additionally, in part E), the original $\omega/p$ scaling is exchanged for mass density as the frequency is constant throughout the measurement.}
	\label{fig13}
\end{figure}

The situation becomes slightly more complicated for the NET-IV since the corresponding continuum model aims to achieve the ET14 level \cite{KovEtal18rg}. Consequently, the dynamic pressure is also present in the state space, $s=s(e, \rho, q_i, \Pi, \Pi_{\langle ij \rangle})$,  and $s=s_{\textrm{eq}}(e, \rho)- \frac{m_1}{2} q_i q_i -\frac{m_2}{2} \Pi_{\langle ij \rangle} \Pi_{\langle ij \rangle} - \frac{m_3}{6} \Pi_{ii} \Pi_{jj}$, where $m_1, m_2, m_3>0$ are positive coefficients and, similarly to the previous cases, they could also be a positive function of the state variables. Regarding the entropy flux, adopting \eqref{ret12} is also possible, leading to further nonlinearities. The potential of these latter (nonlinear) cases are not yet discovered; hence we apply only constant $m_i$ ($i=1,2,3$) coefficients with $J_i = (b_{\langle i j \rangle} + b_{kk}\delta_{ij}/3)q_j$, for which the second law provides the necessary closure to find $b_{\langle i j \rangle}$ and $b_{kk}$, still keeping the possibility to be compatible with \eqref{ret12}. Hence the Onsagerian relations are
\begin{eqnarray} \label{netiv1}
	-\rho m_1 \dot q_i +\frac{1}{3} \partial_i b_{kk}  + \partial_j b_{\langle ji \rangle }&=& n q_i, \\
	-\frac{1}{T} \partial_{\langle i} v_{j \rangle }- \rho m_2 \dot \Pi_{\langle ij \rangle}  &=& l_{11}\Pi_{\langle ij \rangle} + l_{12} \partial_{\langle i} q_{j \rangle }, \\
	b_{\langle i j \rangle } &=&l_{21} \Pi_{\langle ij \rangle} + l_{22} \partial_{\langle i} q_{j \rangle }, \\
	- \frac{1}{T} \partial_j v_j-\rho m_3 \dot \Pi_{ii} &=& k_{11} \frac{\Pi_{ii} }{3} + k_{12} \partial_i q_i, \\
	b_{kk} - \frac{1}{T} &=& k_{21} \frac{\Pi_{ii} }{3} + k_{22} \partial_i q_i. \label{netiv2}
\end{eqnarray}
where the coefficients must fulfill 
\begin{align}
	n\geq0, \quad l_{11} l_{22} - (l_{12}+ l_{21})^2/4 \geq0, \quad  k_{11} k_{22} - (k_{12} +k_{21})^2/4\geq0. \label{netiv3}
\end{align}
to preserve the positive definiteness of the entropy production. Compatibility can be achieved by making the system \eqref{netiv1}-\eqref{netiv2} to be hyperbolic with implying $l_{22}=k_{22}=0$ to eliminate the nonlocal terms. Therefore, to satisfy \eqref{netiv3}, $l_{12} = - l_{21}$ and $k_{12} = - k_{21}$ holds, and yields
\begin{align}
	\tau_q\dot q_i +q_i +\lambda \partial_i T - \alpha_{21} \partial_i \Pi_{kk} - \beta_{21} \partial_i \Pi_{\langle ij \rangle} &= 0,  \label{netiv4} \\ \label{netiv5}
	\tau_S \dot \Pi_{\langle ij \rangle} +\Pi_{\langle ij \rangle} + \mu \partial_{\langle i} v_{j \rangle} + \beta_{12} \partial_{\langle i} q_{j \rangle} &= 0, \\ \label{netiv6}
	\tau_\Pi \dot \Pi_{ii} +\Pi_{ii} + \eta \partial_i v_i + \alpha_{12} \partial_i q_i &= 0, 
\end{align}
for which the coefficients can either be adopted from the ET14 equations or to be fitted, and they constitute the following Onsagerian coefficients:
\begin{align}
	\tau_q = \frac{\rho m_1}{n}, \quad \tau_S = \frac{\rho m_2}{l_{11}}, \quad \tau_\Pi=\frac{3 \rho m_3}{k_{11}}, \quad 
	\lambda=\frac{1}{3 n T^2}, \quad \mu=\frac{1}{T l_{11}}, \quad  \eta=\frac{3}{T k_{11}},
\end{align}
\begin{align}
	\alpha_{12}=\frac{3 k_{12}}{k_{11}} \quad \alpha_{21}=\frac{k_{21}}{9n}, \quad \beta_{12}=\frac{l_{12}}{l_{11}}, \quad \beta_{21}=\frac{l_{21}}{n}.
\end{align}
We want to emphasize again that the NET-IV model \eqref{netiv4}-\eqref{netiv6} are valid only for constant coefficients, and the nonlinear options are not yet discovered. It is straightforward to accept that the NET-IV model offers much freedom, which freedom is not necessary to exploit completely. It remains a matter of choice how the particular modeling task requires, primarily when a measured state dependence of the transport coefficients must be implemented. The compatibility between these approaches is proved and extensively discussed in \cite{KovEtal18rg}. Figure \ref{fig13} presents two benchmarking situations concerning shock structures \cite{TorrStruc04, MadjarEtal14, Madjar15} and speed of sound measurements \cite{Kov18rg, Struc12}, comparing both the kinetic and continuum models to experimental data and detailed Monte Carlo simulations.

\section{Dual-phase-lag concept}

Tzou's original idea was to unify the various heat equations and find their common root on a constitutive level \cite{Tzou95}. Tzou observed the resemblance between the $T$-representation of various heat equations, including two-temperature models, and proposed the DPL concept as
\begin{align}
	\mathbf q(\mathbf x, t+\tau_q) = - \lambda \nabla T (\mathbf x,t+\tau_T), \label{dpl1},
\end{align}
assuming a delay between the heat flux and temperature gradient, contradicting the time homogeneity principle, and that Eq.~\eqref{dpl1} can approximate numerous heat equations, but not in the same way. While the first-order Taylor series expansion of the left hand side yields the MCV constitutive equation, the structure of the GK equation is recovered solely in its $T$-representation. Eq.~\eqref{dpl1} reduces to Fourier's law for $\tau_q=\tau_T=0$ but also reproduces Fourier's solutions for any $\tau_q=\tau_T$, likewise to the Fourier resonance condition \cite{FehEtal21}. Although Tzou mentions the possible microstructural effects behind non-Fourier heat conduction, it is also claimed that the time lags $\tau_q$ and $\tau_T$ are effective parameters, collectively modeling the corresponding microstructural phenomena on a macroscopic level \cite{Tzou95}. Therefore, in the absence of any (thermodynamic or kinetic) theory, it is unattainable to attach direct interpretation to these coefficients \cite{OaneEtal21}.

\subsection{Jeffreys heat equation.} There are two, partially valid approximations of Eq.~\eqref{dpl1}, the MCV equation ($\{1,0\}$-type DPL, i.e., first order in $\mathbf q$, and $\tau_T=0$); the second one is the Jeffreys equation ($\{1,1\}$-type) \cite{Jeffreys24, JosPre89}. The expression `partially' is reasonable since Eq.~\eqref{dpl1} has no such thermodynamic background, which has been developed for these models, and thus there are consequences, but first, let us present the Jeffreys equation. 

Its derivation is more common in NET-IV \cite{Verhas97}, compatible with GENERIC \cite{SzucsEtal21}, and cannot be derived within the framework of RET. The corresponding state space is $s=s(e, \mathbf y)$, where $\mathbf y$ ($\neq \mathbf q$) is an extensive vectorial state variable, and thus $s(e, \mathbf y) = s_{\textrm{eq}} (e) - \frac{m}{2} \mathbf y \cdot \mathbf y$ ($m>0$), and the classical entropy flux is applied ($\mathbf J_s = \mathbf q / T$). The resulting Onsagerian relations are
\begin{align}
	\mathbf q &= l_{11} \nabla \frac{1}{T} - l_{12} m \mathbf y, \\
	\rho \partial_t \mathbf y &= l_{21} \nabla \frac{1}{T} - l_{22} m \mathbf y,
\end{align}
with the requirements on $l_{ij}$,
\begin{align}
	l_{11} \geq 0, \quad l_{22} \geq 0,\quad l_{11} l_{22} - l_{12} l_{21} \geq 0.
\end{align}
After eliminating $\mathbf y$, the Jeffreys-type equation is formed,
\begin{align}
	\tau \frac{\partial \mathbf q}{\partial t}+ \mathbf q = \lambda_1 \nabla \frac{1}{T} + \lambda_2 \frac{\partial}{\partial t} \nabla \frac{1}{T}, \label{jeff1}
\end{align}
with the coefficients
\begin{align}
	\tau = \frac{\rho}{l_{22} m}, \quad \lambda_1 = \frac{\textrm{det} l_{ij}}{l_{22}}, \quad \lambda_2 = \frac{\rho l_{11}}{l_{22} m}=l_{11} \tau, \label{jeff2}
\end{align}
and remarkably, the $\{0,1\}$-type model is excluded in virtue of \eqref{jeff2}, and the MCV equation ($\{1,0\}$-type) is recovered when $l_{11}=0$. Furthermore, the Fourier resonance condition requires that 
\begin{align}
	\lambda_1 = \frac{\lambda_2}{\tau} \Rightarrow \frac{\textrm{det} l_{ij}}{l_{22}} = l_{11} \Rightarrow \frac{l_{12} l_{21}}{l_{22}} =0,
\end{align}
that is, the coupling between $\mathbf y$ and $\mathbf q$ vanishes, and $\mathbf y$ relaxes towards an asymptotically stable equilibrium point. Finally, we note that Rukolaine found nonphysical solutions for the Eq.~\eqref{jeff1} \cite{Ruk14, Ruk17}. However, in that test problem, a particular source term includes a Heaviside step function in time, which could fall from the physically admissible function space for $T$ and $\mathbf q$. Moreover, apparently, the initial conditions ($T=$const., and $\partial_t T=0$) are not thermodynamically compatible for a space-dependent source term, and thus further investigations and experimental testing might be necessary before excluding the Jeffreys heat equation from the physically admissible models. Nevertheless, the DPL concept introduces further disadvantages for higher-order expansions of \eqref{dpl1}.

The $T$-representation of Eq.~\eqref{jeff1} formally resembles to the GK equation in the linear regime, i.e., when $2 \lambda_2 T^{-3} \partial_t T \nabla T$ is omitted from the last term, and all coefficients are constants, thus
\begin{align}
	\tau \partial_{tt} T + \partial_t T = \alpha_1 \Delta T + \alpha_2 \partial_t \Delta T +  \frac{Q_v}{\rho c_v} + \frac{\tau}{\rho c_v} \partial_t Q_v, \quad \alpha_{1,2} = \frac{\hat \lambda_{1,2}}{\rho c_v}, \quad \hat \lambda_{1,2}=\frac{\lambda_{1,2}}{T^2}.
\end{align}
The $q$-representation also bears a similar structure,
\begin{align}
	\tau \partial_{tt} \mathbf q + \partial_t \mathbf q  = \alpha_1 \nabla \nabla \cdot \mathbf q + \alpha_2 \nabla \nabla \cdot \partial_t \mathbf q - \alpha_1 \nabla Q_v - \alpha_2 \nabla \partial_t Q_v.
\end{align}
These forms are valid only for linear models with constant coefficients, otherwise it is not suggested to eliminate any field variables. For higher-order approximations of Eq.~\eqref{dpl1}, the model structure is preserved, only higher-order time derivatives enter either the $T$ or the $q$-representations.

\subsection{Stability conditions.} Although finding a constitutive equation capable of connecting or unifying many other heat equations is interesting, it also has limitations and shortcomings. First, the Taylor series expansion on both sides remains arbitrary, and its convergence is not proven. Even the meaning of the relaxation times can change with the order of expansion, however, every expansion approximates the same model \eqref{dpl1}. Due to the lack of thermodynamic background, its compatibility with the second law remains questionable and could change with the expansion order. 
Further aspects emerge as follows.
\begin{itemize}
	\item Nonlinearities: according to the second law of thermodynamics, the coefficients are not independent. Here, no functional connections are established, therefore, it is not recommended to apply the model for, e.g., $T$-dependent coefficients.
	\item Anisotropy: the tensorial properties of each physical quantity become crucial for anisotropic materials, e.g., the scalar relaxation times become a second-order tensor, together with the thermal conductivity. It could allow further couplings and result in a complex time evolution equation for non-Fourier heat equations. Starting with Eq.~\eqref{dpl1} restricts the modeling capabilities on the isotropic behavior.
	\item Coupled problems: thermodynamics provides a consistent approach in deriving coupled equations such as thermo-diffusion \cite{GrooMaz63non} and thermo-mechanics \cite{IgnOst09b}. Due to the missing background of Eq.~\eqref{dpl1}, the DPL approach could only be helpful for pure heat conduction problems. 
	\item Time shift paradox: sadly, Eq.~\eqref{dpl1} directly contradicts the homogeneity of time; only the difference between the two relaxation times should be essential \cite{KovVan18dpl}. This is in agreement with Rukolaine \cite{Ruk14}, Fabrizio and Franchi \cite{FabFra14}, and it turned out that $\tau_T< \tau_q$ must be satisfied to have a well-posed model. Otherwise, stability and ill-posedness issues arise.
\end{itemize}
Because of these shortcomings, the region of validity of the DPL equations is firmly limited. For instance, many papers present experimental data evaluations using the DPL model without checking the stability conditions. The positivity of relaxation times is not sufficient. The $\{2,1\}$-type DPL, i.e., it is second-order for $\mathbf q$, and first-order for $\nabla T$,
\begin{align}
	\frac{\tau_q^2}{2} \frac{\partial^2 \mathbf q}{\partial t^2} + \tau_q \frac{\partial \mathbf q}{\partial t} + \mathbf q = - \lambda \nabla T -\lambda \tau_T \frac{\partial}{\partial t}\nabla T, \label{dpl2}
\end{align}
has stable solution if \cite{QuinRacke06},
\begin{align}
	\frac{\tau_T}{\tau_q} \geq \frac{1}{2}. \label{dpl3}
\end{align}
For a $\{2,2\}$-type DPL, the stability condition \eqref{dpl3} modifies to  \cite{FabLaz14a, FabEtal16},
\begin{align}
	2 + \sqrt{3}>\frac{\tau_T}{\tau_q} > 2 - \sqrt{3}, \label{dpl4}
\end{align}
in other words, the stability conditions are different for any approximation of Eq.~\eqref{dpl1}, and one must be cautious when choosing a seemingly suitable approximation. The over-diffusive region ($\tau_T>\tau_q$) is preferred by \eqref{dpl3} but begins to be strongly restrained. For further mathematical analysis, we refer to the study of Shen and Zhang \cite{ShenZhang08}. 

Now let us turn our attention to the asymptotic behavior of the DPL model with a series expansion up to arbitrary orders of $n$ and $m$,
\begin{align}
	\mathbf q + \frac{\tau_q}{1!}\partial_t \mathbf q + \frac{\tau_q^2}{2!} \partial_t^2 \mathbf q + \dots+\frac{\tau_q^n}{n!}\partial_t^n \mathbf q = - \lambda \left ( \nabla T + \frac{\tau_T}{1!}\partial_t  \nabla T + \frac{\tau_T^2}{2!} \partial_t^2  \nabla T + \dots + \frac{\tau_T^m}{m!} \partial_t^m  \nabla T \right).
\end{align}
It turned out by investigating the characteristic equations of the resulting operators that if $n$ (or $m$) is $\geq 5$, then there will be at least one pair of complex conjugate roots with positive real parts, i.e., instability persists without any conditions \cite{ChiritaEtal17}. In other words, the Taylor series expansion of such constitutive equations is not convergent.

\begin{figure}[]
	\centering
	\includegraphics[width=11.5cm,height=6.0cm]{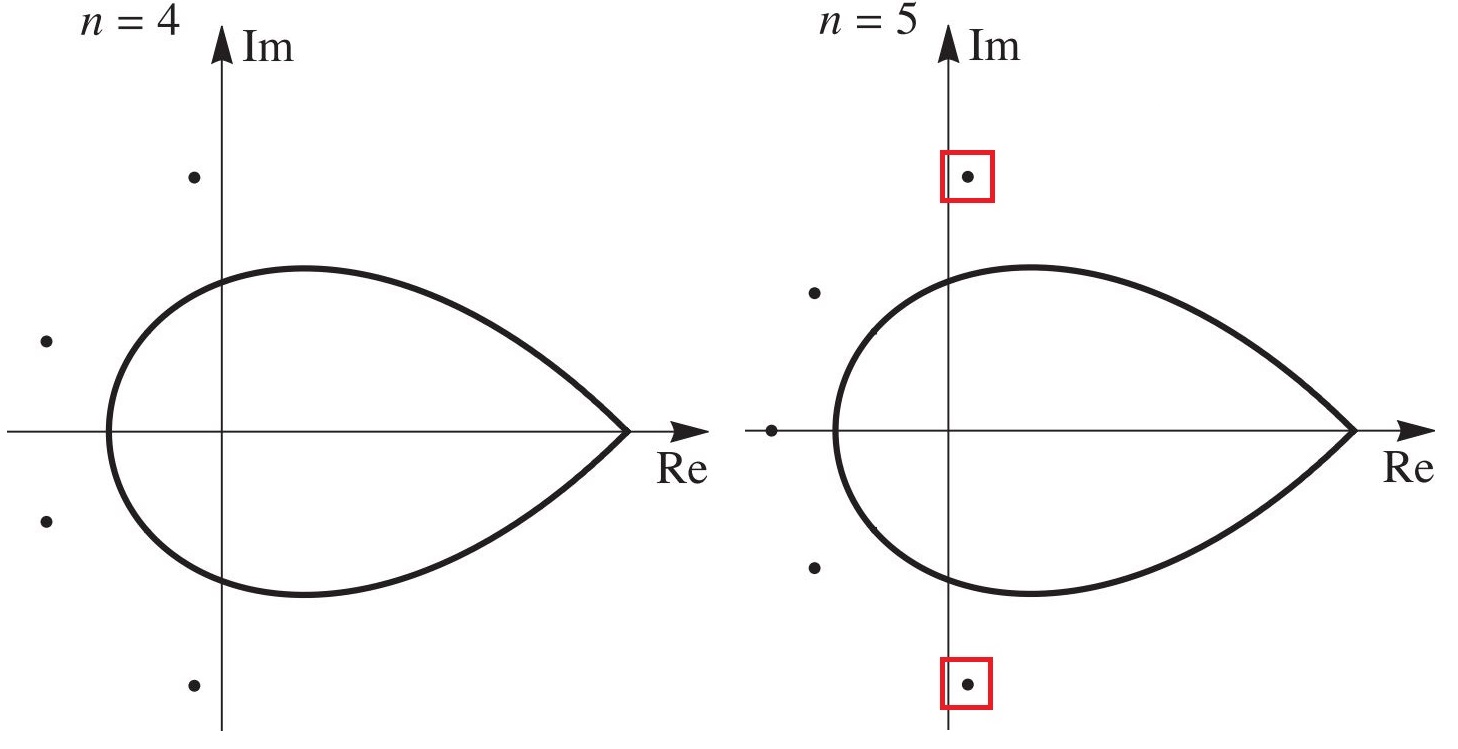}
	\caption{Roots of the characteristic equations. Left: all roots have negative real parts. Right: after $n\geq 5$, a root with a positive real part appears and remains (red squares highlight them), always resulting in instability \cite{ChiritaEtal17}.}
	\label{fig16}
\end{figure}

The further merit of \cite{ChiritaEtal17} emerges by investigating all situations up to the order of \{4,4\}-type DPL from a thermodynamic compatibility point of view, exploiting the fading memory concept of Gurtin and Pipkin \cite{GurPip68}. The $n,m\geq5$ condition does not mean that all combinations of $n,m<5$ are thermodynamically compatible. This is studied in view of the Clausius-Duhem inequality formulated for cycles with period $p=2 \pi/\omega$,
\begin{align}
	\int\displaylimits_{0}^p \mathbf q(t) \cdot \nabla T(t) \textrm{d}t \leq 0, \label{dpl+1}
\end{align}
with $\nabla T(t-s) = \mathbf f \cos\big(\omega(t-s)\big) + \mathbf q \sin\big(\omega(t-s)\big)$, and $\omega>0$, $\mathbf f^2 + \mathbf g^2>0$.

We have already shown that the \{1,0\} and \{1,1\}-types are thermodynamically compatible. Furthermore, the \{0,1\}-type, which can be rewritten as 
\begin{align}
	\nabla T (\mathbf x, t) = - \frac{1}{\tau_T} \int\displaylimits_{0}^\infty e^{-s/\tau_T} \lambda \mathbf q (\mathbf x, t-s) \textrm{d} s,
\end{align}
is still thermodynamically compatible for all $\tau_T>0$ in the light of \eqref{dpl+1}. However, the \{2,0\}-type model cannot satisfy \eqref{dpl+1},
\begin{align}
	\int\displaylimits_{0}^p \mathbf q(t) \cdot \nabla T(t) \textrm{d}t = - \frac{2 \pi \lambda (\mathbf f^2 + \mathbf g^2)}{\omega (\tau_q^4 \omega^4 + 4)} (2- \tau_q^2 \omega^2)
\end{align}
cannot preserve the sign for all $\omega>0$, hence thermodynamically incompatible, similarly to the \{0,2\}-type. Previous stability conditions can be achieved similarly \cite{ChiritaEtal17}. For higher-order models, the stability conditions become more and more complicated. For demonstration, we show only the condition for the \{3,3\}-type, and for the others, we refer to \cite{ChiritaEtal17} for convenience. The stability condition of the \{3,3\}-type model is
\begin{align}
	\sqrt{\left( b_1^2 + 3 a_1 c_1 \right)^3} \leq \frac{1}{2} \left( 27 a_1^2 d_1 - 2b_1^3 - 9a_1 b_1 c_1 \right ), \label{dpl+2}
\end{align}
with the coefficients as a function of $\tau_T/\tau_q$,
\begin{align}
	a_1 = 0.677637 \frac{\tau_T^3}{\tau_q^3}; \quad  b_1=\frac{\tau_T}{\tau_q}\left ( 4.06574 \frac{\tau_T^2}{\tau_q^2} - 6.09877 \frac{\tau_T}{\tau_q} + 4.06582 \right); \nonumber
\end{align}
\begin{align}
	c_1 = 12.1972 \left(\frac{\tau_T}{\tau_q} - 1 \right)^2 + 0.0003; \quad d_1=24.3944. \label{dpl+3}
\end{align}
So the conditions summed up in Table \ref{tab:dpl} are required to achieve thermodynamic compatibility. For well-posedness, proper initial and boundary conditions are also needed, and in that respect, we want to recall the difficulties mentioned concerning the MCV equation in Sec.~\ref{Analytical and numerical solutions.}. 
For a more thorough review of the DPL concept and its applications, we refer to \cite{Zhang09, LiuChen10, ShomEtal22, Tzou96, MaEtal19, YoussefEtal20}, and in regard to the well-posedness and stability properties, we refer to the papers of Quintanilla \cite{QuinRacke06c, Quin08, Quin09}.

\begin{table}[]
	\caption{Summary of the possible DPL models, based on \cite{ChiritaEtal17}, and \xmark \ means that the resulting model is unconditionally incompatible with thermodynamics.}
	\begin{tabular}{ccccc}
		Model type & Compatibility & Condition               & Model type & Compatibility \\
		\{0,0\}    & \checkmark    & -                       & \{2,0\}    & \xmark        \\
		\{1,0\}    & \checkmark    & $\tau_q>0$              & \{0,2\}    & \xmark        \\
		\{0,1\}    & \checkmark    & $\tau_T>0$              & \{3,0\}    & \xmark        \\
		\{1,1\}    & \checkmark    & $\tau_q>0$,  $\tau_T>0$ & \{0,3\}    & \xmark        \\
		\{2,1\}    & \checkmark    & $\tau_T\geq\tau_q/2$    & \{4,0\}    & \xmark        \\
		\{1,2\}    & \checkmark    & $\tau_q\geq\tau_T/2$    & \{0,4\}    & \xmark        \\
		\{2,2\}    & \checkmark    & $2 + \sqrt{3}>\frac{\tau_T}{\tau_q} > 2 - \sqrt{3}$    & \{3,1\}    & \xmark        \\
		\{3,2\}    & \checkmark    &  $ 1.4902 \tau_q \geq \tau_T > 0.28441 \tau_q$                        & \{1,3\}    & \xmark        \\
		\{2,3\}    & \checkmark    &  $ 1.4902 \tau_T \geq \tau_q > 0.28441 \tau_T$                       & \{0,0\}    & \xmark        \\
		\{3,3\}    & \checkmark    &     See Eqs.~\eqref{dpl+2}-\eqref{dpl+3}.                   & \{4,1\}    & \xmark        \\
		\{4,3\}    & \checkmark    &    See \cite{ChiritaEtal17}.                     & \{1,4\}    & \xmark        \\
		\{3,4\}    & \checkmark    &  See \cite{ChiritaEtal17}.    & \{4,2\}    & \xmark        \\
		\{4,4\}    & \checkmark    & See \cite{ChiritaEtal17}.  & \{2,4\}    & \xmark       
	\end{tabular} \label{tab:dpl}
\end{table}

\subsection{Generalized DPL model.} Practically, the DPL concept is the only non-Fourier model applied in biological heat conduction problems, therefore, it is inevitable to summarize the essential aspects here briefly. 
It is also crucial to emphasize that for a macroscopic heterogeneous system in a room temperature environment, such a model should only be interpreted as an effective model, the time lags are due to the interaction of multiple coupled phenomena. Therefore only over-diffusive solutions are expected, thermal waves (under-diffusive) do not exist under such conditions, and that is often misinterpreted in the literature so that $\tau_T > \tau_q$ should be satisfied for physically sound solutions, e.g., for copper, $\tau_T/\tau_q = 163$ \cite{Chand98}.

Let us recall the two-temperature model of Roetzel and Xuan \cite{XuRoe97},
\begin{align}
	\varepsilon \rho_b c_b \left ( \partial_t T_b + \mathbf v_b \cdot \nabla T_b \right) &= \nabla \cdot (\lambda_b \nabla T_b) + h (T_t - T_b), \label{dpl5a} \\
	(1-\varepsilon) \rho_t c_t \partial_t T_t &= \nabla \cdot (\lambda_t \nabla T_t) - h (T_t - T_b) + (1-\varepsilon) q_{\textrm{met}}, \label{dpl5b}
\end{align}
as it serves as the basis for the analysis of Zhang \cite{Zhang09}. After eliminating the blood temperature from Eqs.~\eqref{dpl5a} and \eqref{dpl5b}, with exploiting that there is an additional relaxation process between the tissue and blood following \cite{MinkoVaf99},
\begin{align}
	\frac{\varepsilon \rho_b c_b}{h} \partial_t T_b + T_b = T_t,
\end{align}
they obtain an evolution equation for the tissue,
\begin{align}
	\tau_q \partial_{tt} T_t + \partial_t T_t + \frac{\varepsilon \rho_b c_b}{(\rho c)_{\textrm{eff}}} \mathbf v_b \cdot \nabla T_t = \alpha_{\textrm{eff}} \Big( \Delta T_t + \tau_T \partial_t \Delta T_t \Big) + \left( \frac{1-\varepsilon}{(\rho c)_{\textrm{eff}}} + \frac{\varepsilon \rho_b c_b }{h (\rho c)_{\textrm{eff}}} \partial_t  \right) q_{\textrm{met}} \label{dpl6}
\end{align}
in which metabolic and convective heat transfer sources are included, and the acceleration of the blood flow is omitted. This is called the generalized DPL model. The effective quantities are
\begin{align}
	(\rho c)_{\textrm{eff}} = \varepsilon \rho_b c_b + (1-\varepsilon) \rho_t c_t, \quad \lambda_{\textrm{eff}} = \varepsilon \lambda_b + (1-\varepsilon) \lambda_t, \quad \alpha_{\textrm{eff}} = \frac{\lambda_{\textrm{eff}} }{(\rho c)_{\textrm{eff}} },
\end{align}
for which let us recall that $\varepsilon$ is the known porosity of a given tissue, and the indices refer to the blood and tissue properties. Here, the notable result is that in such a way, $\tau_q$ and $\tau_T$ are found (analogously to the Eq.~\eqref{ttm0}) as
\begin{align}
	\tau_q = \frac{\varepsilon (1-\varepsilon) \rho_b c_b \rho_t c_t }{h (\rho c)_{\textrm{eff}}  }, \quad \textrm{and} \quad \tau_T = \frac{\varepsilon (1-\varepsilon) \rho_b c_b \lambda_t }{h \lambda_{\textrm{eff}} }, \label{dpl7}
\end{align}
which also strengthens the observation about functionally connected coefficients, but -- for instance, in \cite{WangEtal23} -- it is not treated carefully, and therefore the physical consequences are omitted, \cite{WangEtal23b} is a positive counterexample. According to \cite{KhalVaf03}, 
\begin{align}
	- \varepsilon \rho_b c_b \mathbf v_b \cdot \nabla T_t \approx h ( T_b - T_t),
\end{align}
therefore Eq.~\eqref{dpl6} can be further simplified, and one does not need to know the blood velocity field. Zhang tested \cite{Zhang09} the relaxation times of Eq.~\eqref{dpl7} for various sets of realistic parameters and found that $\tau_T > \tau_q$ is satisfied for all cases, also confirmed by \cite{AfrZhang11}. That is a promising result and proves that the DPL concept can be helpful when the equations are handled properly. The model, sadly, is not directly solved and tested on experimental data.

However, $\tau_T > \tau_q$ is not always the case. Hooshmand et al.~\cite{Hoosetal15} implemented Eq.~\eqref{dpl6} (without Eq.~\eqref{dpl7}) to predict the transient temperature history under laser irradiation (the heating is included in a source term) and compared to experimental data. Significant deviation occurred between the observations and predictions. Besides, $\tau_q = 16$ s $>\tau_T = 0.05 \dots 16$ s varied, but $\tau_T$ always kept smaller than $\tau_q$, similarly as \cite{AfrinEtal12}. On the contrary, \cite{LiuChen10} found much better agreement with measurement data with parameters $\tau_T > \tau_q$. If one would continue the Taylor series expansion of that \{1,1\}-type model to \{2,1\} or \{2,2\}-types, then either the problem becomes ill-posed, or the solution becomes unstable. It is of crucial importance to keep clear the physically sound domains of coefficients. 

\section{Green-Naghdi models}
The unique concept of Green and Naghdi \cite{GreenNaghdi91} is based on three pillars. First, they introduce a so-called thermal displacement field $\alpha$ as
\begin{align}
	\alpha = \int\displaylimits_{t_0}^t T(\mathbf r, s) \textrm{d}s + \alpha_0, \quad \alpha_0 = \alpha(t_0),
\end{align}
resembling the fading memory concept of Gurtin and Pipkin, and $\alpha$ must be at least two times continuously differentiable both in space and time, and $\dot \alpha = T$ holds. Second, they modify the heat-entropy flux relation, i.e.,
\begin{align}
	\mathbf{J}_s = \frac{1}{\theta} \mathbf q, \quad \theta=\theta(T, \alpha; \Omega), \quad \Omega=\{a,b,d_1\}\in \mathbb{R}^+,
\end{align}
in which $\theta$ is a positive function and monotonous in $T$, and might include some positive constants as well \cite{BarSte05a} ($c$ is preserved to denote the specific heat). Third, the state space can be constructed in various ways, three of them is included in the original study of Green and Naghdi,
\begin{align}
	\mathbb{S}_{\textrm{I}}=(T, \nabla T), \quad \mathbb{S}_{\textrm{II}}=(\alpha, T, \nabla \alpha), \quad \mathbb{S}_{\textrm{III}}=(\alpha, T, \nabla \alpha, \nabla T),
\end{align}
for which the subscript distinguishes the model type. The type-I includes $\theta=T$, and thus leads to the classical Fourier equation. The type-II is more interesting, as it extends Fourier's law in a particular way,
\begin{align}
	\theta= a+ b \cdot T, \quad \psi = c(\theta - \theta \ln\theta) + \frac{k}{2} \nabla \alpha \cdot \nabla \alpha, \quad \mathbf{J}_s = - \frac{\rho}{\frac{\partial\theta}{\partial T}} \frac{\partial \psi}{\partial \nabla \alpha} = - \frac{\rho k}{b} \nabla \alpha,
\end{align}
where $k>0$ is also a thermal conductivity, and thus the heat flux is obtained through $\mathbf q = \theta \mathbf{J}_s = - (a+ b \cdot T)\frac{\rho k}{b} \nabla \alpha $ utilizing the particular form of Helmholtz free energy $\psi$. When the state space is extended with mechanical variables such as the right Cauchy-Green tensor \cite{BarSte07a, BargFav14}, then that approach is viable to derive a thermo-mechanical model in which the coupling is realized through $\alpha$. The nonlinear term $T \nabla \alpha$ is then omitted, and consequently, the type-II heat equation reads
\begin{align}
	\mathbf q = - \frac{a \rho k}{b} \nabla \alpha, \quad \ddot T = \frac{a k}{c b^2} \Delta T,
\end{align}
appearing as a wave equation for temperature, propagating without dissipation with the characteristic speed of $\sqrt{a k/(c b^2)}$. We emphasize that this occurs as a special case but is not forbidden in this framework and, therefore, cannot be compatible with any previous continuum or kinetic approaches. 
Accordingly, the internal energy density is also modified,
\begin{align}
	e= c \theta + \frac{k}{2} \nabla \alpha \cdot \nabla \alpha.
\end{align}
The setting of $\mathbb{S}_{\textrm{III}}$ includes 
\begin{align}
	\theta= a+ b \cdot T + d_1 \alpha, \quad \psi = \frac{k}{2} \nabla \alpha \cdot \nabla \alpha - \frac{d_2}{2} \alpha^2 - b_2 \alpha T - \frac{b_3}{2} T^2, \quad \mathbf{J}_s = - \frac{\rho}{\frac{\partial\theta}{\partial T}} \frac{\partial \psi}{\partial \nabla \alpha} = - \frac{\rho k}{b} \nabla \alpha
\end{align}
with the restrictions
\begin{align}
	k d_1 \geq 0, \quad b_2 b - b_3 d_1 \geq 0,
\end{align}
resulting in a heat equation
\begin{align}
	\mathbf q = - \kappa_1 \nabla \alpha - \kappa_2 \nabla T, \quad  \frac{\rho a}{b} (b_2 \dot \alpha + b_3 \ddot \alpha) = \kappa_1 \Delta \alpha + \kappa_2 \Delta \dot \alpha, \quad \textrm{or} \quad b_2 \dot T + b_3 \ddot T = \frac{b}{\rho a}(\kappa_1 \Delta T + \kappa_2 \Delta \dot T), \label{gn1}	
\end{align}
with recalling $\dot \alpha =T$, and the transport coefficients, $\kappa_1$ and $\kappa_2$, are positive. Remarkably, Eq.~\eqref{gn1} resembles the GK and Jeffreys equations in its $T$-representation, however, with a completely different constitutive background. These models are found to be well-posed \cite{LazzNibb08}. Regarding its finite element solution methods and further discussion, we refer to \cite{BarSte08, Barg13, GioEtal14}. Theoretically, the type-III model is applicable for second sound modeling and, with the mechanical coupling, could be valid in simulating ballistic heat conduction. In \cite{BarSte08}, a brief comparison with second sound experiments is made, but a thorough investigation of ballistic propagation has not been performed.

\subsubsection{Triple-phase-lag concept}
The idea comes from Choudhuri \cite{Choud07}, combining the Green-Naghdi and DPL models, particularly Eqs.~\eqref{dpl1} and \eqref{gn1}, assuming that three time lags are present,
\begin{align}
	\mathbf q (\mathbf x, t+\tau_q) = - \kappa_1 \nabla \alpha (\mathbf x, t+\tau_\alpha) - \kappa_2 \nabla T (\mathbf x, t+\tau_T), \label{tpl1}
\end{align}
with $0\leq \tau_\alpha < \tau_T <\tau_q$, i.e., immediately restricting the model validity for wave-like propagation. Again, analogously to the DPL model, it also becomes arbitrary at which order we choose to approximate Eq.~\eqref{tpl1} by a Taylor series expansion, therefore, the same stability issues might occur \cite{Quin08b}. 
In more detail, following \cite{Quin08b}, let us assume a first-order expansion in all three time lags, hence the $T$-representation of Eq.~\eqref{tpl1} is
\begin{align}
	\tau_q \rho c_v \partial_{t}^3 T + \rho c_v \partial_{t}^2 T = \kappa_1 \Delta T + \hat \tau_\alpha \partial_t \Delta T + \kappa_2 \tau_T \partial_{t}^2 \Delta T \label{tpl2}
\end{align}
with $\hat \tau_\alpha = \kappa_1 \tau_\alpha + \kappa_2$. If $\hat \tau_\alpha - \kappa_1 \tau_q <0$, unstable solutions emerge for a homogeneous $T=0$ boundary condition. In fact, the sufficient requirement is
\begin{align}
	\phi_1 > \frac{\rho c_v(\tau_q \kappa_1 - \hat \tau_\alpha)}{\kappa_2 \tau_T}, \label{tpl3}
\end{align}
where $\phi_1$ denotes the smallest eigenvalue of the Laplacian $-\Delta$ for the given homogeneous boundary condition. Consequently, the original assumption of $0\leq \tau_\alpha < \tau_T <\tau_q$ should be replaced with Eq.~\eqref{tpl3} for a first order approximation.
Similarly to the DPL model, the situation changes with the order of expansion. If one takes the second-order term as well in $\tau_q$, then a fourth-order term appears as
\begin{align}
	\frac{\tau_q^2}{2} \rho c_v \partial_{t}^4 T  +  \tau_q \rho c_v \partial_{t}^3 T + \rho c_v \partial_{t}^2 T = \kappa_1 \Delta T + \hat \tau_\alpha \partial_t \Delta T + \kappa_2 \tau_T \partial_{t}^2 \Delta T, \label{tpl4}
\end{align}
for which the solution will be always unstable if $\tau_q \hat \tau_\alpha > 2 \kappa_2 \tau_T$. If $\tau_q \hat \tau_\alpha < 2 \kappa_2 \tau_T$, but $\hat \tau_\alpha > \kappa_1 \tau_q$, then unstable solutions may also exist. In order to ensure stability, the necessary condition is
\begin{align}
	\kappa_1 \tau_q < \hat \tau_\alpha < \frac{ 2 \kappa_2 \tau_T  }{\tau_q},
\end{align}
so that the stability condition is far from being trivial, and thus the triple-phase-lag model suffers from the same weaknesses as the DPL approach. For  $\kappa_1=0$, the requirements are reduced to Eq.~\eqref{dpl3} \cite{Quin08b}. For further details on stability, we refer to \cite{ChiritaEtal16}.

To achieve a mechanical coupling, Choudhuri considered the first-order expansion of \eqref{tpl1}, and a thermal expansion source term in the energy balance (similarly to \eqref{emech}), thus Eq.~\eqref{tpl2} is extended as well,
\begin{align}
	\tau_q \rho c_v \dddot T + \rho c_v \ddot T + \gamma T_0 \ddot D + \tau_q \gamma T_0 \dddot D  = \kappa_1 \Delta T + \hat \tau_\alpha  \Delta \dot T + \kappa_2 \tau_T  \Delta \ddot T + \dot Q_v + \tau_q \ddot Q_v, \label{tpl5}
\end{align}
where the partial derivatives are exchanged with material time derivative, and $D$ denotes the dilatation. Eq.~\eqref{tpl5} becomes a mathematically and physically closed system together with the equation of motion, following the formalism of \cite{Choud07}, it reads
\begin{align}
	\rho \ddot {\mathbf u} = \mu \Delta \mathbf u + (\lambda + \mu) \nabla \nabla \cdot \mathbf u - \gamma \nabla T + \rho \mathbf f,
\end{align}
in which $\lambda$ and $\mu$ are the Lame coefficients, $\gamma = (3\lambda + 2\mu) \chi$ (recalling that $\chi$ is the thermal expansion coefficient), and $\mathbf u$ is the displacement field on which a volumetric force $\mathbf f$ might act as well.

For further details on well-posedness and stability, let us refer to the papers \cite{ApiceEtal16, MaganaEtal18, Quin08b}. We note that in \cite{Quin08b, MaganaEtal18}, further modifications are studied based on the work of Gurtin et al. \cite{ChenGurt68, GurtWill67}, namely, the triple-phase-lag model \eqref{tpl1} is combined with a special two-temperature approach. The model is based on the assumption that there are two distinct temperatures, one conductive ($\theta$) and one thermodynamic ($T$), connected through
\begin{align}
	T = \theta - a \Delta \theta, \quad a\geq 0, \label{tpl6}
\end{align} 
and $T$ is in the energy balance \eqref{ebal}, and $\theta$ substitutes $T$ in the constitutive equation \eqref{tpl1}. Moreover, analogously, the thermal displacement $\alpha$ is also exchanged with $v$ in \eqref{tpl1}, and
\begin{align}
	\alpha = v - a \Delta v, \quad a\geq 0 \label{tpl7}
\end{align} 
assumed to be valid. So that Eqs.~\eqref{tpl6} and \eqref{tpl7} are added to the model, resulting in a much more complex approach, and due to the Laplacian terms, these add further higher-order derivatives to the constitutive equation \eqref{tpl1}. For the corresponding stability and well-posedness conditions, we refer to \cite{MaganaEtal18} in which various sub-cases are studied.

\section{Outlook on further concepts}
While in the previous Sections, we systematically went through the models with thermodynamic origins, here we would like to look towards two further concepts having no deeper relationship to and direct compatibility with the second law of thermodynamics. 
These concepts are the thermomass and fractional derivatives, both falling outside the systematic structure of evolution equations, however, it does not exclude the possibility that the resulting model cannot be analogous with any of the previous ones. In parallel, we also emphasize that if a model fits into the systematic generalization of Fourier's law, it does not guarantee its physical admissibility (such as the DPL concept), but undoubtedly eases its interpretation, finding its solution and applications since the resulting model can inherit (at least partially) the existing methods.
Neither concepts of thermomass nor fractional derivatives can inherit any previous properties due to their initial characteristic assumptions from which the evolution equations are obtained, and that attribute distinguishes them from the previous models. Nevertheless, if one accepts their theoretical limitations, these approaches could still provide model equations with acceptable outcomes for specific heat conduction problems.

\subsection{Thermomass concept.}
Based on the work of Nie et al.~\cite{NieEtal20}, and Guo et al.~\cite{Guo2010general, GuoHou10}, the starting point originates from Einstein's mass-energy relation $E=m c^2$, which is assumed to be valid for the internal energy, too. In other words, they assume a thermomass expressed as
\begin{align}
	m_h = \frac{E_h}{c^2},  \quad \Rightarrow \quad \rho_h = \frac{\rho c_v T}{c^2}, \quad \mathbf u_h = \frac{\mathbf q}{\rho c_v T}, \label{tm1}
\end{align}
hence the thermomass gas is characterized with its density $\rho_h$ and the drift velocity $\mathbf u_h$. Analogously with phonons, the thermomass gas is constituted of thermons. Eq.~\eqref{tm1} identifies the mass balance of the thermomass gas directly with the balance of internal energy, 
\begin{align}
	\frac{\partial \rho_h}{\partial t} + \nabla \cdot (\mathbf u_h \rho_h) &= Q_m,\label{tm2}  \\
	\frac{\partial (\rho_h \mathbf u_h)}{\partial t} + \nabla \cdot (\rho_h  \mathbf u_h  \mathbf u_h) &= - \nabla P_h + \mathbf f_h, \label{tm3} \\
	\frac{\partial}{\partial t}\left (\frac{1}{2} \rho_h  \mathbf u_h^2 \right) + \nabla \left ( \mathbf u_h \cdot \frac{1}{2} \rho_h  \mathbf u_h^2 \right ) &= - \nabla P_h \cdot  \mathbf u_h + \mathbf f_h \cdot  \mathbf u_h. \label{tm4}
\end{align}
The pressure-like quantity, $P_h$, appearing in the balances of momentum \eqref{tm3} and kinetic energy \eqref{tm4}, is defined as
\begin{align}
	P_h = \frac{\gamma_h \rho}{c^2} (c_v T)^2, \label{tm5}
\end{align}
with $\gamma_h$ being the so-called Grüneisen constant, which could be used to take the thermo-mechanical effects into account as $\gamma_h$ depends on the thermal expansion coefficient and bulk modulus. However, due to the initial assumption \eqref{tm1}, this concept is restricted only to rigid heat conductors, no coupling is possible, and the mass of the thermomass gas becomes a constitutive quantity. Moreover, any volumetric heat source results in a $Q_m$ source term of the mass balance \eqref{tm2}, i.e., the thermomass cannot be a conserved quantity here.

\subsubsection{Flux limiters and heat choking} Fourier's law can be recovered with the assumption that the so-called resistance $\mathbf f_h$ is proportional to the drift velocity as $\mathbf f_h = - \mu \mathbf u_h$, and the entire left hand side of the momentum balance \eqref{tm3} is zero, therefore
\begin{align}
	\nabla P_h = \mathbf f_h, \quad \Rightarrow \quad \nabla \left ( \frac{\gamma_h \rho}{c^2} (c_v T)^2 \right ) = - \mu \frac{\mathbf q}{\rho c_v T} \quad \Rightarrow \quad \lambda = \frac{\gamma_h \rho^2 c_v^2 T^2}{c^2 \mu} >0, \label{tm6}
\end{align}
and thus Eq.~\eqref{tm6} can be used to determine $\mu$ if $\lambda$ is known. 
Consequently, when $\nabla \cdot (\rho_h  \mathbf u_h  \mathbf u_h)=0$ holds, Eq.~\eqref{tm3} can be transferred to the MCV equation with a relaxation time $\tau = \lambda \rho /(2 \gamma_h c_v^2 T)$. In case of $Q_m=0$, then Eq.~\eqref{tm3} provides the most general form within this framework,
\begin{align}
	\tau \frac{\partial \mathbf q}{\partial t} - c_v \mathbf l \frac{\partial T}{\partial t} + \mathbf l \nabla \mathbf q + \lambda(1-M) \nabla T + \mathbf q = 0, \quad M = \frac{1}{2 \gamma_h \rho^2 c_v^3 T^3} \mathbf q ^2, \quad \mathbf l =\tau \mathbf u_h, \label{tm7}
\end{align}
in which $\mathbf l$ is called the length vector, describing an intrinsic length scale of the conductor, and $M$ is a `thermal Mach number' of the drift velocity relative to the thermal wave speed, also introducing particular nonlinearities into the model, including both temperature and heat flux-dependent terms. One can naturally require that the effective thermal conductivity is strictly positive, 
\begin{align}
	\lambda_{\textrm{eff}} = \lambda(1-M) > 0, \quad \Rightarrow \quad 1-M >0,
\end{align}
therefore 
\begin{align}
	\mathbf q^2 < 2 \gamma_h \rho^2 c_v^3 T^3. \label{tm+1}
\end{align}
Eq.~\eqref{tm+1}, consequently, stands as a substantial condition that limits the flux from above, this is called 'flux limiter' in \cite{SellCimm14} and heat choking in \cite{WangEtal10, LijoEtal12}, expressing the flux dependence of the thermal conductivity. This is experimentally observed in a microchannel gas flow \cite{HaraEtal07}, showing an upper bound for the thermal conductivity. Eq.~\eqref{tm+1} expresses that the heat transfer ability of a nanoscale object under an increasing temperature gradient is limited. 

It is also worth noting that this is not an identical 'flux limiter' condition as the one that appears in the work of Levermore \cite{Levermore84}. In \cite{Levermore84}, similarly to Majumdar's model for nanoscale heat transfer \eqref{maj3}, particular approximations of the radiation transport equation for the specific photon intensity $I({\Omega}, \mathbf r, t)$,
\begin{align}
	\frac{1}{c} \partial_t I + {\Omega} \cdot \nabla I + \sigma I = \frac{c \sigma_a}{4 \pi } B + \sigma_s \int\displaylimits_{0}^{4 \pi} g({\Omega} \cdot {\Omega}') I({\Omega}')\textrm{d} {\Omega}', \label{tm+2}
\end{align}
in investigated, in which $\sigma_s(\mathbf r, t)$ and $\sigma_a(\mathbf r, t)$ are the scattering and absorption coefficients, $\sigma=\sigma_s+ \sigma_a$ holds, and $B(\mathbf r, t)$ is the black body energy density. The last term is called the scattering angular redistribution function. Since Eq.~\eqref{tm+2} is difficult to solve, momentum series expansion becomes suitable, analogously with the Boltzmann equation. Integrating Eq.~\eqref{tm+2} over the entire solid angle $\Omega$ leads to these momentum quantities, and thus such procedure leads again to an infinite hierarchy and demands a closure. If one takes the zeroth and first moments,
\begin{align}
	E (\mathbf r, t) = \frac{1}{c} \int\displaylimits_{0}^{4 \pi} I({\Omega}, \mathbf r, t) \textrm{d} {\Omega}, \quad \textrm{and} \quad F (\mathbf r, t) = \int\displaylimits_{0}^{4 \pi} \Omega I({\Omega}, \mathbf r, t) \textrm{d} {\Omega}, \label{tm+3}
\end{align}
the energy density $E (\mathbf r, t)$ and its flux $F (\mathbf r, t)$, they result in a diffusion equation
\begin{align}
	\partial_t E + \nabla \cdot F + c\sigma_a(E-B) = 0,
\end{align}
for which one needs to connect $E$ and $F$ in a thermodynamically compatible way. The simplest nontrivial relation is called Eddington approximation \cite{Levermore84, Eddington88b}, and reads
\begin{align}
	F = - k \nabla E, \quad k>0,
\end{align}
that is a diffusion approximation for the radiation transport equation \eqref{tm+2}. This does not necessarily satisfy the relation
\begin{align}
	|F| \leq c E, \label{tm+4}
\end{align}
following from Eq.~\eqref{tm+3}. If a diffusion equation satisfies \eqref{tm+4}, it is called flux-limited diffusion \cite{Levermore84}. Comparing \eqref{tm+4} to \eqref{tm+1}, it becomes clear that these flux-limited properties emerge from a rather different background and also influence the evolution equations differently. On the one hand, Eq.~\eqref{tm+1} is a nonlinear requirement concerning the state variables, also making the evolution equation nonlinear. On the other, \eqref{tm+4} is a linear requirement even for linear equations. Besides, we must keep in mind that \eqref{tm+4} follows from an approximation procedure (momentum expansion) of the photon transport model contrary to the thermomass model.

\subsubsection{Continuum approach} It is insightful to mention the attempt of Sellitto and Cimmelli \cite{SellCimm15} to find a continuum thermodynamic version of the thermomass equation \eqref{tm7} within the framework of EIT. They assumed that the classical state space is extended with a vectorial internal variable $\boldsymbol \chi$, and its time evolution equation obeys
\begin{align}
	\dot {\boldsymbol \chi} = \nabla \cdot \mathbf J_{\boldsymbol \chi} + \mathbf Q_{\boldsymbol \chi},
\end{align}
where $\mathbf J_{\boldsymbol \chi}$ and $Q_{\boldsymbol \chi}$ denote its flux and production terms, and
\begin{align}
	\mathbf Q_{\boldsymbol \chi} &= \Gamma_0(T) \boldsymbol \chi, \\
	\mathbf J_{\boldsymbol \chi} & = \Gamma_1(T) \mathbf I + \Gamma_2(T) \left (\frac{1}{2} |\boldsymbol \chi|^2 \mathbf I + \boldsymbol \chi \otimes \boldsymbol \chi \right ),
\end{align}
for which the regular scalar $T$-dependent functions $\Gamma_i$ ($i=0,1,2$) are found when comparing the resulting evolution equation to Eq.~\eqref{tm7}. Furthermore, they relate $\boldsymbol \chi$ to the heat flux through 
\begin{align}
	\mathbf q = g(T, |\boldsymbol \chi|^2) \boldsymbol \chi, \quad \Rightarrow \quad \mathbf J_s = \hat g (T, |\boldsymbol \chi|^2) \boldsymbol \chi,
\end{align}
where the functions $g$ and $\hat g$ are restricted by the second law of thermodynamics.
Although they could find the continuum counterpart of the thermomass mass equation \eqref{tm7} under certain assumptions, they concluded that it does not fit into the systematic structure of heat equations beyond the MCV equation. Such a model also lacks the ability to reproduce the size-dependent thermal conductivity properties for nanosystems found with the GK equation. Moreover, the effective thermal conductivity $\hat \lambda = \lambda(1-M)$ is also characteristic of the thermomass model. The heat flux dependence from $M$ of $\hat \lambda$ would result in further nonlinear terms in a continuum model, leading to $\tau=\tau(|\mathbf q|^2)$ on the contrary to Eq.~\eqref{tm7}. Finally, let us note that the thermomass model would have a much stronger background if the  balances were expressed by four-divergences of the four-densities of the extensive physical quantities, at least in a Galilean-relativistic framework \cite{Van17gal}. Hence the elements of the state space could be found as the corresponding time-, and space-like parts of a higher-order tensor. This work has begun recently \cite{SuGuo22}. For further reading and connection with kinetic theory, let us refer to \cite{DongEtal11}.

\subsection{Fractional derivative concept.}

Another way to generalize Fourier's law, Eq.~\eqref{f1}, is to use the fractional derivative approach \cite{ZecTer15, SierEtal13}, which formalism enjoys growing interest and has been found to be helpful in various situations. Despite its popularity, the physical heat and mass transport models are ad hoc as the derivative terms are arbitrarily transformed to fractional ones, and the compatibility with basic physical principles is not proved. For that reason, these models could be ill-posed or suffer from instability, as seen for the particular approximations of the DPL equation earlier. 

One of the simplest but enlightening problems is related to units. As units are represented by one-dimensional oriented vector spaces in mathematics \cite{Mato04b}, hence they must be handled with the same care as any other quantity, and it restricts both the physical and mathematical possibilities. Using (arbitrarily) fractional units for a fractional heat equation is inevitable, too. Otherwise, quantities from different vector spaces will be added, which is an apparent mathematical contradiction. However, physics cannot correctly interpret fractional thermal conductivity or time units. This is an apparent and unsolved contradiction that appears in recent papers of Vázquez et al.~\cite{Vazquez11, VazEtal11}, Carillo et al.~\cite{CarrilloEtal17}, and others \cite{SierociukEtal15}.

Fractional heat equations can either be fractional in space or time. As an example of the spatially fractional heat equation, the Fourier heat equation can be modified with a fractional Laplacian:
\begin{align}
	\frac{\partial T}{\partial t} = \alpha \frac{\partial^\beta T}{\partial x^\beta}, \quad 1<\beta\leq 2, \label{fd0}
\end{align}
however, for a $\beta^\textrm{th}$ order derivative, $\alpha^\beta$ would be more appropriate with unit of $m^{2 \beta}/s$, but it is no more thermal diffusivity. Thermodynamics constrains the interval on which $\beta$ is allowed, i.e., $\beta=1$ would mean a conservative, hyperbolic equation, therefore, that case is prohibited. Eq.~\eqref{fd0} is difficult to handle for practical engineering processes, and that model can be better understood as a heat equation with modified length scales on the analogy of the nonlocal DPL equation \cite{Tzou11, Zhmakin23b},
\begin{align}
	\mathbf q(\mathbf x+l_q, t+\tau_q) = - \lambda \nabla T(\mathbf x, t+\tau_T), 
\end{align}
or even 
\begin{align}
	\mathbf q(\mathbf x+l_q, t+\tau_q) = - \lambda \nabla T(\mathbf x + l_T, t+\tau_T), 
\end{align}
in which $l_q$ and $l_T$ are the additional length scales, further models are derived using series expansions again.

Nonlocality is present in the approach of Mongioví and Zingales \cite{MonZin13} in a unique way. They assume that the temperature evolution at a given spatial point is influenced by all others of the body and thus modify the energy balance,
\begin{align}
	\partial_t e(\mathbf x, t) = -\frac{1}{\rho (\mathbf x)} \nabla \cdot \mathbf q(\mathbf x, t) + \int\displaylimits_{V} \psi (\mathbf x, \mathbf y, t) \rho (\mathbf y) \textrm{d} V, \label{zin1}
\end{align}
in which the integral is meant to be along the coordinates $\mathbf y$ of the entire body, and the function $\psi$ is a long-range contribution field and is proportional to the temperature difference,
\begin{align}
	\psi( \mathbf x, \mathbf y, t) = \kappa_\alpha g(|| \mathbf x - \mathbf y||) [T(\mathbf y, t) - T(\mathbf x,t)],
\end{align}
where $\kappa_\alpha$ is assumed to be a material parameter, and $g$ is a spatially-decaying function in order to weaken the contribution with distance. Furthermore, while they allow the mass density to be space-dependent, it is assumed to be constant in time, however, it does not affect the outcome as later, they simplify with homogeneous mass distribution. Eq.~\eqref{zin1} is solvable for a given $g$, they applied
\begin{align}
	g(|| \mathbf x - \mathbf y||) = \frac{1}{d_n(\alpha)} \frac{1}{|| \mathbf x - \mathbf y||^{n+\alpha}}, \quad \alpha \in \mathbb{R}, \quad n\in\mathbb{N}
\end{align}
in which $d_n(\alpha)$ is a normalization constant depending on $\alpha$, i.e., on the order of differentiation. They also argue that any such strictly positive decaying function would be compatible with the second law of thermodynamics, resulting in positive entropy production.
That particular heat transfer term can be understood as a fractional-order source term, which keeps all the other derivatives unchanged. They also provide a solution technique \cite{MonZin13}.

Its counterpart, the time-fractional heat equation, actually modifies the time scales. Various modifications of Fourier and MCV equations can be found in the literature, arbitrarily depending on which fractional derivative type seems more suitable \cite{PodThi98b}.
For instance, one modification of the MCV equation is
\begin{align}
	q + \tau^\gamma \frac{\partial^\gamma q}{\partial t^\gamma} = - \lambda \frac{\partial T}{\partial x}, \label{fd3}
\end{align}
where $\gamma$ is called the anomalous diffusion exponent \cite{ComptMetz97}. It results in
\begin{align}
	\frac{\partial T}{\partial t} + \tau^\gamma \frac{\partial^{1+\gamma} T}{\partial t^{1+\gamma}} = D \frac{\partial^2 T}{\partial x^2} + \hat Q + \tau_\gamma \frac{\partial^\gamma Q}{\partial t^\gamma}. \label{fd4}
\end{align}

As expected, Eq.~\eqref{fd3} reduces to Eq.~\eqref{mcv1} at $\gamma \rightarrow 1$. 
The two-temperature models also have a fractional modification, proposed by Shen et al.~\cite{ShenEtal20},
\begin{align}
	C_\textrm{e} \tau_{\textrm{ph}}^{\gamma-1} \frac{\partial^\gamma T_\textrm{e}}{\partial t^\gamma} &=\lambda_\textrm{e} \frac{\partial^2 T_\textrm{e}}{\partial x^2} + \alpha_{\textrm{e,ph}}(T_{\textrm{ph}}-T_\textrm{e})+S, \label{fd1} \\
	C_{\textrm{ph}} \tau^{\gamma-1}_{\textrm{ph}} \frac{\partial^\gamma T_{\textrm{ph}}}{\partial t^\gamma} &=-\alpha_{\textrm{e,ph}}(T_{\textrm{ph}}-T_\textrm{e}) \label{fd2}
\end{align}
where $\tau_{\textrm{ph}}$ is the phonon mean free time and $\partial^\gamma/\partial t^\gamma$ denotes the Caputo-type derivative with fractional order $0<\gamma\leq1$ and the indices e and ph are for the electron and phonon carriers, respectively. Interestingly, the same relaxation time $\tau_{\textrm{ph}}$ is artificially added, however, with inappropriate power. Additionally, regarding the numerical analysis of such equations, \cite{ShenEtal20} provides a comprehensive basis, including stability analysis.
Eventually, any model can be transformed into a fractional order, implementing the possibility of making a model more general as the order of the derivative is found through a fitting procedure. 
While the Caputo-type derivative is more convenient to model fatigue or plasticity, the Caputo-Fabrizio-type derivative is more advantageous for viscoelasticity and heat conduction, avoiding singular kernels \cite{CapFab15}, similarly to the Atangana–Baleanu derivative \cite{AtanBal16}. These definitions proved to be more satisfactory and enjoying great interest recently. We also refer to \cite{FaiZin20} for a recent review of fractional models and their applications.

\section{Summary and further perspectives}
\begin{figure}[]
	\centering
	\includegraphics[width=16.5cm,height=9.0cm]{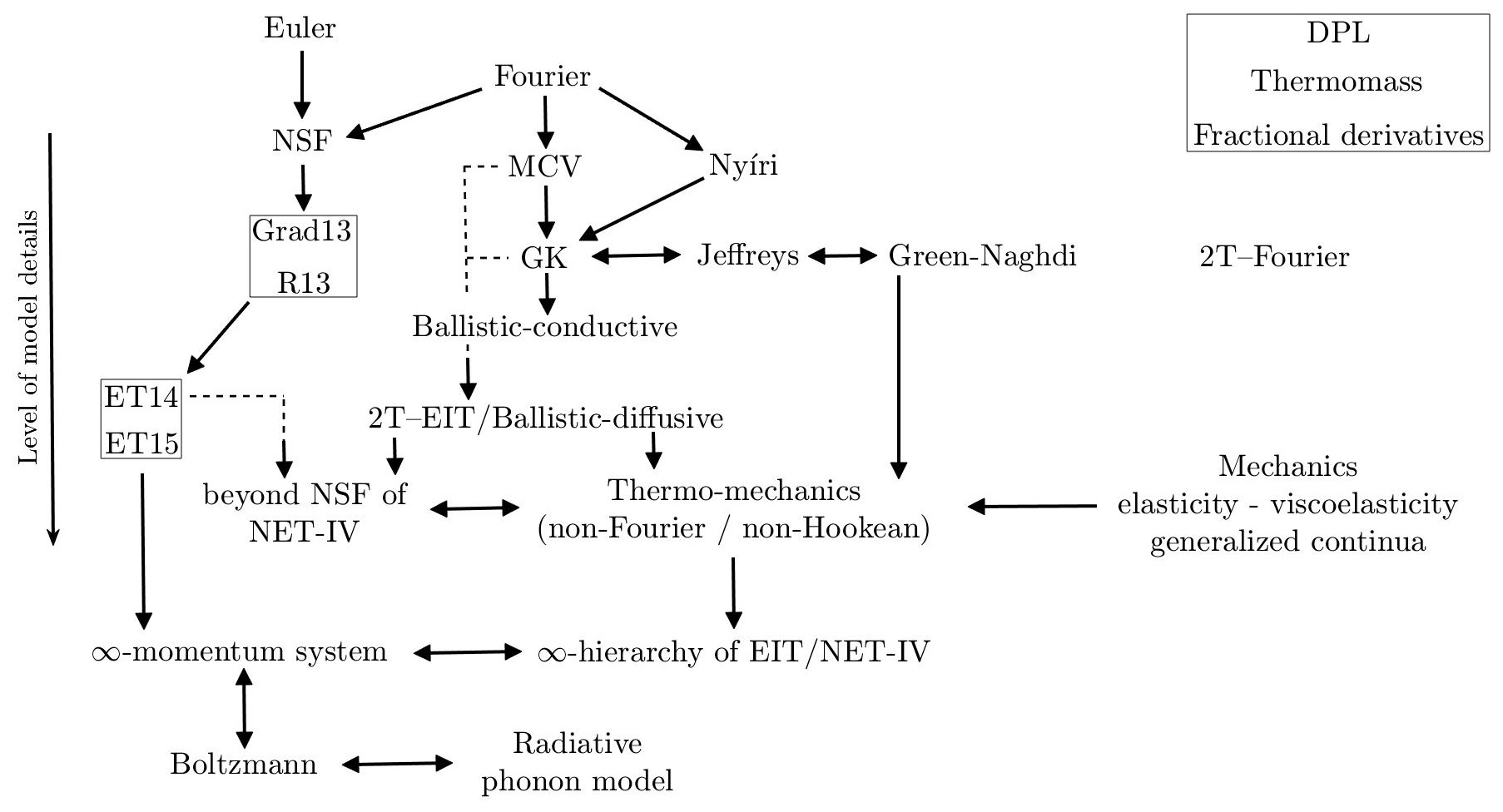}
	\caption{Hierarchy of evolution equations.}
	\label{fig11}
\end{figure}

After about 90 years of the pioneering works of Tisza and Landau \cite{Tisza38, Lan47}, the field of non-Fourier heat conduction significantly improved and severely influenced both thermodynamic theories and applications. In parallel, the emergence of various thermodynamic approaches made this field more colorful, and the number of possibilities is practically (countably) infinite. It also makes this research field challenging to overview and follow. The diverse branches of thermodynamics are evolving more or less independently, nevertheless, they all work with the same model structures. That structural compatibility allowed us to build up that systematic overview and proceed step by step from the simplest model to the most complicated approaches. Figure \ref{fig11} wants to reflect how we have gone through so far and provide a helpful guide to find the proper model we need for a particular problem. This field of research needs unification, and in that respect, we refer to the Special Issue 'Fundamental aspects of nonequilibrium thermodynamics' \cite{VanSI20}, which attempted to gather and overview the efforts towards a uniform framework, however, in a much more fundamental context than this review can show.
Heat conduction serves excellent problems for benchmarking to test various techniques and recognize which thermodynamic theory can be employed more efficiently with fewer assumptions. Indeed, an extensive benchmarking of the existing thermodynamic approaches would significantly facilitate the comparisons and unification, and that should be a comprehensive research program.

Naturally, not all models are highlighted in Fig.~\ref{fig11}, we focused on the well-known ones, and the arrows reflect their natural order with respect to model details. This is not identical with accuracy, e.g., the Boltzmann equation cannot solve every problem as it has solid limitations for validity, such as any other model. Although deciding which model is 'more detailed' is challenging, we obeyed a relatively simple principle to build that hierarchy: what features can be considered on a constitutive level. From that point of view, the Euler equation is the simplest one as it omits the constitutive relations of Fourier and Newton but still requires the equations of state. Then Fourier's law can be extended into three directions: adding the fluid equations leads to the classical NSF model, adding memory (or inertia) results in the MCV equation, and adding nonlocality leads to the Nyíri equation. Combining these builds the GK equation, which can be analogous to the Jeffreys model up to a certain extent since they differ on the constitutive level. Furthermore, the GK equation is included in the R13 model as a special sub-case and has such systematic generalization that can be compatible with ET14/15 equations. We have also seen that the Green-Naghdi approach can be compatible with Jeffreys/GK $T$-representation, which can easily be coupled to mechanics. 
The 2T-model with doubling and coupling Fourier's law, however, stands here as relatively independent due to the limitations of the Fourier equation but incorporates much more details than the single Fourier heat equation and effectively (in its $T$-representation), appears to be more general than the GK equation, but cannot be reduced to.
In this respect, the DPL, thermomass, and fractional derivative models do not fit into this structure by lacking a proper thermodynamic background. This is also the case for the triple-phase-lag models, suffering from the same stability issues as the DPL equations and strictly restricted to linear isotropic situations due to the missing constructive derivation and background.

We further extend that principle with nonlinearity aspects. While the kinetic models provide a rather strict prediction about the state dependence of transport coefficients, they remain limited as they require additional insight into molecular-level behavior. On the other hand, any measured nonlinear function can be implemented into a continuum model, and adding that the coefficients are free (i.e., can be fitted to experiments), the continuum models possess considerable potential to widen the applicability limits. This does not mean that a continuum model is always the best choice, as they might need to fit many more coefficients than the kinetic models and cannot predict their prior behavior. Hence one must choose carefully for the given problem, but the best case would be, again, a unification as much as possible. 

The next level is reached when the complete background of continuum mechanics is added to the thermal models. This points much beyond elasticity (although thermal expansion is the most 'fundamental' phenomenon for many problems) but incorporates the various generalizations of the classical continuum mechanics such as Cosserat media \cite{Mindlin65, Forest98, AltEre12, MauMet10}, and viscoelasticity (or rheology) \cite{Christensen82, FulEta14m1, Verhas97}. Adding that a conducting medium can be anisotropic (or even a specific phenomenon demands the presence of anisotropy such as piezoelectricity), it could either result in further compatibility conditions between different approaches or reveal further limits of the validity.

\begin{figure}[]
	\centering
	\includegraphics[width=11cm,height=7.0cm]{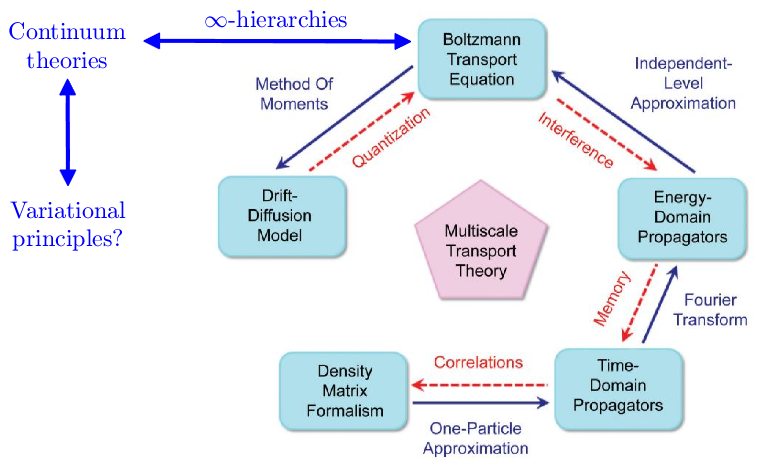}
	\caption{Formalism of multiscale transport theories, extended with continuum theories by virtue of Fig.~\ref{fig11} \cite{Walczak18}.}
	\label{fig15}
\end{figure}

Overall, apparently, the most advanced model would be a coupled, generalized thermo-mechanical(-electrical-chemical-diffusive) approach with the possibility to implement any nonlinearity one would need. On the one hand, that model would have immense application potential and might exclude numerous generalizations of constitutive equations. However, on the other hand, it would also be challenging to handle, even for the most straightforward problems. Hence, the proper one should be selected for a specific task, but the unification would highlight crucial compatibility properties. None of these approaches are negligible, but we must balance them and choose the proper formalism.
In that regard, it is worth showing the Figure \ref{fig15} from Walczak \cite{Walczak18}, extended with continuum theories by virtue of Fig.~\ref{fig11}, and quoting the closing thoughts of \cite{Walczak18}: "The ultimate goal of such multiscale transport theory (MTT) is unification of all the heat conduction mechanisms and microscopic scattering phenomena within the same conceptual framework [...]. Since nanoscale and macroscale systems differ from each other by length, energy, and time scales, the MTT should link the macroscopically measured quantities to the microscopic response of the system onto an external perturbation (like temperature difference)." Concerning the compatibility between various approaches and multiscale theories, the Boltzmann equation can serve as a common point, and compatibility with kinetic theory can ease that unification process. These thoughts are also expressed in the recent paper of Grmela \cite{Grmela23}.

Furthermore, we want to call attention to the underlying principles used to generate the model equations. Variational principles, although they play a fundamental role in physics, they work well only for non-dissipative systems. The emergence of dissipation, however, makes the development of a universal variational principle incredibly challenging. With the methodologies of modern thermodynamic approaches, surprisingly, one can derive Euler–Lagrange form and symplectic structure of the evolution equations for non-dissipative processes \cite{VanKov20a}. Additionally, the symplectic structure can ease the numerical implementation of evolution equations, too \cite{PortEtal17, Romero10I, Romero10II, ShanOtt20}.

In parallel, applying complex models requires a thorough investigation of the mathematical properties, especially well-posedness, stability, and, in connection, the initial and boundary conditions. Exploiting the second law helps to fulfill stability and well-posedness; moreover, it could also be helpful for numerical methods to track the stability properties of a system under time evolution. It is associated with the thermodynamically compatible definitions of initial and boundary conditions. While the tensorial quantities seem easy to implement in a one-dimensional setting, this could be incredibly difficult to handle in a two-, or even a three-dimensional setting. For instance, the nonlocal terms in the GK equation reduce to a single second-order spatial derivative, thus, do not seem to have difficulties. Still, it becomes challenging for a general three-dimensional problem. Similarly to the objectivity properties, in a one-dimensional setting, one might not realize that significant difficulties can emerge later.
It can be even more problematic for higher-order momentum expansions as the tensorial order continuously increases. It is not straightforward what type and how many (independent) boundary conditions they need and how to implement all of them in a thermodynamically compatible way. The situation is similar for initial conditions, and it is still an open question how to initialize a system from a non-equilibrium state. Adding that there is a need for an efficient and reliable numerical technique (more probably based on multi-field finite elements), these stand as substantial barriers in front of transferring any non-Fourier equation into the engineering practice. 

Concerning stability, we must mention again the relativistic dissipative fluids -- the relativistic generalizations of the NSF system -- as the existence of stable equilibrium stands as a central issue here and is closely connected to the hyperbolic and causality attributes \cite{Olson90}. The first relativistic generalization by Eckart \cite{Eckart40} is found to be unstable, predicting a blow-up for the temperature history \cite{Wodarzik82, HisLin85a, HisLin88a}. This issue can be resolved, at least partially, with the approach of Israel and Stewart \cite{IsrSte79}, the relativistic version of Extended Irreversible Thermodynamics, where the linearized equations are hyperbolic. However, the resulting properties notably depend on what flow frame one chooses \cite{BiroEtal08, Van09a}, because the stability is strongly connected to the form of thermal dissipation. The simple heuristic nonrelativistic ideas are not valid in the relativistic framework. It is demonstrated clearly in the solution of the paradoxical question of relativistic temperatures: energy and momentum cannot be separated in a covariant theory \cite{BirVan10a}. Using the Eckart frame, hence assuming that the four-velocity is parallel with the particle flow, leads to a restricted domain of hyperbolicity. Beyond that limit, the model becomes parabolic without a stable equilibrium solution \cite{HisLind87, HisOls89}. On the other hand, the energy frame, called the Landau-Lifshitz frame, behaves favorably without a heat flux, causality and stability properties are conditionally restored \cite{Olson90}. Eventually, that topic alone would deserve a separate review paper since all modern thermodynamic schools have also developed their methodology to handle relativistic problems. Without elaborating the details, let us refer the reader to the following literature: for RET \cite{RuggSug21b, Rugg23, SalZan23}, EIT \cite{PavonEtal80, PavonEtal82}, NET-IV \cite{Van17gal, VanBir12a}, GENERIC \cite{Ottinger98, GavAnt23}, thermomass \cite{SuGuo22}, SHTC \cite{RomenEtal20}, and \cite{DenRisc22b} provides a deep insight into the methods and theories related to relativistic fluids and the relativistic Boltzmann equation.

Despite the numerous differences between the modern thermodynamic frameworks, much was accomplished in the previous decades, and non-Fourier heat equations point far beyond the first low-temperature observations on liquid helium. Both the macro and nanoscale directions are promising and continuously developing, and all approaches consist of valuable elements and open the way to new ideas. As we can see, the golden age of non-Fourier heat equations is still ahead of us, nonetheless, we stress that unification (and hence agreement between the thermodynamic schools) and uniform concepts in relativistic-nonrelativistic, and quantum-classical frameworks along decisive benchmark problems will be inevitable, and we hope that the present review can aid that process.

\section{Acknowledgement}
The author is thankful for Péter Ván and Mátyás Szücs for the useful and fruitful discussions. The author is also thankful for the Editorial Office for their help in the submission and review process.

All figures and tables included in the present paper are used with permissions confirmed by the publishers and/or by the authors.

\section{Funding}
Project no. TKP-6-6/PALY-2021 has been implemented with the support provided by the Ministry of Culture and Innovation of Hungary from the National Research, Development and Innovation Fund, financed under the TKP2021-NVA funding scheme. The research was funded by the Sustainable Development and Technologies National Programme of the Hungarian Academy of Sciences (FFT NP FTA), also supported by the János Bolyai Research Scholarship of the Hungarian Academy of Sciences, and by the National Research, Development and Innovation Office-NKFIH FK 134277.


\bibliographystyle{unsrt}
\footnotesize

\end{document}